\documentstyle[preprint,aps,floats,epsf]{revtex}

\newcommand{\ttbar}{t\overline{t}}
\newcommand{\qqbar}{q\overline{q}}
\newcommand{\bcc}{\begin{center}}
\newcommand{\ecc}{\end{center}}
\newcommand{\beqn}{\begin{eqnarray}}
\newcommand{\eeqn}{\end{eqnarray}}
\newcommand{\beqy}{\begin{equation}}
\newcommand{\eeqy}{\end{equation}}
\newcommand{\mt}{m_t}
\newcommand{\mste}{m_{\tilde t_1}}
\newcommand{\mstz}{m_{\tilde t_2}}
\newcommand{\ste}{\tilde t_1}
\newcommand{\stz}{\tilde t_2}

\newcommand{\stl}{\tilde t_L}
\newcommand{\str}{\tilde t_R}
\newcommand{\msbe}{m_{\tilde b_1}}

\newcommand{\astop}{A_{\tilde t}}
\newcommand{\phist}{\Phi_{\tilde t}}

\newcommand{\tanb}{\tan{\beta}}
\newcommand{\ma}{M_{A^0}}
\newcommand{\mkh}{M_{h^0}}
\newcommand{\mgh}{M_{H^0}}
\newcommand{\mhp}{M_{H^\pm}}

\newcommand{\mchari}{M_{\tilde{\chi}^\pm_i}}
\newcommand{\mneut}{M_{\tilde{\chi}^0}}
\newcommand{\mneuti}{M_{\tilde{\chi}^0_i}}
\newcommand{\qqa}{\mbox{$q\overline{q}$ annihilation}}
\newcommand{\ggf}{\mbox{gluon fusion}}

\newcommand{\shit}[2]{\epsfxsize=#1 \epsfbox[10 30 560 790]{./#2}}

\begin{document}


\preprint{$
\begin{array}{l}
\mbox{Fermilab-Pub-97/164-T}\\[-3mm]
\mbox{KA-TP-9-1997}\\[-3mm]
\mbox{May~1997} \\ [1cm]
\end{array}
$}

\title{Top Pair Production at Hadron Colliders in non-minimal 
Standard Models\\[1cm]}

\author{W.~Hollik$^a$, W.M.~M\"osle$^a$
and D.~Wackeroth$^b$\footnote{E-mail:~dow@fnth09.fnal.gov}\\[0.5cm]}

\address{$^a$Institut f\"ur Theoretische Physik, Universit\"at Karlsruhe,
D-76128 Karlsruhe, Germany\\
$^b$Theory Group,
Fermi National Accelerator Laboratory, Batavia, IL 60510, USA\\[2cm]}
\maketitle
\tightenlines
\begin{abstract}
The cross section of top pair production in hadronic collisions
to ${\cal O}(\alpha\alpha_s^2)$ is calculated
within the General 2-Higgs-Doublet Model
and the Minimal Supersymmetric Standard Model.
At the parton level the ${\cal O}(\alpha)$ one-loop corrections
to the main production mechanisms, 
$\qqbar\rightarrow \ttbar$ and $gg\rightarrow \ttbar$,
significantly modify the Born-cross sections:
in the threshold region $\sqrt{\hat s}\stackrel{>}{\sim}2 m_t$
they are enhanced
up to $50 \%$ and with increasing cm energy $\sqrt{\hat s}$,
they can be reduced by up to the same order of magnitude.
In a wide range of the parameter space
of the models under consideration 
the observable hadronic cross sections for top pair
production at the Tevatron $p\overline{p} \rightarrow \ttbar X$
and at the LHC $p p \rightarrow \ttbar X$
are typically reduced by several percent ($\stackrel{<}{\sim}10 \%$)
compared to the lowest order result.
In special regions of the parameter space,
that is in the vicinity of the threshold for the top quark
decay $t\rightarrow b  \; H^+$ ; $\tilde t \; \tilde \chi^0$,
the radiative corrections are considerably enhanced,
comparable in size to QCD effects.
\end{abstract}


\newpage


%
\section{Introduction}
The discovery of the top quark in 1995 by the
CDF~\cite{cdf} and D\O\ \cite{dzero} experiments
at the Fermilab Tevatron once again impressively confirmed the
Minimal Standard Model (MSM)~\cite{msm,hig} as a valid description of 
electroweak particle interactions up to presently accessible energies.
In the meantime, experimental effort has been concentrated on
detailed studies of the top quark properties.
Presently, the top quark mass is known to be (world average)~\cite{wine}  
\[ m_t = 175.6 \pm 5.5 \; \mbox{GeV} \; ,\] 
which is consistent with the
MSM prediction obtained by performing a global MSM-fit to all
available electroweak precision data~\cite{ewwg}
\[ \mt = 172.7 \pm 5.4\; \mbox{GeV} \; .\]
The measurement of the top pair production cross section $\sigma_{\ttbar}$
at hadron colliders is a significant test of the Standard Model.
The observation of deviations from the Standard Model 
predictions, including electroweak and QCD corrections,
could indicate new non-standard production or decay
mechanisms. The measurement of $\sigma_{\ttbar}$ performed
at the Tevatron~\cite{cdfnew,wine}
\begin{eqnarray}
\mbox{CDF} &:&  \sigma_{\ttbar}(m_t= 175\;\mbox{GeV})
 = 7.5^{+1.9}_{-1.6} \; \mbox{pb}
\nonumber\\
\mbox{D\O\ } &:&  \sigma_{\ttbar}(m_t=170 \; \mbox{GeV})
= 5.77 \pm 1.76 \; \mbox{pb} \nonumber
\end{eqnarray}
is largely 
in good agreement with the theoretical QCD prediction~\cite{berg,cat,laen}
\begin{eqnarray}
\sigma_{\ttbar}(m_t=175 \; \mbox{GeV}) = 5.52^{+0.07}_{-0.42} \; \mbox{pb}
&,&
\sigma_{\ttbar}(m_t=170 \;\mbox{GeV}) =  6.48^{+0.09}_{-0.48} 
\; \mbox{pb}
\nonumber\\
\sigma_{\ttbar}(m_t=175 \; \mbox{GeV}) &=& 4.75^{+0.63}_{-0.68}
  \;  \mbox{pb}
\nonumber\\
\sigma_{\ttbar}(m_t=170 \; \mbox{GeV}) &=& 5.83^{+0.85}_{-0.51}
  \;  \mbox{pb}   \; .
\nonumber
\end{eqnarray}
The complete MSM electroweak one-loop corrections to
top pair production at hadron colliders are calculated as well~\cite{diplpubsm}
but have only little impact on $\sigma_{t\overline t}$ at the
Tevatron ($\sim 1\%$).
The value of $\sigma_{\ttbar}$ measured by the
CDF collaboration is slightly higher than the theoretical prediction but
the study of systematic uncertainties and the combination of
different decay channels is still in progress~\cite{cdfnew}.
The current large experimental uncertainty, however, still
leaves room for non-standard physics effects through
the virtual presence of new particles, for instance.
At the upgraded Tevatron with ${\cal L} = 10 fb^{-1}$
a measurement of $\mt$
and of the total $\ttbar$ production rate with a
precision of $\delta m_t=2$ GeV and of
$\delta \sigma_{\ttbar}/\sigma_{\ttbar}=6\%$, respectively,
is within reach~\cite{topacc}.
At the LHC the mass measurement is expected to be 
more accurate due to higher statistics
and the cross section measurement will at least be
as precise as that performed at the Tevatron since
the Tevatron-estimate is already to a large extent 
limited by systematic uncertainties~\cite{topacc}.

The envisaged high precisions open a new rich field of top quark
phenomenology. Here we are going to concentrate on the
implications of non-standard electroweak-like
radiative corrections, for the top pair
production cross section in hadronic collisions.  

Since the mechanism which introduces
gauge boson masses in a gauge invariant way
by spontaneously breaking the electroweak symmetry~\cite{hig}
is the least experimentally 
explored sector of the MSM, possible extensions of the Higgs-sector
are of particular interest.
There, the consideration of a second Higgs-doublet plays a 
special role~\cite{hunters}:
\begin{itemize}
\item
the extension to two Higgs-doublets
represents a minimal extension of the MSM Higgs-sector
as far as the number of newly introduced parameters is concerned,
\item
as a new physical phenomenon, charged Higgs-bosons occur,
\item
the requirement $\rho=\frac{M_W^2}{c_W^2 M_Z^2}=1$
at Born-level is still fulfilled and
\item
two Higgs-doublets are required in a minimal 
supersymmetric extension of the MSM. 
\end{itemize}
Supersymmetry (SUSY)~\cite{susy} represents an additional symmetry between 
fermions and bosons which implemented in the MSM 
solves such MSM deficiencies like the
hierarchy problem, the necessity of fine tuning 
and the non-occurrence of gauge coupling
unification at high energies. 

Past studies of non-minimal Standard Model implications on
the top pair production cross section at hadron colliders
comprised the calculation of
\begin{itemize}
\item
the ${\cal O}(\alpha)$ one-loop corrections within the
General 2-Higgs-Doublet Model (G2HDM) to both the $\qqa$ and
gluon fusion subprocesses
with numerical results for the Tevatron~\cite{ttmssm} and the
LHC~\cite{ttthdm,dipl},
\item
the SUSY QCD ${\cal O}(\alpha_s)$ contribution
to the $\qqa$ and gluon fusion subprocesses with numerical results for
the Tevatron~\cite{susyqcd,qcdew} and the LHC~\cite{sqcdlhc}, and
\item
the SUSY electroweak-like (EW-like) one-loop corrections
to the $\qqa$ subprocess
with numerical results for the Tevatron~\cite{qcdew,ewlike}.
\end{itemize}
The electroweak radiative corrections are of special interest due to the
strong Yukawa-couplings of the top quark to the Higgs-bosons.
While the MSM prediction of the electroweak one-loop contribution
does not exceed $\sim 3\%$ of the Born-cross section~\cite{diplpubsm,ttsm}
the models discussed here involve the interesting possibility of
an enhancement of the Yukawa-couplings and of additional 
contributions through the virtual presence of supersymmetric particles.  
Thus, we give a complete description of
the top pair production cross sections to ${\cal O}(\alpha \alpha_s^2)$ of
both main production mechanisms, $\qqa$ and gluon fusion,
within the G2HDM (= Standard Model with two Higgs-doublets but without
imposing SUSY constraints) and the Minimal Supersymmetric
Standard Model (MSSM) which involves the following
one-loop contributions:
\begin{itemize}
\item
the electroweak gauge boson contributions ($W$ and $Z$ boson exchange),
\item
the Higgs-boson contributions within the G2HDM,
\item
the EW-like MSSM contribution, where the
contribution of the supersymmetric Higgs-sector
and the SUSY EW-like one-loop corrections are discussed separately.
\end{itemize}
We provide explicit analytical expressions for the
form factors which parametrize the one-loop
modifications of the $g\ttbar$-vertex,
present numerical results at the parton level and
give a detailed discussion of their numerical significance  
at the upgraded Tevatron with $\sqrt{S}=2$ TeV and at the LHC
with $\sqrt{S}=14$ TeV.
\section{Top pair production in non-minimal Standard Models}
The main production mechanism for $\ttbar$ production
at the Tevatron is
the annihilation of a quark-antiquark pair 
\[ q(p_4) + \overline{q}(p_3) \rightarrow t(p_2) + \overline{t}(p_1) \]
whereas at the LHC the top quark pairs are mainly produced
via the fusion of two gluons
\[ g(p_4) + g(p_3) \rightarrow t(p_2) + \overline{t}(p_1) \; .\]
At the parton level, the corresponding differential cross sections 
to order ${\cal O}(\alpha \alpha_s^2)$ 
are obtained by contracting the
matrix elements describing the ${\cal O}(\alpha)$ contribution
to these subprocesses $\delta {\cal M}_i, i=\qqbar, gg$ 
with the Born-matrix elements ${\cal M}_B^i$
\begin{equation}\label{parton}
\delta \frac{d \hat \sigma_i(\hat t,\hat s)}{d \hat t}
= \frac{1}{16 \pi^2 \hat s} 2 {\cal R}e \; \overline{\sum}
(\delta {\cal M}_i \times {\cal M}_B^{i*}) \; ,
\end{equation}
where $\hat t=(p_3-p_1)^2,\hat s=(p_3+p_4)^2$ are Mandelstam variables.
The explicit representations of $\delta {\cal M}_i$ as well as ${\cal M}_B^i$
are given in~\cite{diplpubsm} (Eqs.(3.34,3.39)), where
the ${\cal O}(\alpha)$ contribution within the MSM
to both production mechanisms $\qqa$ and $\ggf$
has been studied. Throughout this paper we
closely follow the notation of~\cite{diplpubsm}
and refer to it for more details.
Here we study the modification of the form factors introduced
in~\cite{diplpubsm}
by non-minimal Standard Model ${\cal O}(\alpha)$ contributions.
In order to reveal the numerical effect of these corrections
on the top pair production cross sections and to study the dependence on
the parameters of the underlying model
we introduce a relative correction $\Delta_i$ at the parton level
\begin{equation}\label{deltapart}
\hat \sigma_i(\hat s)=\hat \sigma_B^i(\hat s)+\delta \hat \sigma_i(\hat s) = 
\sigma_B^i (1+\Delta_i) \; ,
\end{equation}
where we already carried out the $\hat t$-integration.
To compare with non-standard results we show in Fig.~1
the relative corrections $\Delta_{\qqbar}$ and $\Delta_{gg}$ obtained
within the MSM. There, the  mass of the MSM Higgs-boson $M_H$ 
is assumed to be within
bounds resulting from the negative searches at LEP
$M_H \ge 66$ GeV~\cite{higexp} 
and from theoretical arguments based on unitarity 
$M_H \le 1$ TeV~\cite{higtheo}.
In the numerical evaluation the electroweak MSM parameters 
are chosen to be~\cite{pdg,ewwg,wmass}:
\[ \mt = 175 \, \mbox{GeV}, m_b=4.7 \, \mbox{GeV} , M_W = 80.356 \, \mbox{GeV}
, M_Z= 91.1863 \, \mbox{GeV} , \alpha^{-1} = 137.035989 \]
\begin{figure}[htb]
\begin{center}
\setlength{\unitlength}{1cm}
\setlength{\fboxsep}{0cm}
\begin{picture}(16,7)
\put(-1.75,0){\shit{7cm}{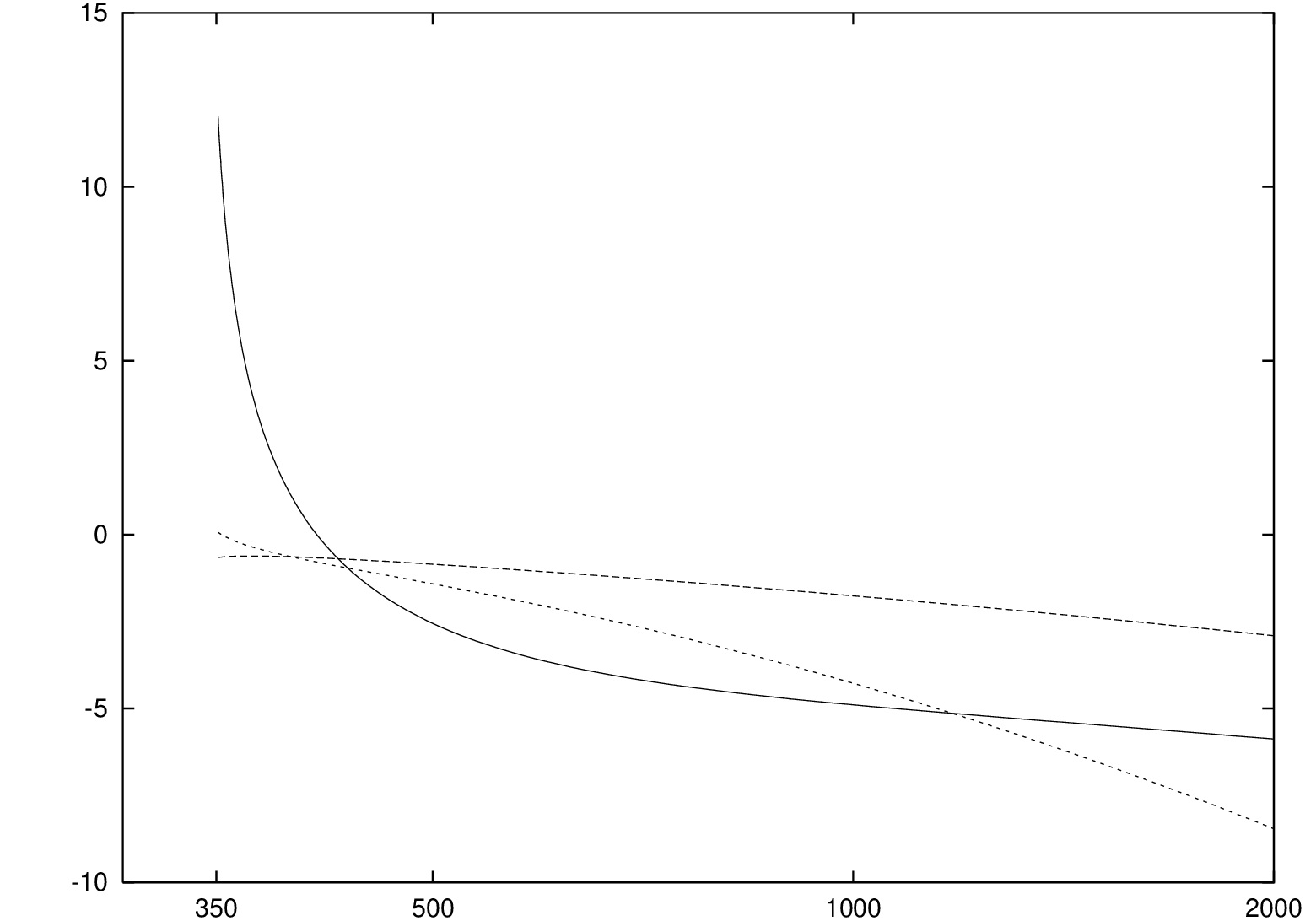}}
\put(0,7){\makebox(0,0){$\Delta_{\qqbar} , \%$}}
\put(3.25,0){\makebox(0,0){$\sqrt{\hat s}$ , GeV}}
\put(7.,0){\shit{7cm}{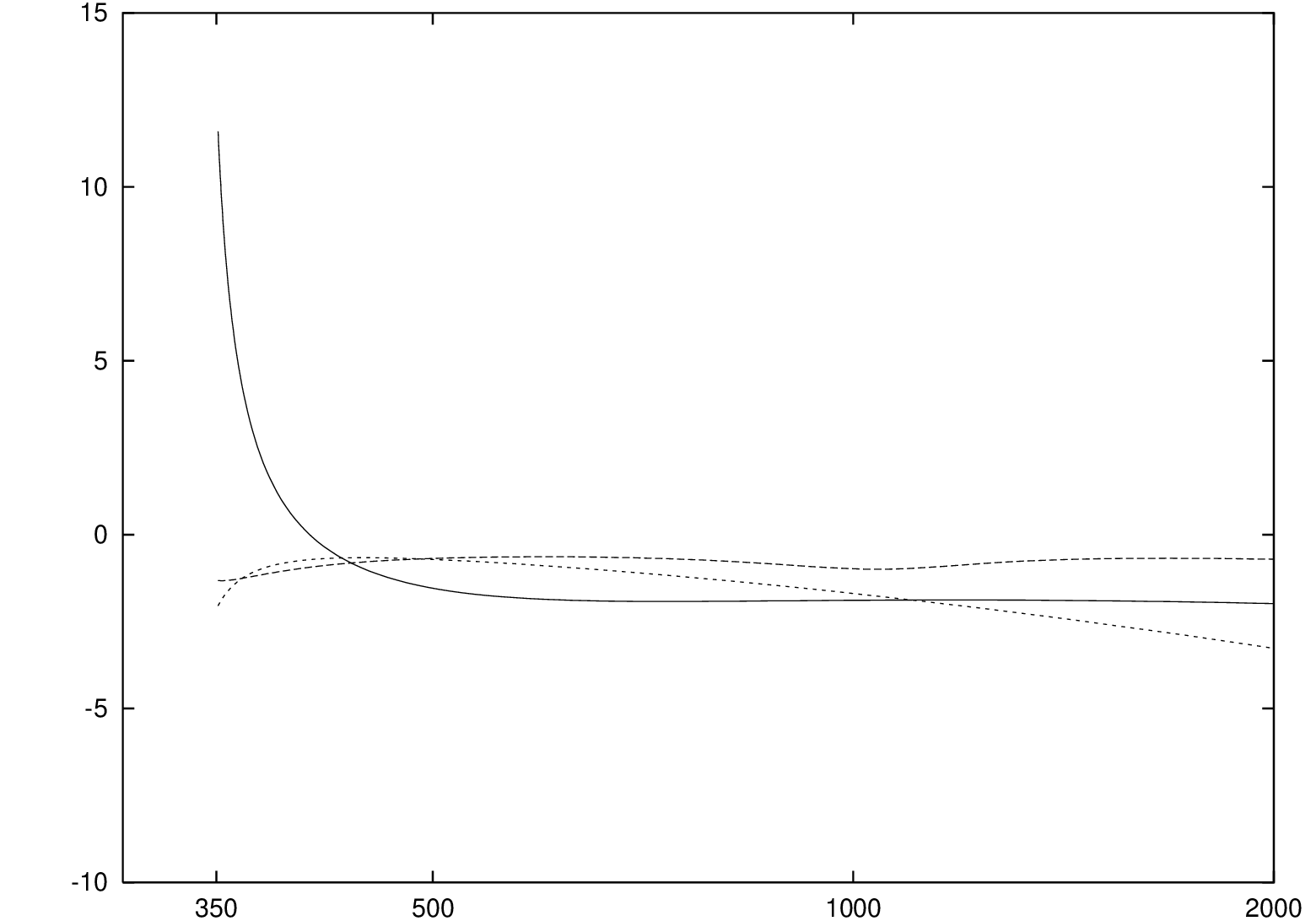}}
\put(8.5,7){\makebox(0,0){$\Delta_{gg} , \%$}}
\put(12.5,0){\makebox(0,0){$\sqrt{\hat s}$ , GeV}}
\end{picture}
\end{center}
\caption{The relative corrections $\Delta_{\qqbar}$ and
$\Delta_{gg}$ within the MSM.
The ${\cal O}(\alpha)$ contribution of the electroweak
gauge bosons (dotted line) and of the Higgs-sector are shown separately
(solid line: $M_H=65$ GeV, dashed line: $M_H=1$ TeV).}
\end{figure}
In the following we introduce the relevant features of
the models under consideration,
present the analytical expression for the form factors to
$\qqa$ and gluon fusion and study their numerical impact.
\subsection{The General 2-Higgs-Doublet Model}
The General 2-Higgs-Doublet Model (G2HDM)~\cite{hunters}
introduces six additional
parameters into the theory: four masses and two mixing angles.
In standard notation they read: $\ma,\mkh,\mgh,\mhp,\tanb$ and 
$\alpha$. Since we are working in the 't Hooft-Feynman gauge
we also have to include the contribution of the 
Higgs-ghosts $G^0$ and $G^{\pm}$.
The G2HDM one-loop corrections to the 
top pair production processes, $\qqa$  and $\ggf$,
can be easily obtained from~\cite{diplpubsm} when the
Yukawa-couplings written in the generalized form of Fig.~2
are replaced with the 2-Higgs-doublet Yukawa-couplings of Tab.~\ref{coup}.
\begin{figure}[htb]
\begin{center}
\setlength{\unitlength}{1cm}
\begin{picture}(15,2.5)
\put(0,0){\epsfbox{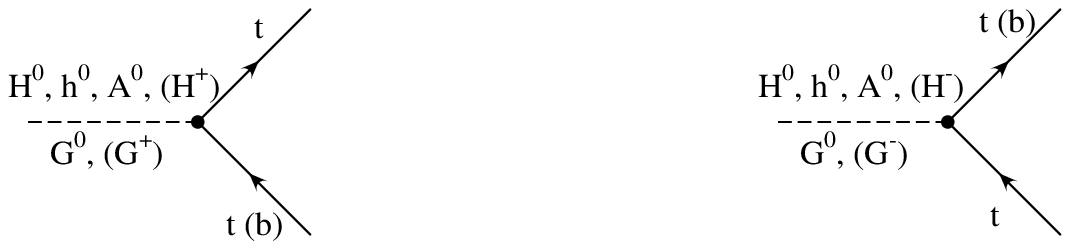}}
\put(4.75,1.5){\makebox(0,0){: $\frac{ie m_t}{2 s_W M_W}(c_s-c_p\gamma_5)$}}
\put(12.5,1.5){\makebox(0,0){: $\frac{ie m_t}{2 s_W M_W}(c_s+c_p' \gamma_5)$}}
\end{picture}
\end{center}
\caption{The Feynman-rules for the $H$-$t$-$t(b)$-vertex within the
2-Higgs-Doublet Model
written in generalized form. The coupling parameters
$c_s,c_p,c_p'$ are explicitly given in Tab.~I.}
\end{figure}
\renewcommand{\arraystretch}{2}
\begin{table}[htb]\centering
\caption{The top-Yukawa-couplings 
within the 2-Higgs-Doublet Model
($m_b$: bottom quark mass, $\alpha$: mixing angle, $\tanb$: ratio of
the vacuum expectation values of the two Higgs-doublets).}
\begin{tabular}{|ccccccc|}
 & $H^0$ & $h^0$
 & $A^0$ & $H^{\pm}$ & $G^0$ & $G^{\pm}$ \\ \hline
$c_s$  & $-\sin\alpha/\sin\beta$ & $-\cos\alpha/\sin\beta$
       & 0 & $\frac{1}{\sqrt{2}}(\cot\beta+\frac{m_b}{\mt} \tanb)$
       & 0 & $\frac{1}{\sqrt{2}}(1-\frac{m_b}{\mt})$ \\ \hline
$c_p$  & 0 & 0 & $-i \cot\beta$ & $\frac{1}{\sqrt{2}}(\cot\beta-
\frac{m_b}{\mt} \tanb)$ & $-i$ &
$\frac{1}{\sqrt{2}}(1+\frac{m_b}{\mt})$ \\ \hline
$c_p'$ & 0 & 0 & $-c_p$ & $c_p$ & $-c_p$ & $c_p$  \\
\end{tabular}
\label{coup}
\end{table}
\renewcommand{\arraystretch}{1}
\noindent
Additionally, the summation over the MSM Higgs-bosons in~\cite{diplpubsm}
has to be extended to 
\[\sum_{S=\eta,\chi,\Phi^{\pm}}\rightarrow
\sum_{S=H^0,h^0,A^0,H^{\pm},G^0,G^{\pm}}\]
and now two neutral scalars $H^0,h^0$
contribute to the $s$-channel Higgs-exchange diagram in the gluon fusion
subprocess.

In order to demonstrate the effect of the convolution
with parton distribution functions and to study the dependence
on the free parameters we present
numerical results obtained within the G2HDM also at the parton level.
For the numerical evaluation of the electroweak one-loop corrections
within the G2HDM we chose the mixing angle $\alpha$ to be $\alpha=\pi/2$.
Then the heavy Higgs-boson $H^0$ develops MSM-like Yukawa-couplings
for very large values of $\tanb$, and $h^0$ decouples.
For $\alpha = 0$ the neutral scalars $H^0$ and $h^0$
just switch roles and for values in between 
the sum of their contribution leads to smaller relative corrections.
The pseudo scalar $A^0$ is expected to contribute noticeably
only for small values of $\tanb$. It turns out, that 
there is very little variation ($<2 \%$) of the relative corrections with $\ma$
for $\tanb=0.7$ and
no dependence on $\ma$ can be observed for $\tanb$=70.
Thus, throughout the following numerical discussion within the G2HDM
we chose a representative value of $\ma=50$ GeV.
A study of the dependence on the charged Higgs-boson mass $\mhp$
shows that there is little variation of the
partonic cross sections with $\mhp$, only
for large values of $\tanb$ due to the 
enhancement of the suppression factor $m_b/\mt$ in the Yukawa-coupling
a small dependence can be observed.
Thus, in the following discussion we chose a representative value of
$\mhp=50$ GeV.
The only exception occurs in a very special region
in the G2HDM parameter space: $m_t \approx \mhp+m_b$,
where due to a discontinuity in the
derivative of the $B$-functions the  
radiative corrections can be considerably enhanced.
\begin{figure}
\begin{center}
\setlength{\unitlength}{1cm}
\setlength{\fboxsep}{0cm}
\begin{picture}(16,20)
\put(-1.75,14.5){\shit{7cm}{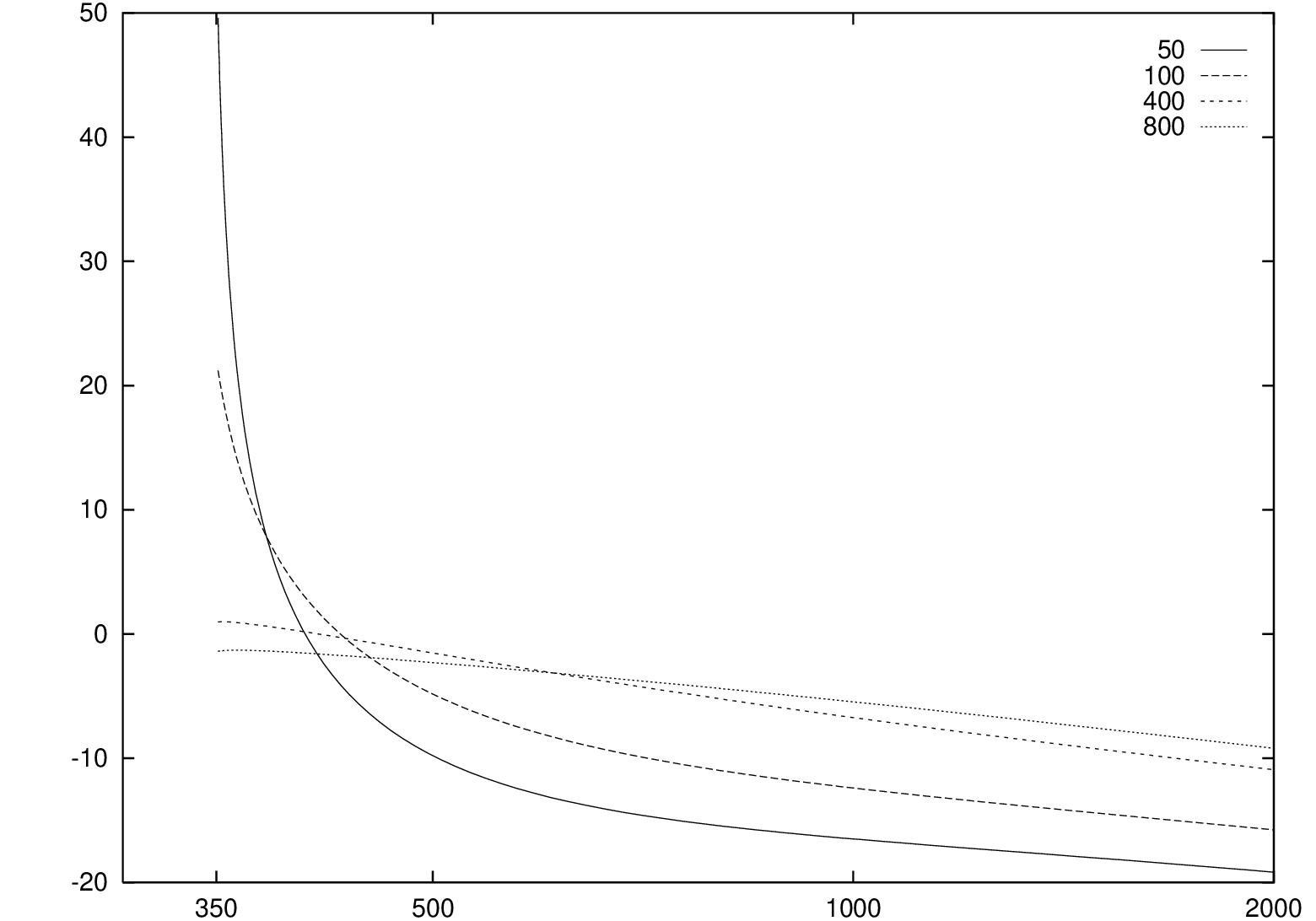}}
\put(0,21.5){\makebox(0,0){$\Delta_{\qqbar} , \%$}}
\put(3.25,14.5){\makebox(0,0){$\sqrt{\hat s}$ , GeV}}
\put(3.25,20.5){\makebox(0,0){$\tanb=0.7$}}
\put(7.,14.5){\shit{7cm}{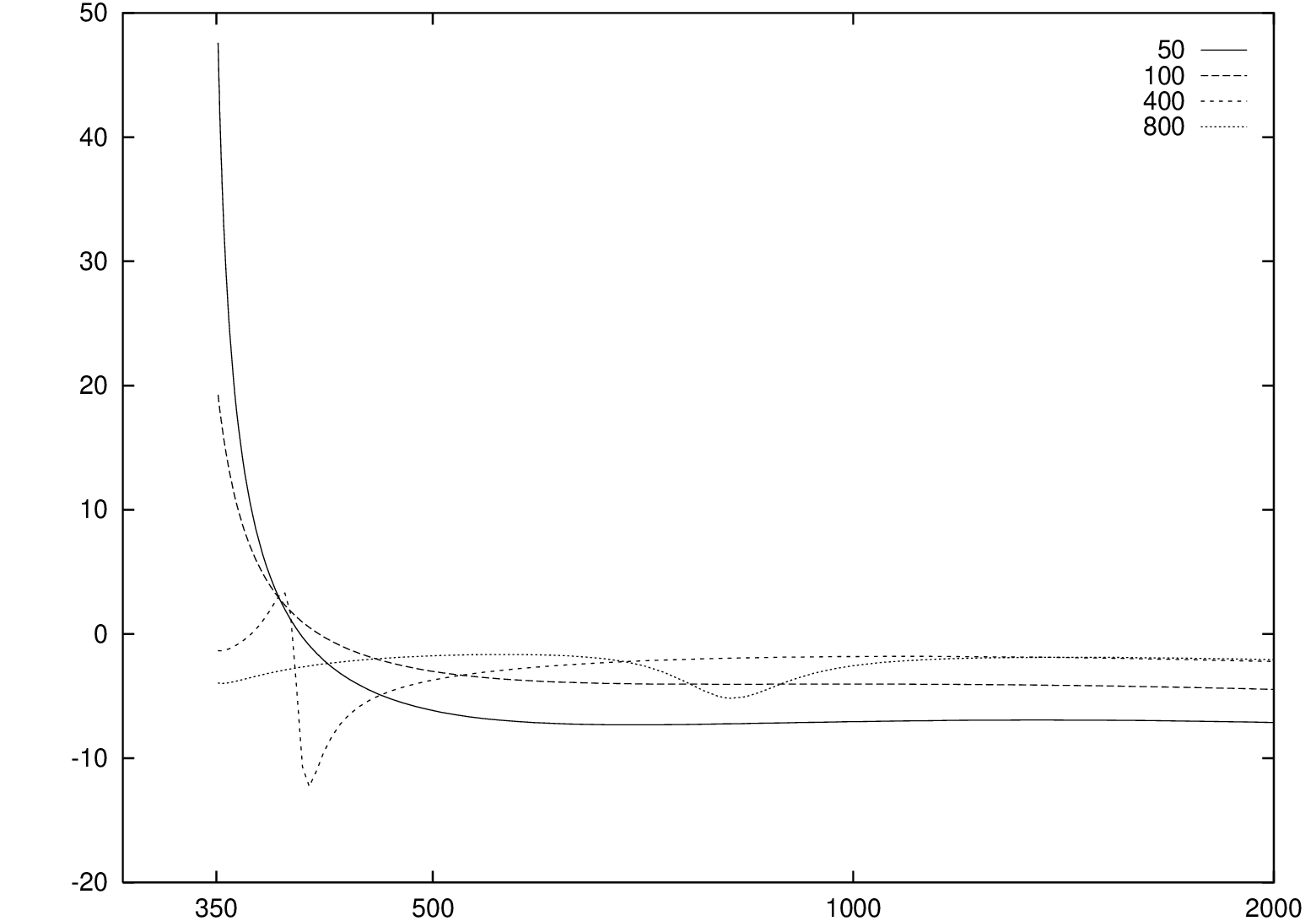}}
\put(8.5,21.5){\makebox(0,0){$\Delta_{gg} , \%$}}
\put(12.5,14.5){\makebox(0,0){$\sqrt{\hat s}$ , GeV}}
\put(12.5,20.5){\makebox(0,0){$\tanb=0.7$}}
\put(-1.75,7.25){\shit{7cm}{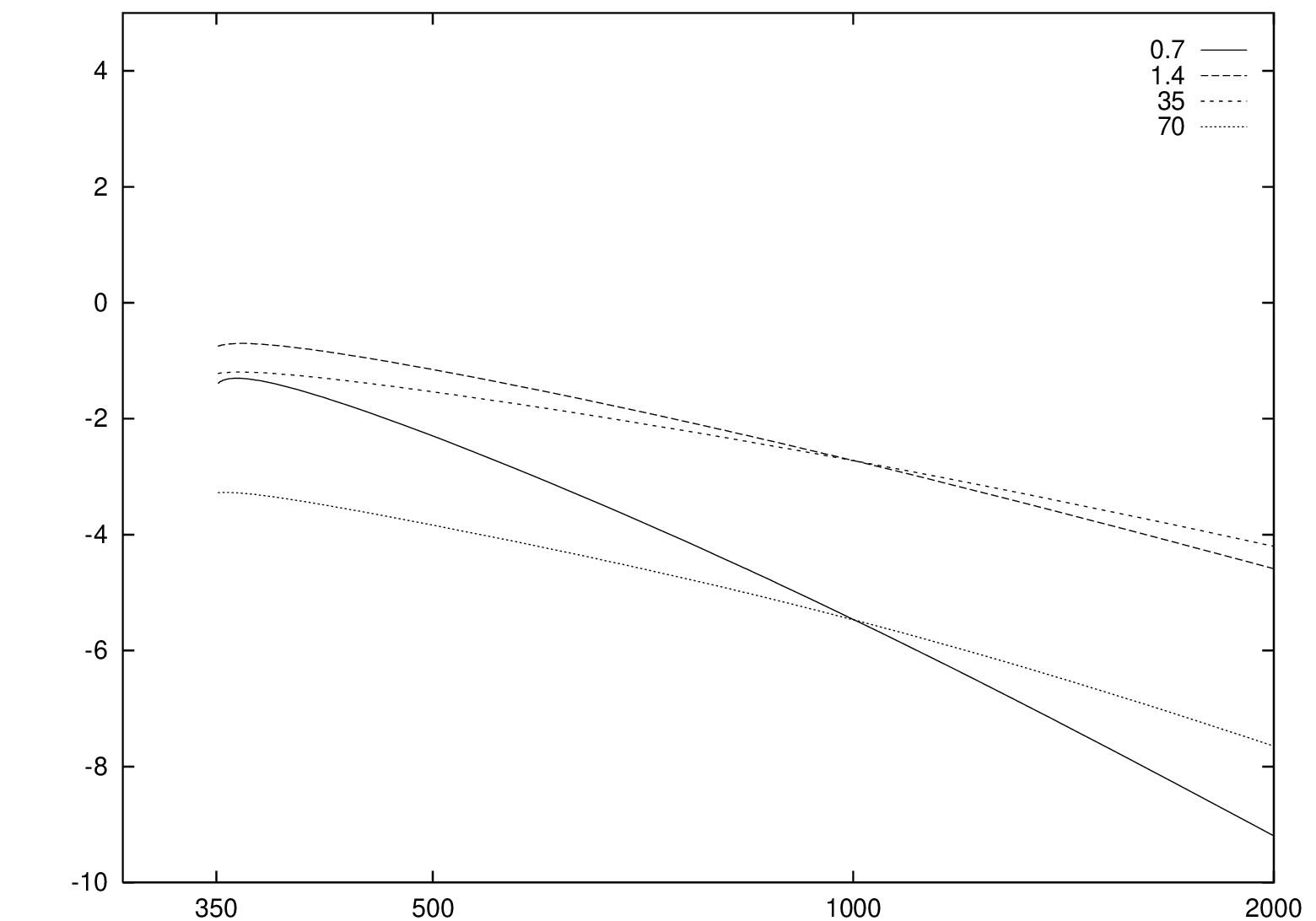}}
\put(3.25,7.25){\makebox(0,0){$\sqrt{\hat s}$ , GeV}}
\put(3.25,8.25){\makebox(0,0){$\mgh=800$ GeV}}
\put(7.,7.25){\shit{7cm}{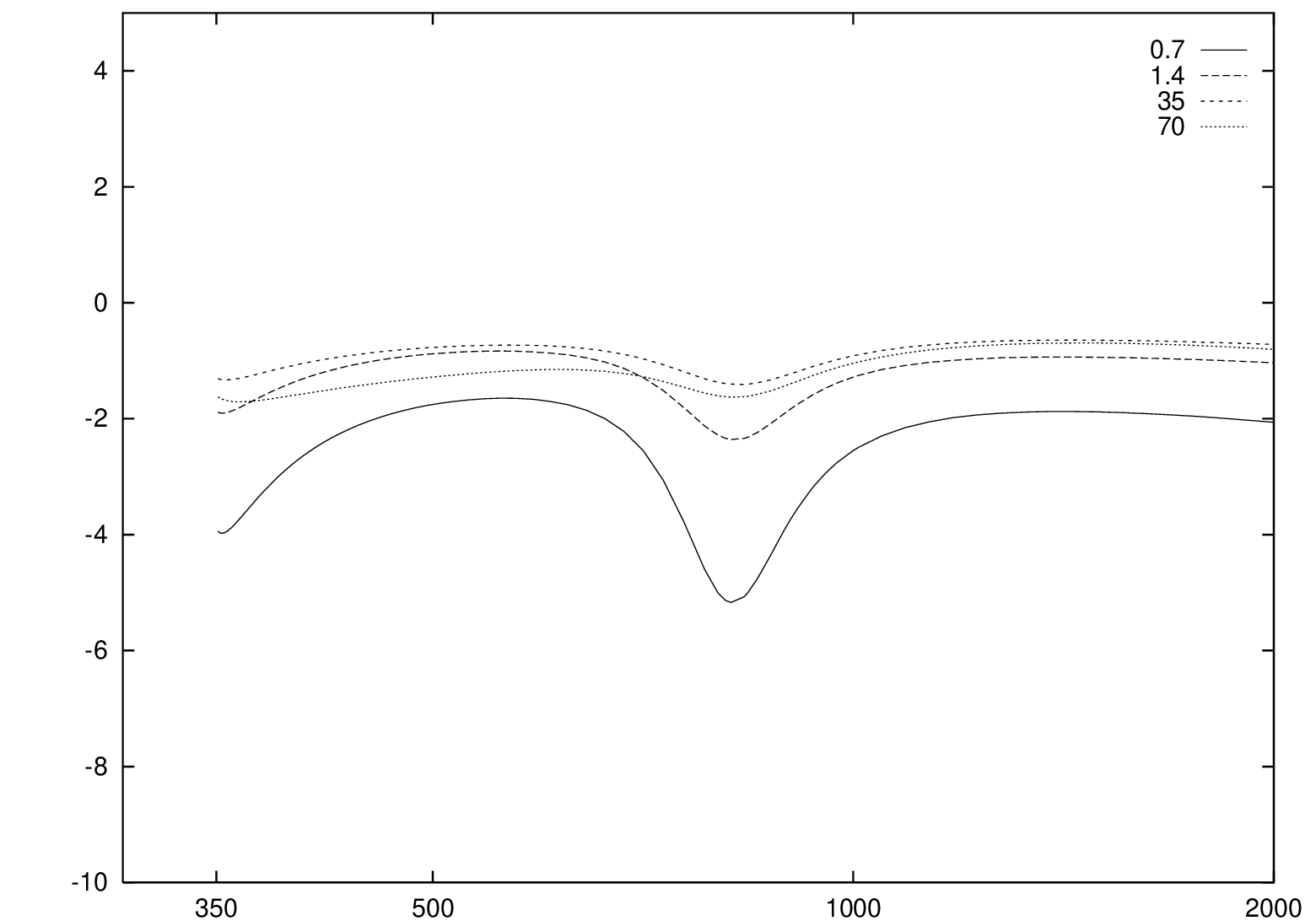}}
\put(12.5,7.25){\makebox(0,0){$\sqrt{\hat s}$ , GeV}}
\put(13,8.25){\makebox(0,0){$\mgh=800$ GeV}}
\put(-1.75,0){\shit{7cm}{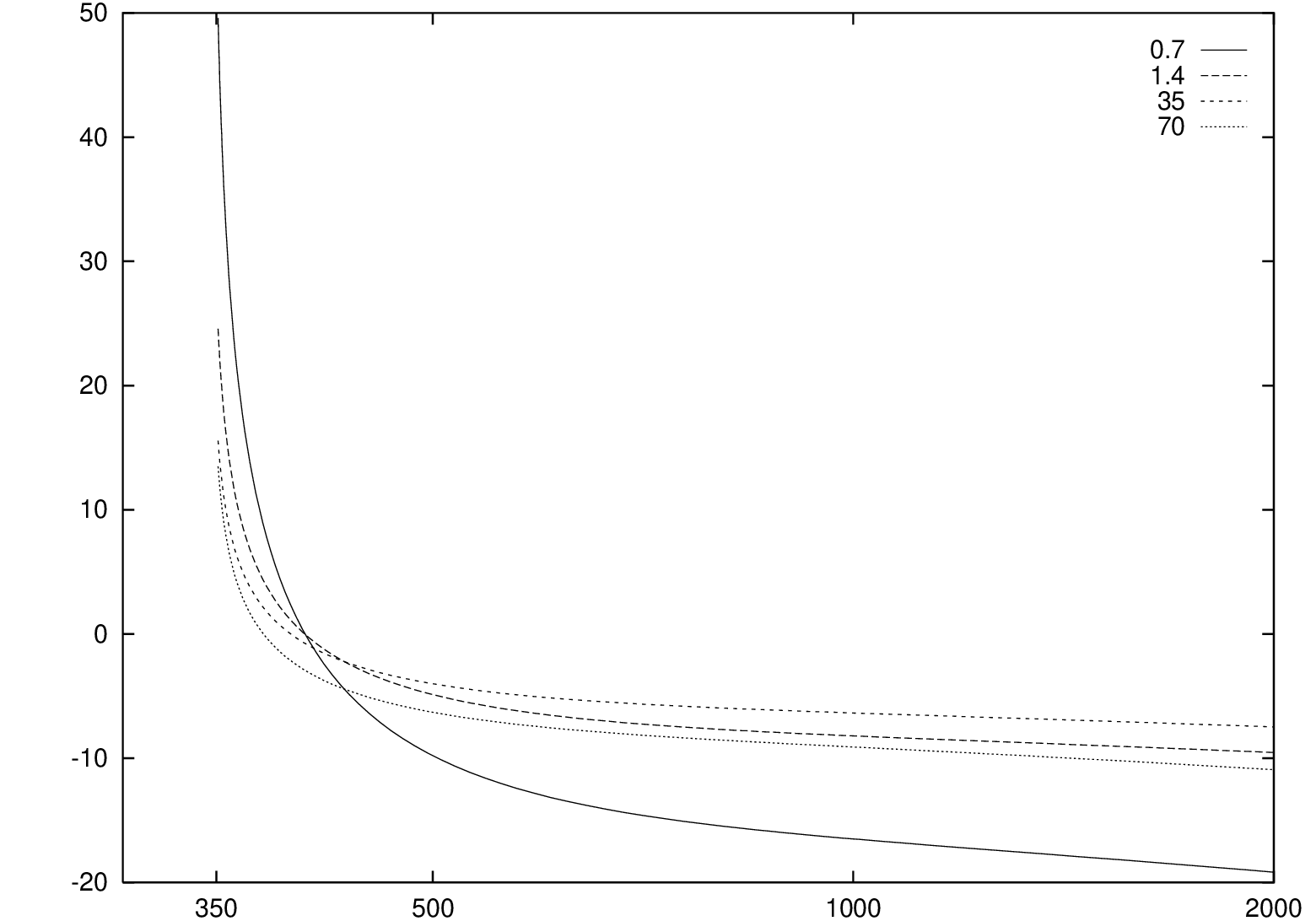}}
\put(3.25,0){\makebox(0,0){$\sqrt{\hat s}$ , GeV}}
\put(3.25,6){\makebox(0,0){$\mgh=50$ GeV}}
\put(7.,0){\shit{7cm}{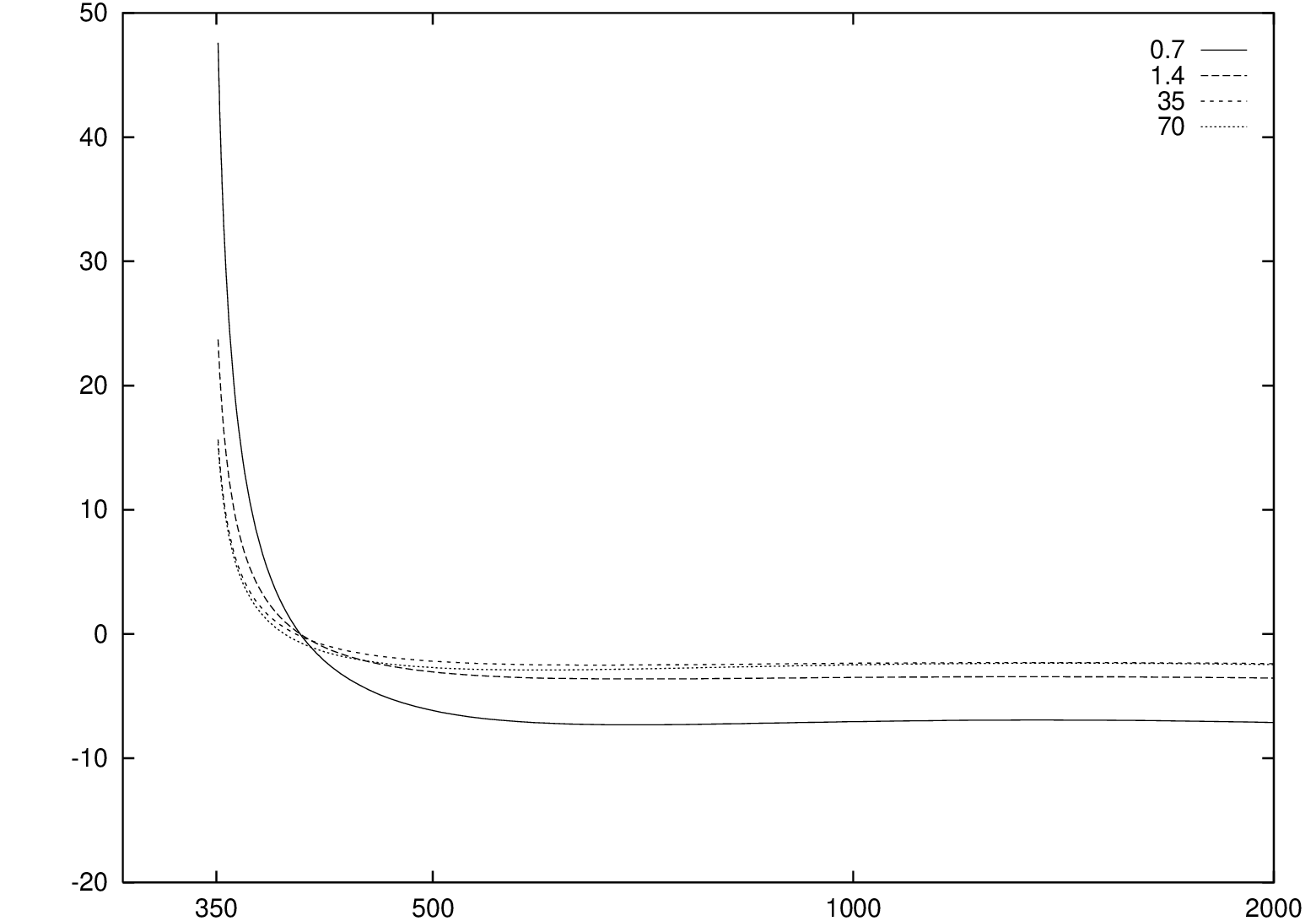}}
\put(12.5,0){\makebox(0,0){$\sqrt{\hat s}$ , GeV}}
\put(12.5,6){\makebox(0,0){$\mgh=50$ GeV}}
\end{picture}
\end{center}
\caption{The variation of the relative corrections $\Delta_{\qqbar}$ and
$\Delta_{gg}$ with $\mgh$ and $\tanb$ within the G2HDM (with $\mkh=45$ GeV,
$\ma=50$ GeV, $\mhp=50$ GeV and $\alpha=\pi/2$).}
\end{figure}
\noindent
In Fig.~3 the dependence of the ${\cal O}(\alpha)$
Higgs-sector contribution to the $\qqa$ and gluon fusion 
subprocesses within the G2HDM 
on $\mgh$ and $\tanb$ can be studied.
At the parton level, very large corrections up to $+50\%$
arise in the threshold region $\sqrt{\hat s}\approx 2 \mt$
assuming $H^0$ is very light and $\tanb$ very small.
As expected due to the structure of the
Yukawa-couplings, the relative corrections decrease for increasing values of
$\tanb$. The increase of the relative corrections for very large $\tanb$
is due to the charged Higgs-boson contribution as discussed earlier.
In the case of gluon fusion a Breit-Wigner-resonance
structure can be observed originating from the
$s$-channel Higgs-exchange diagrams
when $\sqrt{\hat s} \approx M_{H^0,h^0}>2 \mt$. 

\subsection{The Higgs-sector of the MSSM}

The requirements of supersymmetry~\cite{susy,haka}
lead to the existence of (at least) one additional
Higgs-doublet where the parameters of the Higgs-potential
are tightly correlated. At one-loop level, this translates
into the improved physical masses and the
effective mixing angle $\alpha_{eff}$ as follows~\cite{mrel}:
\beqn\label{massrel}
\mhp^2 & = & M_W^2+\ma^2
\nonumber \\ 
M_{H^0,h^0}^2 & = & \frac{1}{2}(\ma^2+M_Z^2+\Omega+\Sigma \pm \sqrt{R})
\nonumber \\
\tan \alpha_{eff} & = & \frac{\Lambda-\sin{\beta} \cos{\beta}
(\ma^2+M_Z^2)}{M_Z^2 \cos^2\beta+\ma^2
\sin^2\beta+\Sigma -\mkh^2} 
\eeqn
with 
\beqn
R & = & (\ma^2+M_Z^2)^2-4\ma^2 M_Z^2\cos^2{2\beta}
+(\Omega-\Sigma) (\Omega-\Sigma+2 (\ma^2-M_Z^2) \cos{2\beta})
\nonumber \\
&+ & 4 \Lambda (\Lambda - (\ma^2+M_Z^2) \sin{2 \beta}) \;.
\end{eqnarray}
In the approximation where only the dominating $\mt^4$-terms
have been kept, the one-loop corrections
parametrized in terms of $\Lambda, \Sigma$ and $\Omega$
read as follows
(at tree-level: $\Lambda=\Sigma=\Omega=0$):
\begin{eqnarray}
\Omega & = & v
 \left[\log\left(\frac{\mste \mstz}{m_t^2}\right) +
\astop S (2 k+\astop S g)\right]
\nonumber \\
\Sigma &=& v g (\mu S)^2
\nonumber \\
\Lambda &= &- v \mu S 2 (k+\astop S g)
\eeqn
with
\beqn
v &=&  \frac{3 G_F}{\sqrt{2} \pi^2} \frac{m_t^4}{\sin^2{\beta}}
\nonumber \\
k &= &\frac{1}{\mste^2-\mstz^2} \log\left(\frac{\mste}{\mstz}\right)
\nonumber\\
g &= &\frac{1-k (\mste^2+\mstz^2)}{(\mste^2-\mstz^2)^2}
\nonumber\\
S & = & \astop-\frac{\mu}{\tanb} \; .
\eeqn 
Note, that our $\mu$-sign convention is different from the one used
in~\cite{mrel}.
We neglect the impact of radiative corrections on the
charged Higgs-boson mass $\mhp$ since they do not exceed
10 GeV in a wide range of the parameter space~\cite{mcharged} and 
as pointed out earlier
the relative corrections do not significantly depend on
the value of $\mhp$.
The Higgs-mixing parameter $\mu$ of the superpotential,
the soft supersymmetry-breaking squark masses
$M_{\tilde Q},M_{\tilde U}$ and the trilinear soft supersymmetry-breaking
parameter $\astop$ are treated as free parameters.
They only enter the supersymmetric Higgs-sector via
radiative corrections. 
In Eq.~(\ref{massrel}) we took into account that the superpartners to the
left- and right-handed top quarks, $\stl$ and $\str$, 
are not necessarily mass eigenstates
since the mass matrix ${\cal M}$ is of non-diagonal form
\begin{equation}
(\begin{array}{cc}\tilde t_L & \str \end{array}) \;
 {\cal M} \left(\begin{array}{c} \stl \\ \str
\end{array}\right) 
\end{equation}
with
\begin{equation}\label{sfermimix}
{\cal M} = \left(\begin{array}{cc}
M_{\tilde{Q}}^2+m_t^2-\cos{2\beta} 
(M_Z^2-4 M_W^2)/6 & m_t (A_{\tilde{t}}-\mu/\tanb)\\
m_t (A_{\tilde{t}}-\mu/\tanb) & M_{\tilde{U}}^2+m_t^2+2\cos{2\beta}
(M_Z^2-M_W^2)/3 \end{array}\right) \; .
\end{equation}
Thus, $\stl,\str$ can mix so that the physical 
mass eigenstates $\ste,\stz$ are model-dependent linear combinations of these
states. The latter are obtained by diagonalizing 
the mass matrix by performing the transformation
\begin{eqnarray}\label{trafo}
\ste & = & \cos{\phist} \; \stl+\sin{\phist} \; \str
\nonumber \\
\stz & = & -\sin{\phist} \; \stl+\cos{\phist} \; \str \; .
\end{eqnarray}
Since the off-diagonal elements of the sfermion mass matrices 
are proportional to the fermion masses we only consider
$L,R$-mixing in the stop quark sector.

We choose the light stop quark mass $\mstz$ and
the mixing angle $\phist$ to be input parameters, so that 
the remaining parameters of the stop quark sector are determined by
equations which relate the elements of ${\cal M}$
\begin{equation}
{\cal M} = \left(\begin{array}{cc}
a & b \\
b & d \end{array}\right)
\end{equation}
with its eigenvalues and the mixing angle $\phist$
\begin{eqnarray}
m^2_{\tilde t_1,\tilde t_2} & = & \frac{a+d}{2}\pm\sqrt{\frac{(a-d)^2}{4}+b^2}
\nonumber \\
m^2_{\tilde t_1} & = & \cos^2{\phist} \;  a
+\sin^2{\phist} \;  d + 2\cos{\phist}\sin{\phist}\;  b
\nonumber \\
m^2_{\tilde t_2} & = & \sin^2{\phist} \; a
+\cos^2{\phist} \; d - 2\cos{\phist}\sin{\phist} \; b \;.
\end{eqnarray}
The soft-supersymmetry-breaking parameter $M_{\tilde{Q}}$ (and thus 
$a$) is already fixed by the mass of the left-handed sbottom quark and
$\tanb$ (no $\tilde b_L$-$\tilde b_R$-mixing
and $m_{\tilde b_L}=m_{\tilde b_R}$)
\begin{equation}
M_{\tilde{Q}}^2 = m_{\tilde b_L}^2-m_b^2+\frac{\cos{2\beta}}{6}
(M_Z^2+2 M_W^2) \; . 
\end{equation}
Consequently, with the choice of $\phist$, $\mstz$,
$m_{\tilde b_L}$ and $\tanb$ the off-diagonal element $b$ and the
heavy stop quark mass $\mste$ are fixed.
To summarize, the predictions of the
supersymmetric 2-Higgs-Doublet Model depend on two parameters
of the Higgs-sector which are conventionally chosen to be
$\ma$ and $\tanb$ and when considering radiative corrections
to the supersymmetric mass relations also on
$m_{\tilde b_L},\mu,\phist$ and $\mstz$.
In the special case of no $L,R$-mixing ($\phist$=0, $S=0$) 
there is no $\mu$-dependence and $\Sigma=\Lambda=0$.

\subsection{The SUSY EW-like one-loop corrections}

In the following we present the ${\cal O}(\alpha)$ contribution of the
gaugino-Higgsino-sector within the MSSM
to the $\qqa$ and gluon fusion subprocesses.
At one-loop level the $g\ttbar$-vertex
is modified due to the exchange of two charginos $\tilde \chi_{i=1,2}^{\pm}$
and four neutralinos $\tilde \chi^0_{i=1\cdots 4}$. 
They are linear combinations of the supersymmetric partners of the 
electroweak gauge bosons and Higgs-bosons, the gauginos and Higgsinos,
and are obtained by diagonalizing the corresponding mixing matrices $X,Y$
(mass eigenstates $\tilde \chi^0,\tilde \chi^{\pm}$
in four-component notation)~\cite{haka}:
\begin{equation}
{\cal L}_m  = - \frac{1}{2}
\sum_i \bar{\tilde \chi}_i^0 (N^* Y N^{-1})_{ii} \tilde \chi_i^0 
\end{equation}
with
\begin{equation}
Y = \left(\begin{array}{cccc} M_1 & 0 & -M_Z s_W \cos{\beta} &
M_Z s_W \sin{\beta} \\
 0 & M_2 & M_Z c_W \cos{\beta} & -M_Z c_W \sin{\beta} \\
-M_Z s_W \cos{\beta} & M_Z c_W \cos{\beta} & 0 & -\mu \\
M_Z s_W \sin{\beta} & -M_Z c_W \sin{\beta} & -\mu & 0 
\end{array}\right)
\end{equation}
and the mass term for the charginos
\begin{equation}
{\cal L}_m = - \sum_i \bar{\tilde\chi}_i^+ 
(U^* X V^{-1})_{ii} \tilde \chi_i^+ 
\end{equation}
with
\begin{equation}
X=\left(\begin{array}{cc} M_2 & M_W \sqrt{2} \sin{\beta} \\
 M_W \sqrt{2} \cos{\beta} & \mu 
\end{array}\right) \; .
\end{equation}
As a result the mass eigenstates $\tilde \chi_i^0,\tilde \chi_i^{\pm}$
are model-dependent with masses $\mneuti,\mchari$
depending on the SUSY parameters $\mu,M_1,M_2$ and on $\tanb$.
In order to reduce the number of independent parameters
we assume the $SU(2)\times U(1)$ theory being embedded in a grand unified
theory so that the following relation becomes valid~\cite{haka}:
\begin{equation}
M_1=\frac{5 s_W}{3 c_W} M_2 \; .
\end{equation}
$\mchari$ and $\mneuti$ are consequently fixed by the choice of $\mu, M_2$
and $\tanb$.
The diagonalizing matrices for the charginos $U,V$ are real and can be easily 
chosen so that only positive mass eigenvalues $M_{\tilde \chi_i^{\pm}}$
occur. Explicit expressions can be found in~\cite{haka}.
Different from the notation of~\cite{guka,haka} we chose
the diagonalizing matrix $N$ to be real and allow negative
mass eigenvalues $M_{\tilde \chi_i^0}$. The according
Feynman-rules involving a neutralino can be retrieved from~\cite{guka,haka}
by performing the transformation described in Appendix A.3 of~\cite{guka}
(here the real matrix $Z$ used in~\cite{guka} is denoted by $N$).

The Feynman-diagrams describing the SUSY EW-like
one-loop corrections to the
$\qqa$ and the gluon fusion subprocesses are shown in Fig.~4 and
Fig.~5, respectively.
\begin{figure}[t]
\begin{center}
\setlength{\unitlength}{1cm}
\begin{picture}(15,2.5)
\put(0,0){\epsfbox{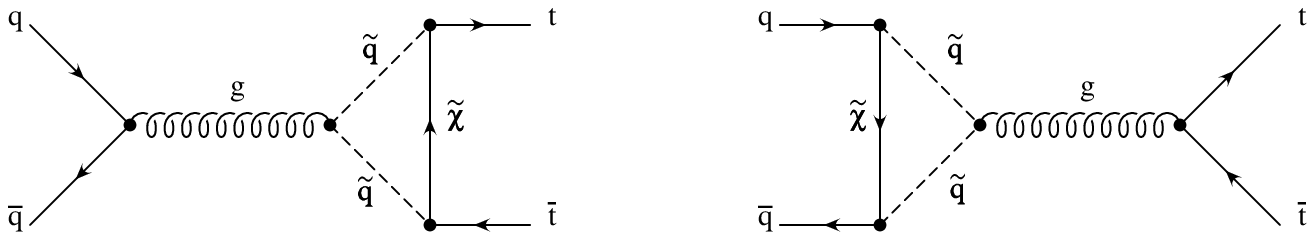}}
\end{picture}
\end{center}
\caption{The Feynman-diagrams for the SUSY EW-like contributions to the
$\qqa$ subprocess.}
\end{figure}
\begin{figure}
\begin{center}
\setlength{\unitlength}{1cm}
\begin{picture}(15,16)
\put(0,0){\epsfbox{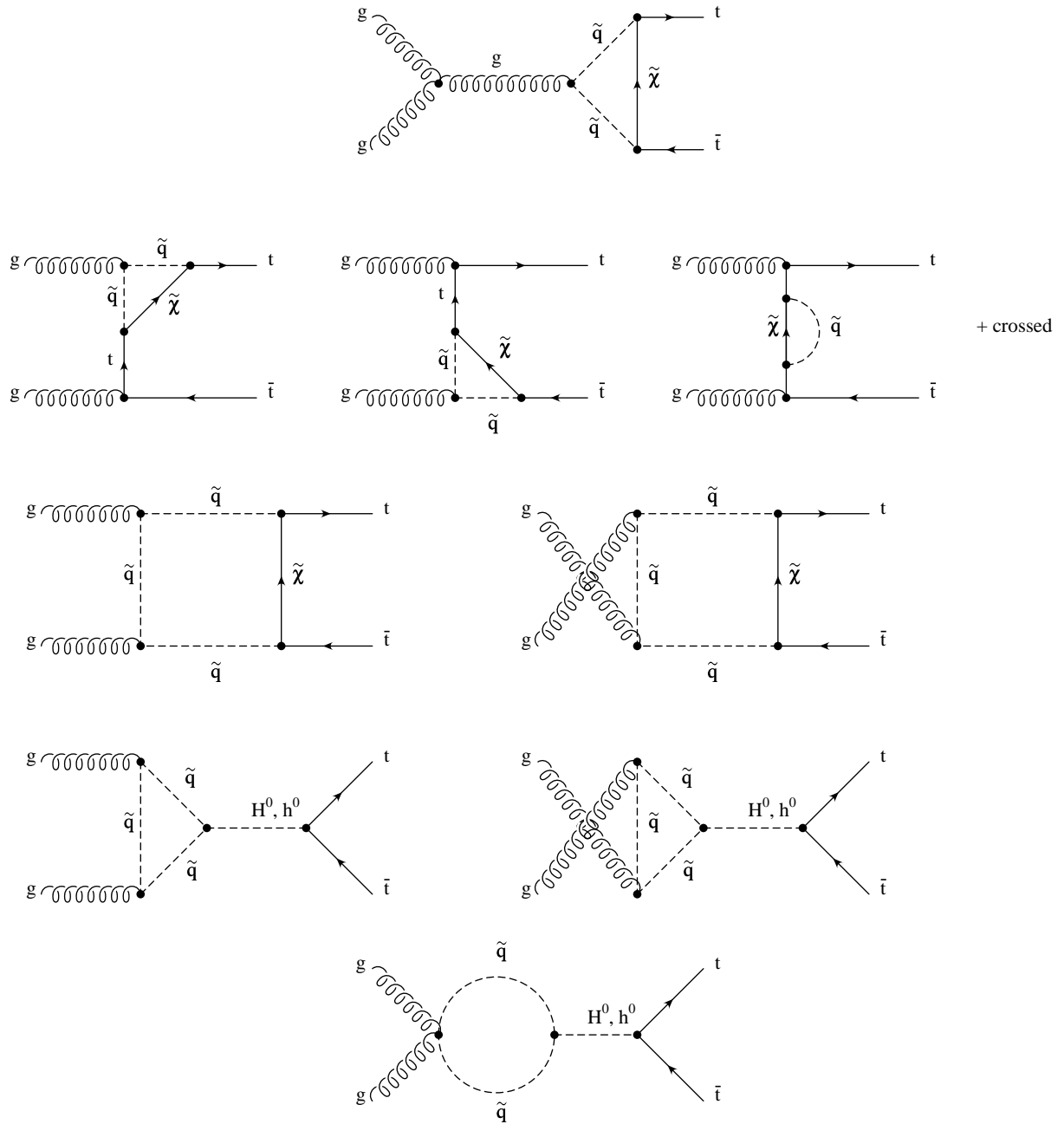}}
\end{picture}
\end{center}
\caption{The Feynman-diagrams for the SUSY EW-like contributions to the
gluon fusion subprocess.}
\end{figure}
\noindent
The Feynman-rules for the 
arising triple and quartic squark-gluon interactions are given in 
Fig.~6~\cite{haka}.
\begin{figure}[htb]
\begin{center}
\setlength{\unitlength}{1cm}
\begin{picture}(15,2.5)
\put(0,0){\epsfbox{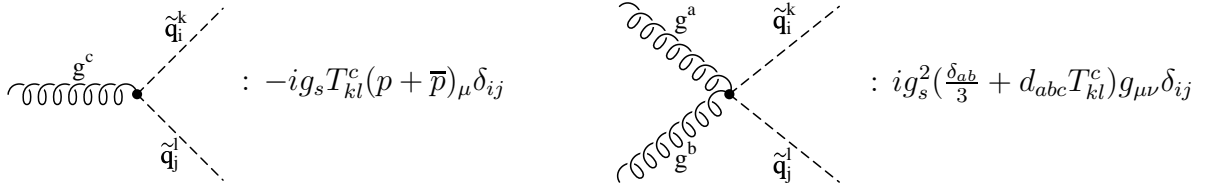}}
\put(5,1.5){\makebox(0,0){: $-i g_s T^c_{kl} (p+\overline p)_{\mu}
\delta_{ij}$}}
\put(13.75,1.5){\makebox(0,0){: $i g_s^2 (\frac{\delta_{ab}}{3}+d_{abc} T^c_{kl})
g_{\mu\nu}\delta_{ij}$}}
\end{picture}
\end{center}
\caption{The Feynman-rules for the triple and quartic squark-gluon
interactions. 
$a,b,c=1\cdots 8$ and $k,l=1\cdots 3$ are color indices and $i,j=L,R$.
$g_s^2=4\pi\alpha_s$ denotes the strong coupling parameter and
$T^c=\lambda^c/2$ with the Gell-Mann-matrices $\lambda^c$
satisfying the $SU(3)$ anticommutation relation $2d_{abc} \lambda^c
 = \{\lambda_a,\lambda_b\}-4/3 \delta_{ab}$.}
\end{figure}	
\noindent
By introducing the coupling parameters $g_{s,p}^j$ with $j=L,R$
(without mixing) and $j=1,2$ (with mixing) the 
neutralino(chargino)-stop(sbottom)-top-vertex
can again be written in a generalized form as shown in Fig.~7.
\begin{figure}
\begin{center}
\setlength{\unitlength}{1cm}
\begin{picture}(15,3)
\put(0,0){\epsfbox{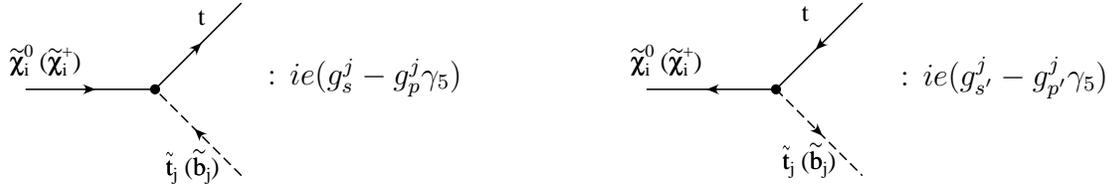}}
\put(4.75,1.5){\makebox(0,0){: $ie (g_s^j-g_p^j \gamma_5)$}}
\put(13.25,1.5){\makebox(0,0){: $ie (g_{s'}^j - g_{p'}^j \gamma_5)$}}
\end{picture}
\end{center}
\caption{The $\chi_i^{0,\pm}$-$\tilde t_j (\tilde b_j)$-$t$ - vertex
written in generalized form.
The coupling parameters $g_{s,p}^j$ are explicitly given in Tab.~II.}
\end{figure}
When considering $\stl$-$\str$-mixing the interaction eigenstates 
are replaced by the mass eigenstates by using the transformation
of Eq.~(\ref{trafo})
in the interaction Lagrangian, which has the following impact on the 
stop quark couplings of Tab.~\ref{susycoup}:
\begin{eqnarray}
g_s^{1,2} & = & \cos{\phist} \; g_s^{L,R}\pm \sin{\phist} \; g_s^{R,L}
\nonumber\\
g_p^{1,2} & = & \cos{\phist} \; g_p^{L,R}\pm\sin{\phist} \; g_p^{R,L} \;.
\end{eqnarray}

With these Feynman-rules and
after taking into account the counter terms of the
{\em on-shell} renormalization procedure
the SUSY EW-like one-loop corrections to the
$\qqa$ subprocess can be described
by means of finite form factors $F_V,F_M$ (introduced in~\cite{diplpubsm})
which modify the $g\ttbar$-vertex
\begin{equation}
F_{V,M}(\hat s) = \sum_{j=1}^2 \left(\sum_{i=1}^4
S_{V,M}(m_{\tilde t_j},\mneuti)
+ \sum_{i=1}^2 S_{V,M}(m_{\tilde b_j},\mchari)\right)
\end{equation}
with
\begin{eqnarray}
S_V(m_1,m_2) & = & \lambda_j^+ 2 C_2^0(\hat s,m_1,m_1,m_2)+\delta Z_V
\\
S_M(m_1,m_2) &=& \lambda_j^+ (-4 \mt^2)
(C_1^- + 2 C_2^-)(\hat s,m_1,m_1,m_2)
\nonumber\\
&+ & \lambda_j^- 2  \mt m_2  (2 C_1^- + C_0)(\hat s,m_1,m_1,m_2) \; ,
\end{eqnarray}
where we used the abbreviation $\lambda_j^{\pm}=(g_s^j)^2\pm (g_p^j)^2$.
\renewcommand{\arraystretch}{2}
\begin{table}[htb]\centering
\caption{The scalar and pseudo-scalar couplings $g_{s}^{L,R}$ and $g_{p}^{L,R}$
describing the neutralino(chargino)-left 
(right)handed stop(sbottom)-top-vertex.
For convenience a matrix $N'$ with 
$N'_{i1,i2}=c_W N_{i1,i2}\pm s_W N_{i2,i1}$ has been introduced.}
\begin{tabular}{|ccc|}
\multicolumn{3}{|c|}{$g_{s,p}^j = a\pm b \; ; \; g_{s'}^j = g_s^j
\; ; \; g_{p'}^j = -g_p^j$}  \\ \hline \hline
      & $a$ & $b$ \\ \hline 
$\tilde \chi_i^0 \tilde t_L t$ & 
$-N_{i4} m_t /(2\sqrt{2}s_W M_W \sin\beta)$ &
$-\frac{1}{\sqrt{2}} (\frac{2}{3} N'_{i1}+\frac{1-\frac{4}{3}s_W^2}{2c_Ws_W} N'_{i2})$
\\ \hline
$\tilde \chi_i^0 \tilde t_R t$ & 
$\frac{1}{\sqrt{2}} \frac{2}{3} (N'_{i1}-\frac{s_W}{c_W} N'_{i2})$ &
$-N_{i4} m_t / (2\sqrt{2} s_W M_W \sin\beta)$
\\ \hline
$\tilde \chi_i^+ \tilde b_L t$ & 
$V_{i2} m_t /( 2 \sqrt{2} s_W M_W \sin\beta)$ &
$-\frac{1}{2 s_W} U_{i1}$
\\ \hline 
$\tilde \chi_i^+ \tilde b_R t$ &  0 &
$U_{i2} m_b / (2 \sqrt{2} s_W M_W \cos\beta)$\\
\end{tabular}
\label{susycoup}
\end{table}
\noindent
The renormalization constant $\delta Z_V$ is determined by the
top quark self-energy $\Sigma_{V,S}$
\begin{equation}
\label{zrenorm}
\delta Z_V = -\Sigma_V(p^2=\mt^2)-
2 \mt^2 \frac{\partial}{\partial p^2}(\Sigma_V+\Sigma_S)_{p^2=\mt^2}
\end{equation}
with
\begin{eqnarray}
\label{topself}
\Sigma_V(p^2) & = &- \lambda_j^+ B_1(p^2,m_2,m_1)
\nonumber\\
\Sigma_S(p^2) & = & \lambda_j^- \frac{m_2}{\mt} B_0(p^2,m_2,m_1) \; .
\end{eqnarray}

The Feynman-diagrams of Fig.~5 represent the SUSY EW-like
${\cal O}(\alpha)$ contribution to the gluon fusion
subprocess consisting of 
the modification of the $g\ttbar$-vertex in the
$s$- and $t$-production channel described by
the form factors $F_V,F_M$ and $\rho_i^V$, respectively,
the self-energy insertion to the off-shell 
top quark ($\rho_i^{\Sigma}$), the UV-finite box diagrams ($\rho_i^{\Box}$)
and the UV-finite $s$-channel Higgs-exchange ($\rho_i^{\triangle}$).
These contributions together with the counter terms
modify the gluon fusion form factors
in~\cite{diplpubsm} as follows:
\begin{equation}
\rho_i^{(V,\Sigma,\Box),(t,u)} =
 \sum_{j=1}^2 \left(\sum_{i=1}^4 T_i^{(V,\Sigma,\Box),(t,u)}
(m_{\tilde t_j},\mneuti) 
+ \sum_{i=1}^2 T_i^{(V,\Sigma,\Box),(t,u)}
(m_{\tilde b_j},\mchari)\right)
\end{equation}
with\\
\underline{Vertex corrections:}
\begin{eqnarray}
T_1^{V,t}(m_1,m_2) & = & 4 \lambda_j^+ C_2^0(\hat t,m_1,m_1,m_2)
+2 \delta Z_V
\nonumber\\
T_4^{V,t}(m_1,m_2) & = & 2 \lambda_j^+ (\hat t-\mt^2)
(C_2^2-C_2^{12}+C_1^2)(\hat t,m_1,m_1,m_2)
\nonumber\\
T_{11}^{V,t}(m_1,m_2) & = & - T_{1}^{V,t}(m_1,m_2)
\nonumber\\
T_{14}^{V,t}(m_1,m_2) & = & 2 \lambda_j^+
(-C_2^1-C_2^2+2 C_2^{12}-C_1^1-C_1^2)(\hat t,m_1,m_1,m_2)
\nonumber\\
&+ & 2 \lambda_j^- \frac{m_2}{\mt}
(C_1^1+C_1^2+C_0)(\hat t,m_1,m_1,m_2)
\nonumber\\
T_{16}^{V,t}(m_1,m_2) & = & -4 T_{14}^{V,t}(m_1,m_2)
\end{eqnarray}
\underline{Top quark self-energy insertion:}
\begin{eqnarray}
T_1^{\Sigma,t}(m_1,m_2) & = & -(\hat t+\mt^2) (\Sigma_V(\hat t,m_2,m_1)
+\delta Z_V)
\nonumber\\
&- & 2 \mt^2 (\Sigma_S(\hat t)-\delta Z_V-
\Sigma_S(\mt^2)-\Sigma_V(\mt^2))
\nonumber\\
T_{11}^{\Sigma,t}(m_1,m_2) & = & 2 \hat t (\Sigma_V(\hat t,m_2,m_1)
+\delta Z_V)
\nonumber\\
&+ &  (\hat t+\mt^2) (\Sigma_S(\hat t)-\delta Z_V-
\Sigma_S(\mt^2)-\Sigma_V(\mt^2))
\end{eqnarray}
\underline{Box contribution:}
\begin{eqnarray}
T_2^{\Box,t}(m_1,m_2) & = & 2\lambda_j^+ D_3^{02}(\hat t,m_1,m_1,m_1,m_2)
\nonumber\\
T_4^{\Box,t}(m_1,m_2) & = & 4 \lambda_j^+
(D_2^0+2 D_3^{01}+D_3^{02})(\hat t,m_1,m_1,m_1,m_2)
\nonumber\\
T_6^{\Box,t}(m_1,m_2) & = & -2 \lambda_j^+
(D_1^2+D_3^2+2 (2 D_2^{12}+D_2^2+D_3^{12}+D_3^{123}-2 D_3^{21}))
(\hat t,m_1,m_1,m_1,m_2)
\nonumber\\
T_{12}^{\Box,t}(m_1,m_2) & = & -4 \lambda_j^+
(2 D_3^{01}+D_3^{02})(\hat t,m_1,m_1,m_1,m_2)
+4 \lambda_j^- \frac{m_2}{\mt}D_2^0(\hat t,m_1,m_1,m_1,m_2)
\nonumber\\
T_{16}^{\Box,t}(m_1,m_2) & = & 4 \lambda_j^+
(D_1^2+D_3^2+2 (D_1^1+2D_2^1+4 D_2^{12}+2 D_2^{13}+D_2^2+D_3^1
\nonumber\\
&+ &  3 D_3^{12}+3 D_3^{123}+3 D_3^{13}+3 D_3^{21}))
(\hat t,m_1,m_1,m_1,m_2)
\nonumber\\
&- & 4 \lambda_j^- \frac{m_2}{\mt} (D_0+D_2^2+2 (2 D_1^1+D_1^2
+D_2^1+2 D_2^{12}+D_2^{13}))(\hat t,m_1,m_1,m_1,m_2) \; .
\nonumber\\
& &
\end{eqnarray}
$\delta Z_V$ and $\Sigma_{V,S}(p^2)$ are given by
Eq.~\ref{zrenorm} and Eq.~\ref{topself}, respectively.
The $u$-channel contribution $\rho_i^{(V,\Sigma,\Box),u}$ can
be obtained from the $t$-channel form factors
by replacing $\hat t$ with $\hat u$.
For the notations concerning the $B,C,D$-functions see~\cite{diplpubsm}.
The $s$-channel Higgs-exchange contribution is described by:
\begin{eqnarray}\label{triaform}
\rho_{12}^{\triangle} & = & \frac{\mt}{2 s_W M_W}\sum_{S=H^0,h^0}
\sum_{j=1,2}\sum_{q} \frac{c_s c_{s,j}^{susy}}{\mt} (\hat s-M_S^2-i M_S \Gamma_S)
\nonumber \\
& & 
[-2 m_{\tilde q_j}^2 C_0(\hat s,m_{\tilde q_j},m_{\tilde q_j},m_{\tilde q_j})-1]
\end{eqnarray}
with~\cite{hunters}
\begin{eqnarray}
C_0(\hat s,m_{\tilde q_j},m_{\tilde q_j},m_{\tilde q_j})
& = & \left\{
\begin{array}{c}
\frac{1}{2\hat s} \left(\log\frac{1+\sqrt{1-\tau}}{1-\sqrt{1-\tau}}
-i \pi\right)^2 ; \tau=\frac{4m_{\tilde q_j}^2}{\hat s}< 1 \\
-\frac{2}{\hat s} \left(\sin^{-1}\sqrt{\frac{1}{\tau}}\right)^2 ; \tau \geq 1
\end{array} \right. \; ,
\end{eqnarray}
where $c_s$ is taken from
Tab.~\ref{coup} and $c_{s,j}^{susy}$ from Tab.~\ref{susy}.
The tree level decay widths of the light and heavy neutral
Higgs-bosons $\Gamma_{h^0,H^0}$ within the MSSM 
are given in~\cite{hunters}.
\begin{table}[htb]\centering
\caption{Higgs-boson couplings to left and right-handed squarks within the MSSM
(with $c_+ = \cos(\alpha+\beta)$, $s_+ = \sin(\alpha+\beta)$ and
$A_{\tilde f\neq \tilde t}=\mu/ \tanb$)}
\begin{tabular}{|ccc|}
\multicolumn{3}{|c|}{$c_{s,(LL,RR,LR)}^{susy}$} \\ \hline \hline
 &   $T_3^f>0 , Q_f=2/3$ & $T_3^f<0, Q_f = -1/3$  \\ \hline
$H^0 \tilde f_L \tilde f_L$ & $-\frac{M_Z}{c_W s_W}(T_3^f-Q_f s_W^2)
c_+ -\frac{m_f^2}{M_W s_W} \frac{\sin\alpha}{\sin\beta}$ &
$-\frac{M_Z}{c_W s_W}(T_3^f-Q_f s_W^2)
c_+ -\frac{m_f^2}{M_W s_W} \frac{\cos\alpha}{\cos\beta}$
\\ \hline
$H^0 \tilde f_R \tilde f_R$ & $-\frac{M_Z}{c_W} s_W Q_f 
c_+ -\frac{m_f^2}{M_W s_W} \frac{\sin\alpha}{\sin\beta}$ &
-$\frac{M_Z}{c_W} s_W Q_f
c_+ -\frac{m_f^2}{M_W s_W} \frac{\cos\alpha}{\cos\beta}$ 
\\ \hline
$H^0 \tilde f_L \tilde f_R$ & $-\frac{m_f}{2 M_W s_W\sin\beta}(-\mu \cos\alpha
+A_{\tilde f} \sin\alpha)$ &
$-\frac{m_f}{2 M_W s_W\cos\beta}(-\mu \sin\alpha
+A_{\tilde f} \cos\alpha)$
\\ \hline
$h^0 \tilde f_L \tilde f_L$ & $\frac{M_Z}{c_W s_W}(T_3^f-Q_f s_W^2)
s_+ -\frac{m_f^2}{M_W s_W} \frac{\cos\alpha}{\sin\beta}$ &
$\frac{M_Z}{c_W s_W}(T_3^f-Q_f s_W^2)
s_+ +\frac{m_f^2}{M_W s_W} \frac{\sin\alpha}{\cos\beta}$
\\ \hline
$h^0 \tilde f_R \tilde f_R$ & $\frac{M_Z}{c_W} s_W Q_f 
s_+ -\frac{m_f^2}{M_W s_W} \frac{\cos\alpha}{\sin\beta}$ &
$\frac{M_Z}{c_W} s_W Q_f
s_+ -\frac{m_f^2}{M_W s_W} \frac{\sin\alpha}{\cos\beta}$
\\ \hline 
$h^0 \tilde f_L \tilde f_R$ & $-\frac{m_f}{2 M_W s_W \sin\beta}(\mu \sin\alpha
+A_{\tilde f} \cos\alpha)$ &
$-\frac{m_f}{2 M_W s_W\cos\beta}(-\mu \cos\alpha
-A_{\tilde f} \sin\alpha)$ \\
\end{tabular}
\label{susy}
\end{table}
In the stop quark sector
when taking into account $\tilde t_L$-$\tilde t_R$-mixing
$c_{s,(LL,RR)}^{susy}$ is replaced with
\begin{equation}
c_{s,(1,2)}^{susy} = \cos^2\phist \; c_{s,(LL,RR)}^{susy}
+\sin^2\phist \; c_{s,(RR,LL)}^{susy}\pm 2 \sin\phist \cos\phist
\; c_{s,LR}^{susy} \; .
\end{equation}
The masses of the supersymmetric partners to the light quarks
in the unmixed up-type
and down-type quark sector are again fixed by choosing 
$m_{\tilde b_L}$ and assuming that $m_{\tilde q_L}=m_{\tilde q_R}$.
The corresponding light quark masses are considered to be zero.

In order to study the effects of the SUSY EW-like contribution and
what remains after including the contribution from the
supersymmetric Higgs-sector with special
emphasis on the $s$-channel Higgs-exchange diagrams
we discuss in Fig.~8 and Fig.~9 the
SUSY EW-like one-loop corrections
and the full ${\cal O}(\alpha)$ MSSM contribution, respectively, to
the gluon fusion subprocess.
Since the SUSY $s$-channel Higgs-exchange contribution 
is proportional to the quark masses 
when the squark masses are considered to be degenerate in mass,
only the stop quark-loop contributes significantly.
The characteristic structure around $\sqrt{\hat s}=\mgh$ in the Figs.~8,9
originates from the $s$-channel Higgs-exchange diagrams. 
Since $\mkh \ll \sqrt{\hat s}$ we only observe the
resonance structure due to the exchange of the heavy neutral Higgs-boson $H^0$.
As can be seen when comparing the SUSY EW-like with the full
MSSM contribution 
the top quark-loop contribution to the $s$-channel Higgs-exchange
diagrams dominates. 
Only when the SUSY Higgs-decay width
$\Gamma_{H^0 \rightarrow \tilde q \tilde q,
\tilde \chi \tilde \chi}$ is large the resonance structure
arising from the top quark-loop contribution is also suppressed as
observed in Fig.~8 and Fig.~9 for $\phist=\pi/4$
and $\mstz=75$ GeV, respectively. 
In the following we chose $\tanb=0.7$ and $\ma=450$ GeV ($\Rightarrow \mgh$).
Concerning the $s$-channel Higgs-exchange
diagrams the dependence on $\tanb$ and $\mgh$ is essentially not 
different from what we observed within the G2HDM.

In Fig.~8 we study the dependence on the light stop quark mass $\mstz$ and
on the mixing angle $\phist$. Both parameters affect the
resonance structure via the decay width $\Gamma_{H^0}$
and also via the Higgs-boson mass $\mgh$ when
the described radiative corrections to the supersymmetric
Higgs-mass relations are taken into account.
Moreover, in the resonance region the negative form factor 
$\rho_{12}^{\triangle}$ of Eq.~(\ref{triaform}) increases with $\mstz$ 
and thus decreases the otherwise positive SUSY EW-like contribution.
When $\phist \neq 0$ is chosen the $L,R$-mixing term in the
coupling parameter $c_{s,(1,2)}^{susy}$ contributes and depending on the sign 
of $\phist$ (more accurate: the sign of $\mu \phist$)
either diminishes or enhances the SUSY $s$-channel Higgs-exchange contribution.
In Fig.~9 the dependence on $\mu$ and $\msbe$ is shown.
Again the choice of $\mu$ effects the resonance structure by changing
$\mgh$. Also in Fig.~9 the enhancement due to the threshold effect in 
the vicinity of $\mt=\mstz+\mneut$ can be observed 
for $\mstz=75$ GeV and $\mu =-90$ GeV.
The dependence on $\mu$ is much less significant for larger values of
$\mstz$. Finally, we discuss the effects of varying the
sbottom quark mass $\msbe$. The dependence of the 
$s$-channel Higgs-exchange contribution
on $\msbe$ mainly originates from its impact on the
value of $\mgh$ and the contribution of the sbottom
quark to the decay width $\Gamma_{H^0\rightarrow \tilde q \tilde q}$.
\begin{figure}
\begin{center}
\setlength{\unitlength}{1cm}
\setlength{\fboxsep}{0cm}
\begin{picture}(16,20)
\put(-1.75,14.5){\shit{7cm}{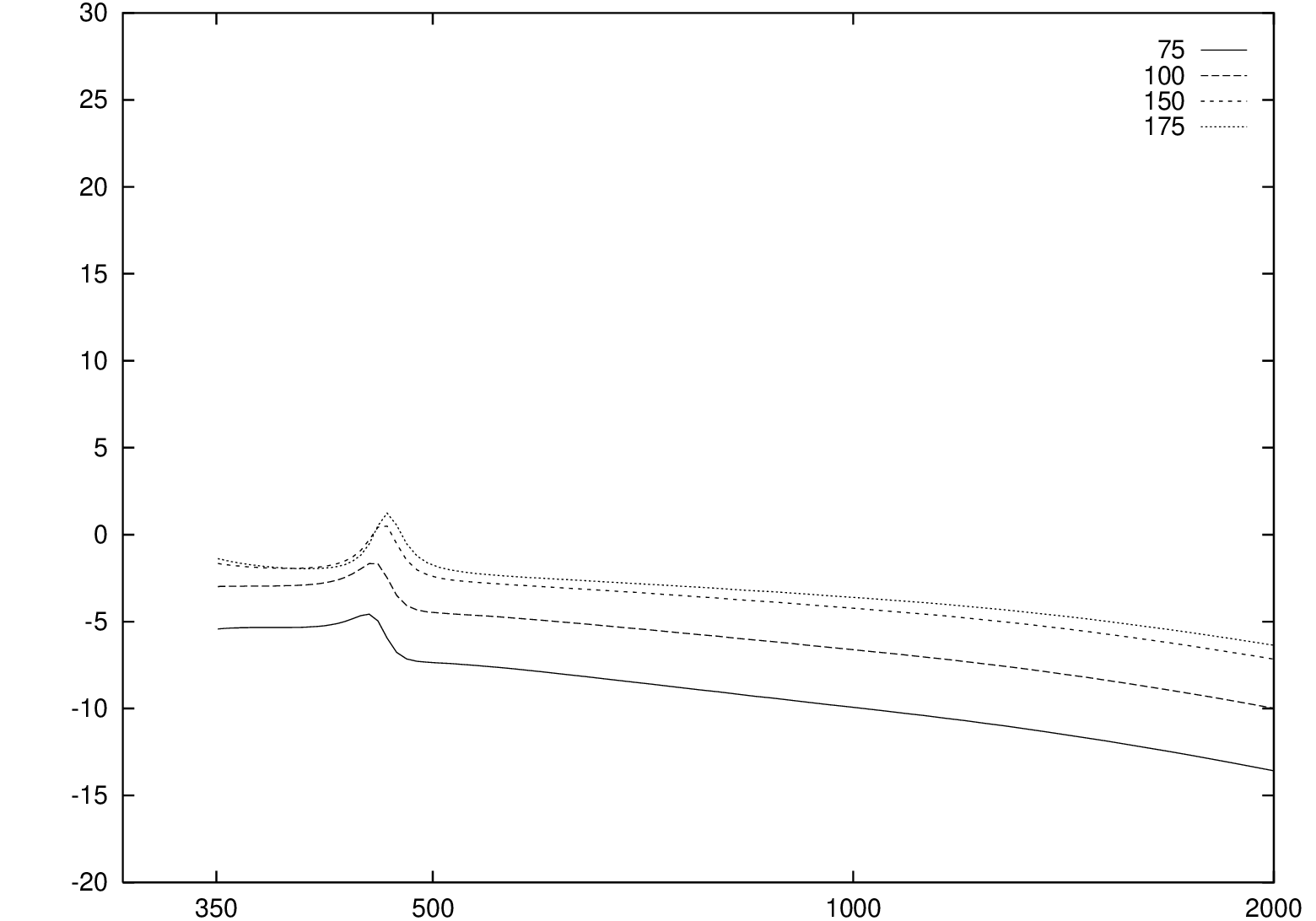}}
\put(1,21.5){\makebox(0,0){$\Delta_{gg} , \%$ SUSY EW-like}}
\put(3.25,14.5){\makebox(0,0){$\sqrt{\hat s}$ , GeV}}
\put(3.25,20.5){\makebox(0,0){$\phist=0$}}
\put(7.,14.5){\shit{7cm}{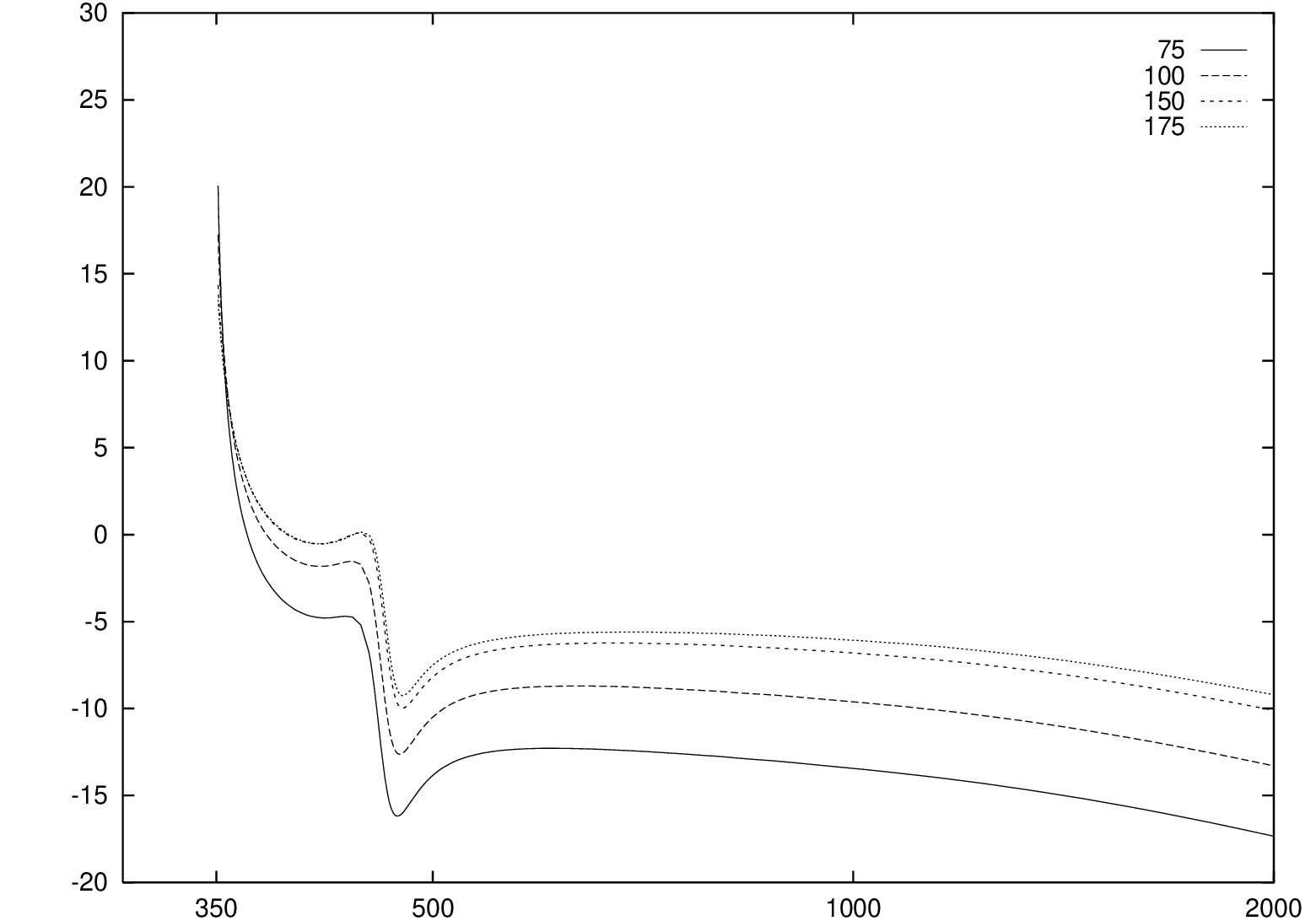}}
\put(9,21.5){\makebox(0,0){$\Delta_{gg} , \%$ MSSM}}
\put(12.5,14.5){\makebox(0,0){$\sqrt{\hat s}$ , GeV}}
\put(12.5,20.5){\makebox(0,0){$\phist=0$}}
\put(-1.75,7.25){\shit{7cm}{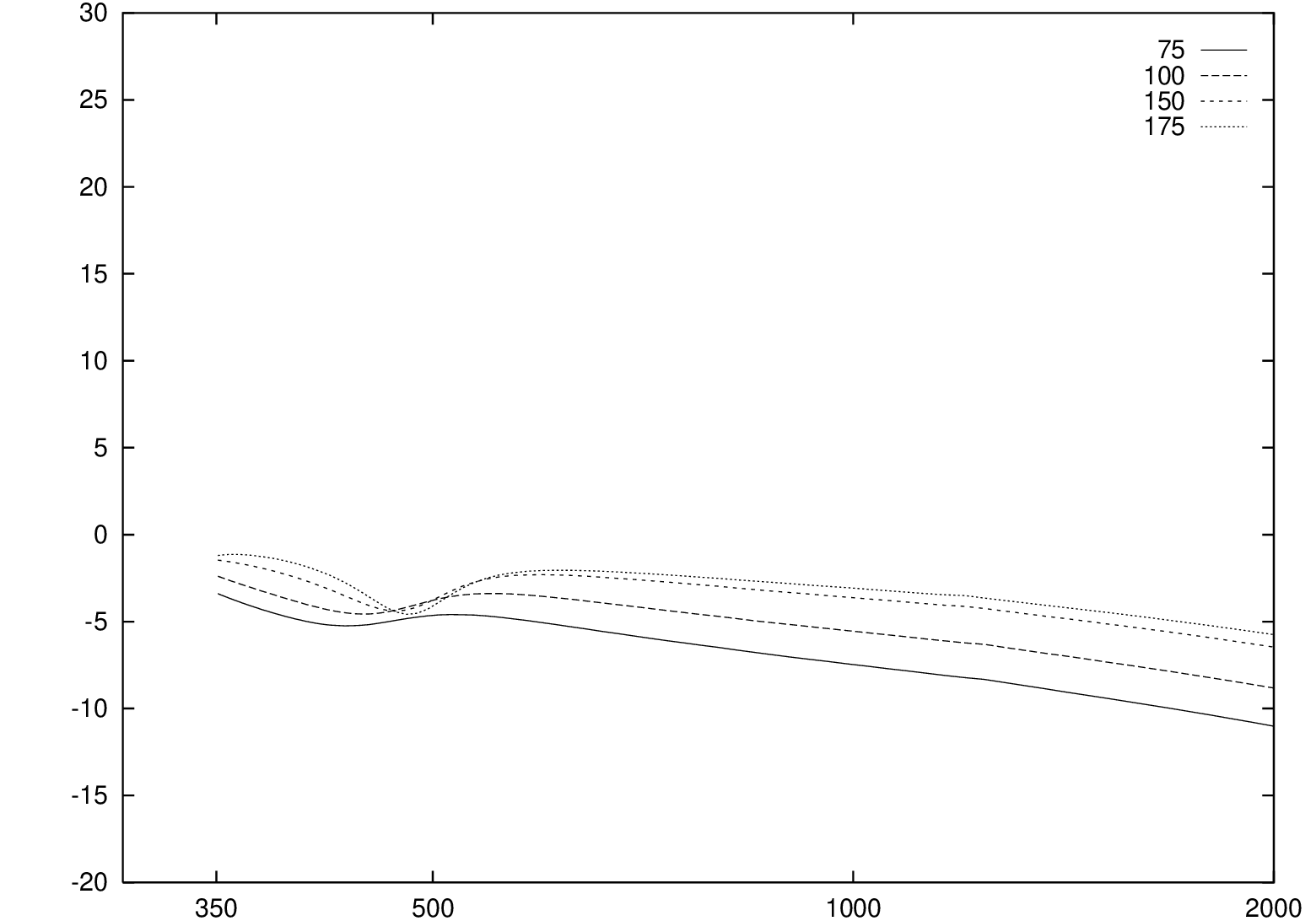}}
\put(3.25,7.25){\makebox(0,0){$\sqrt{\hat s}$ , GeV}}
\put(3.25,13.25){\makebox(0,0){$\phist=\pi/4$}}
\put(7.,7.25){\shit{7cm}{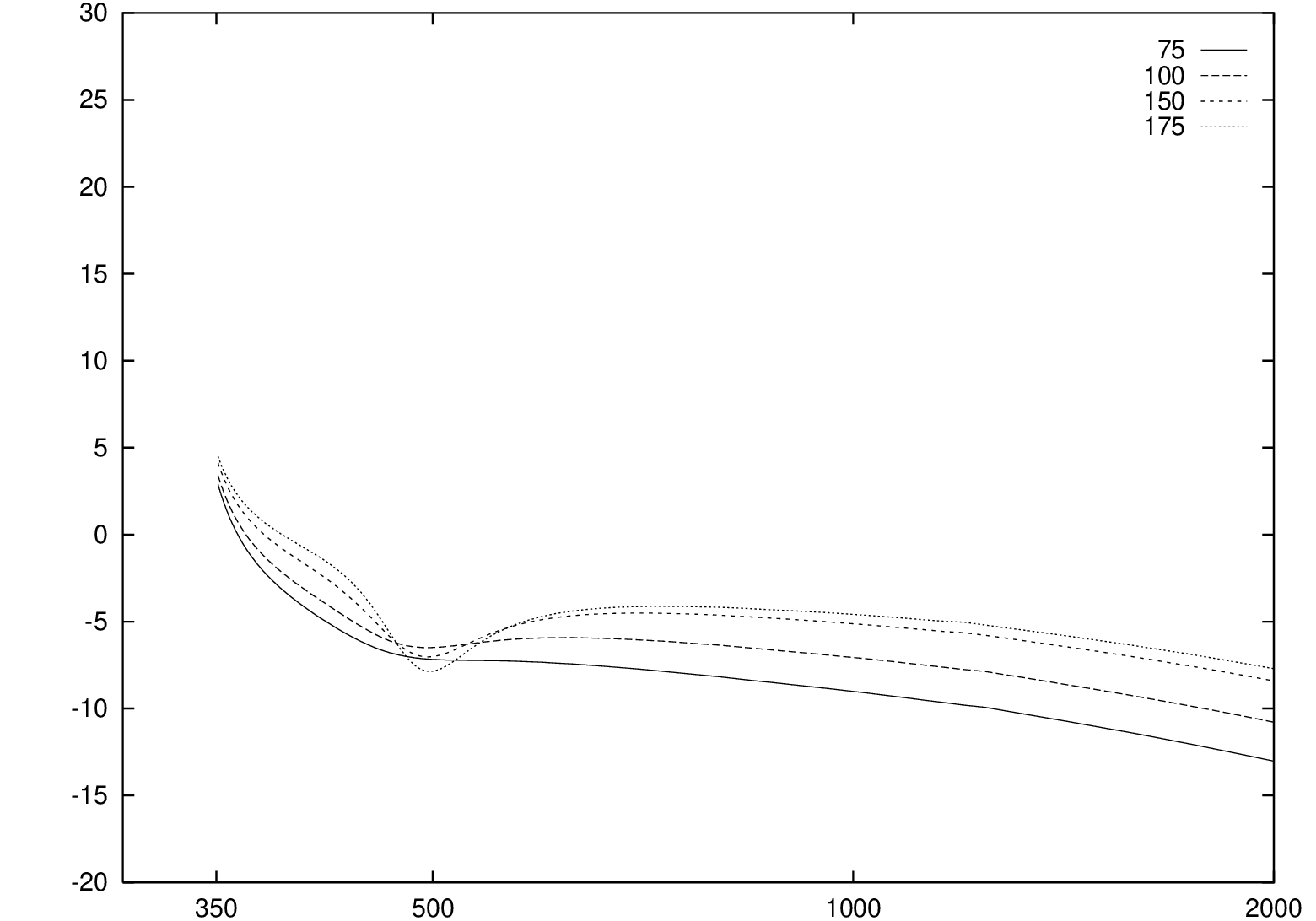}}
\put(12.5,7.25){\makebox(0,0){$\sqrt{\hat s}$ , GeV}}
\put(13,13.25){\makebox(0,0){$\phist=\pi/4$}}
\put(-1.75,0){\shit{7cm}{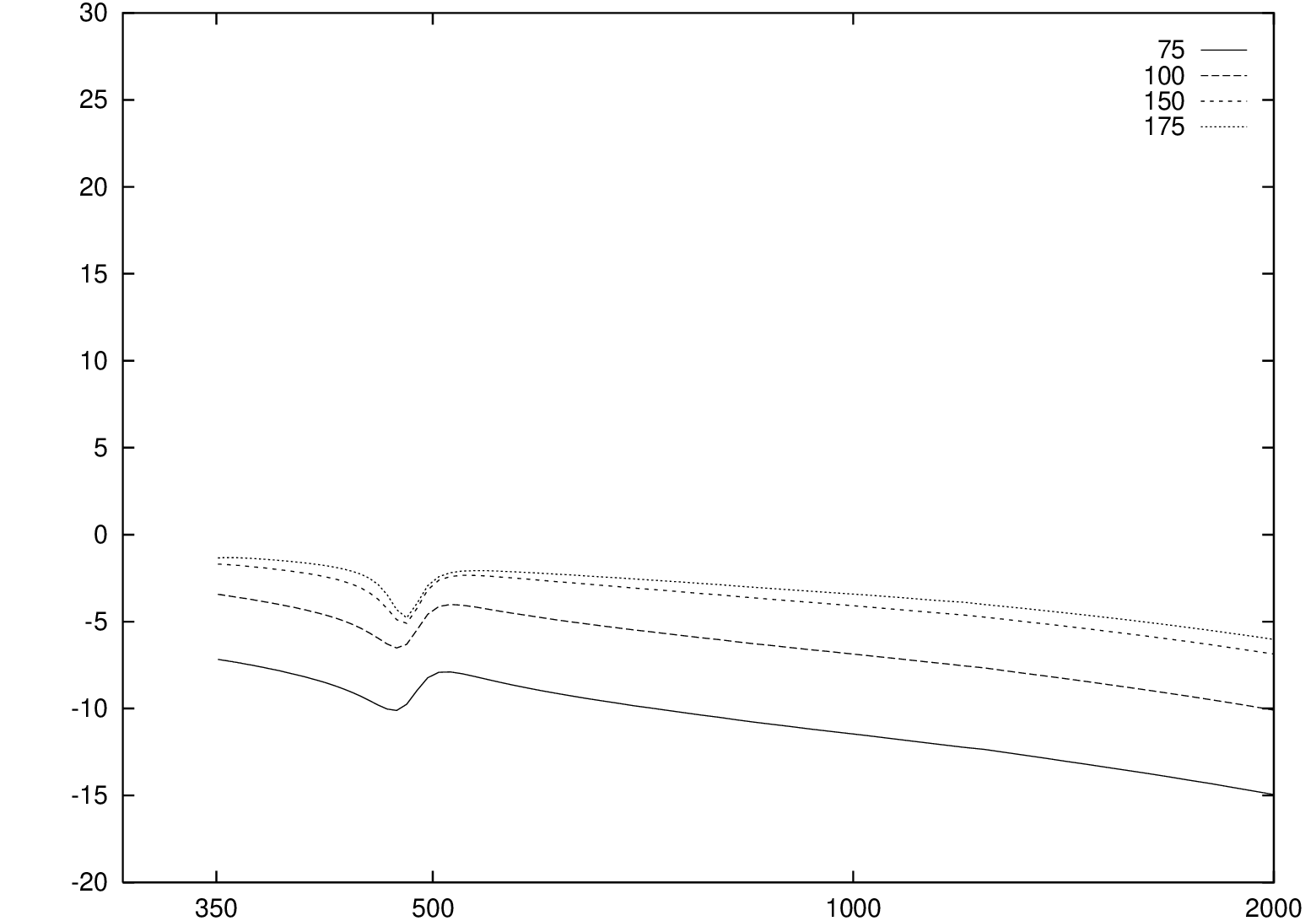}}
\put(3.25,0){\makebox(0,0){$\sqrt{\hat s}$ , GeV}}
\put(3.25,6){\makebox(0,0){$\phist=-\pi/4$}}
\put(7.,0){\shit{7cm}{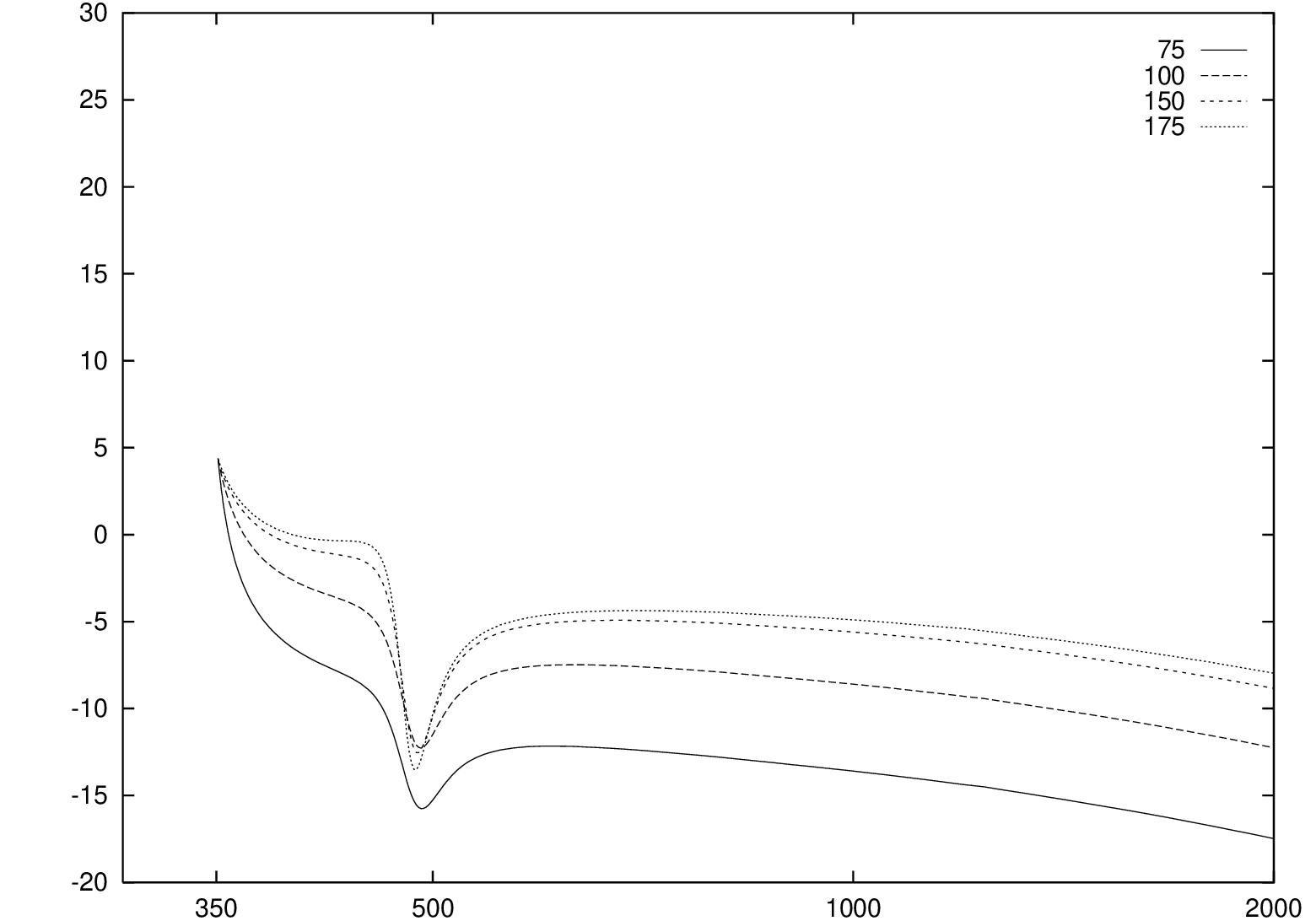}}
\put(12.5,0){\makebox(0,0){$\sqrt{\hat s}$ , GeV}}
\put(12.5,6){\makebox(0,0){$\phist=-\pi/4$}}
\end{picture}
\end{center}
\caption{The variation of the relative correction
$\Delta_{gg}$ with $\mstz$ for different values of
the mixing angle $\phist$ (with $\ma=450$ GeV, $\tanb=0.7$, 
$\msbe=400$ GeV, $\mu=150$ GeV and $M_2=3 |\mu|$).}
\end{figure}
\begin{figure}
\begin{center}
\setlength{\unitlength}{1cm}
\setlength{\fboxsep}{0cm}
\begin{picture}(16,20)
\put(-1.75,14.5){\shit{7cm}{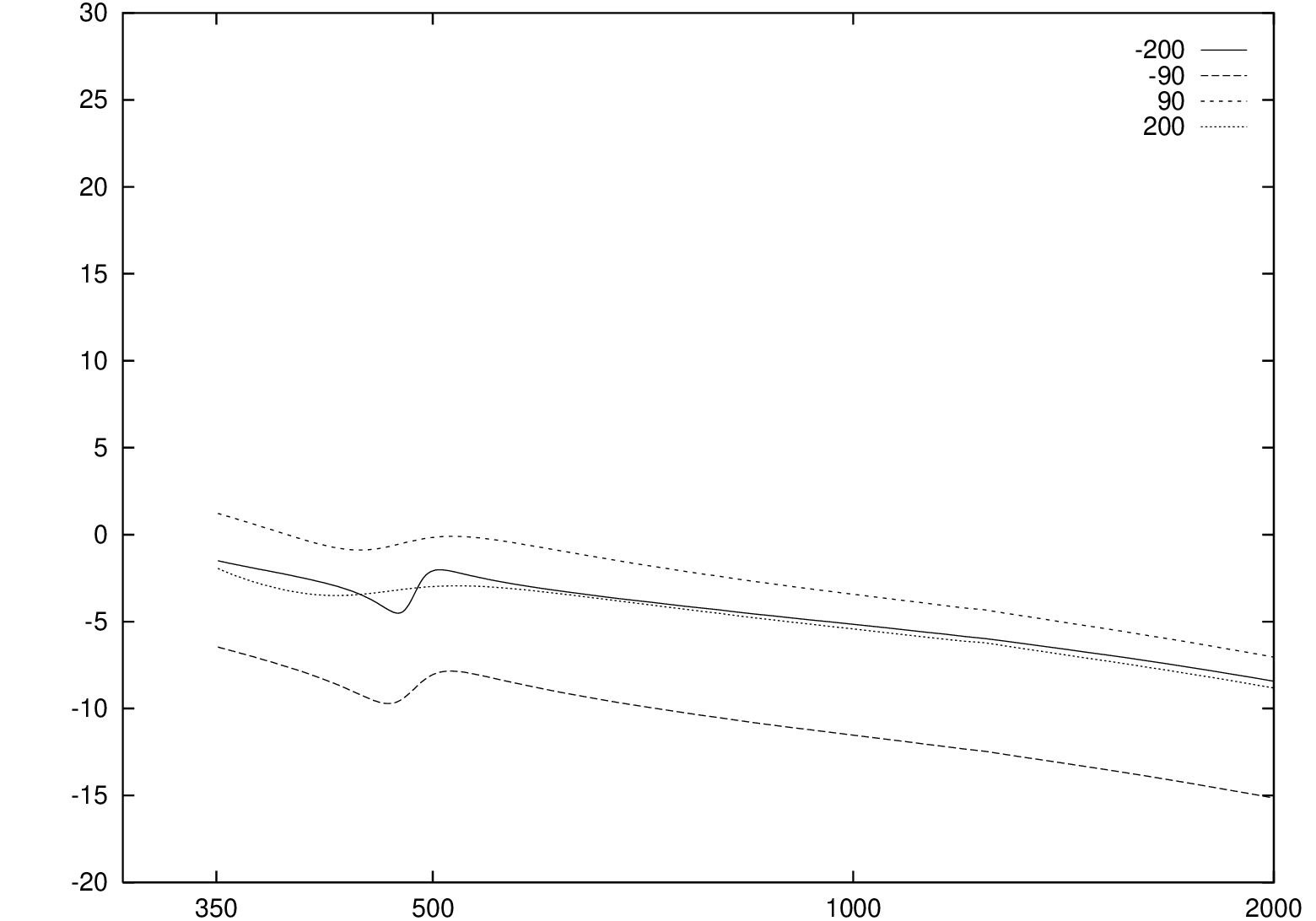}}
\put(1,21.5){\makebox(0,0){$\Delta_{gg} , \%$ SUSY EW-like}}
\put(3.25,14.5){\makebox(0,0){$\sqrt{\hat s}$ , GeV}}
\put(5.9,20.5){\makebox(0,0){$\mu:$}}
\put(3.25,20.5){\makebox(0,0){$\mstz=75$ GeV}}
\put(7.,14.5){\shit{7cm}{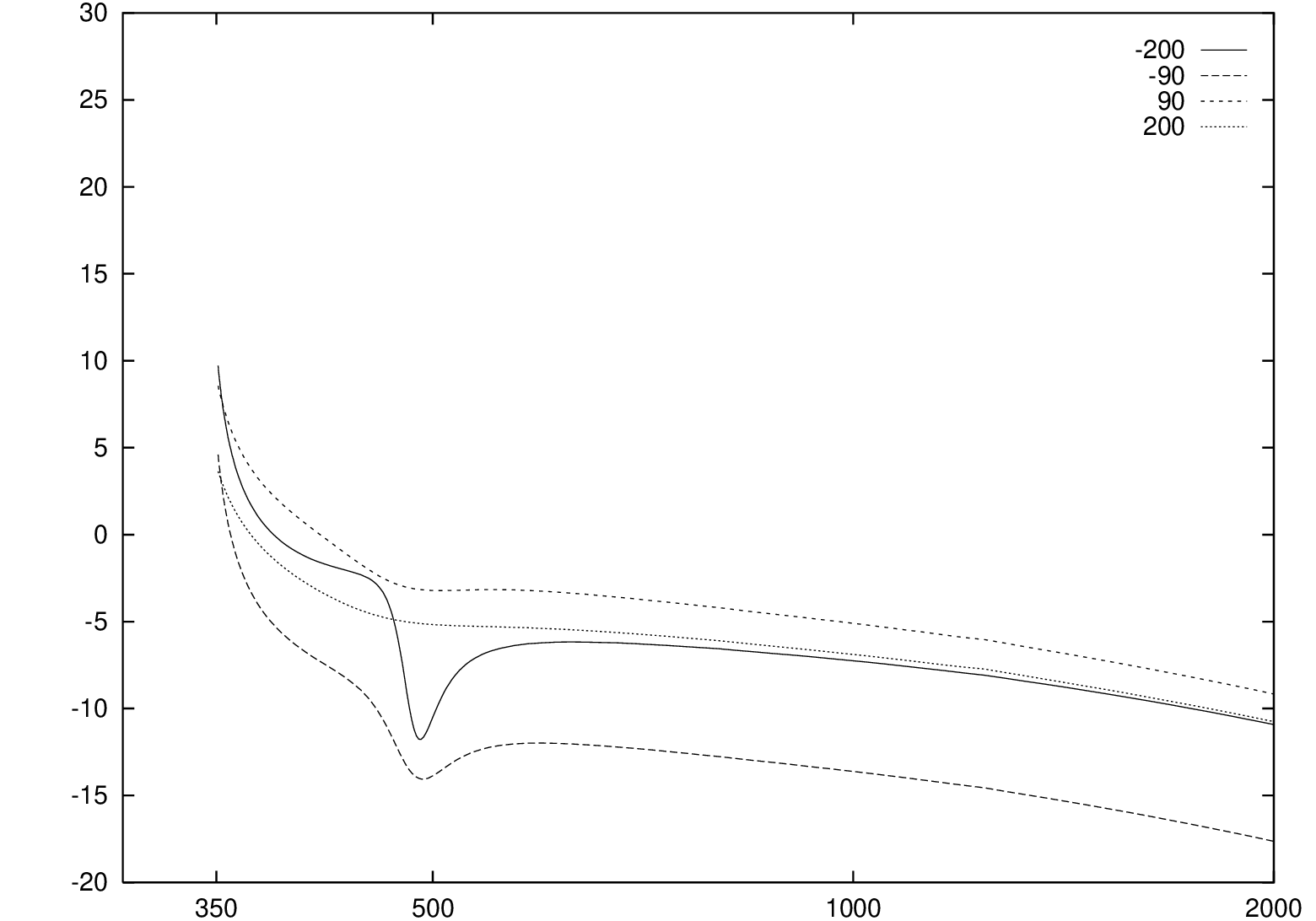}}
\put(9,21.5){\makebox(0,0){$\Delta_{gg} , \%$ MSSM}}
\put(12.5,14.5){\makebox(0,0){$\sqrt{\hat s}$ , GeV}}
\put(14.9,20.5){\makebox(0,0){$\mu:$}}
\put(12.5,20.5){\makebox(0,0){$\mstz=75$ GeV}}
\put(-1.75,7.25){\shit{7cm}{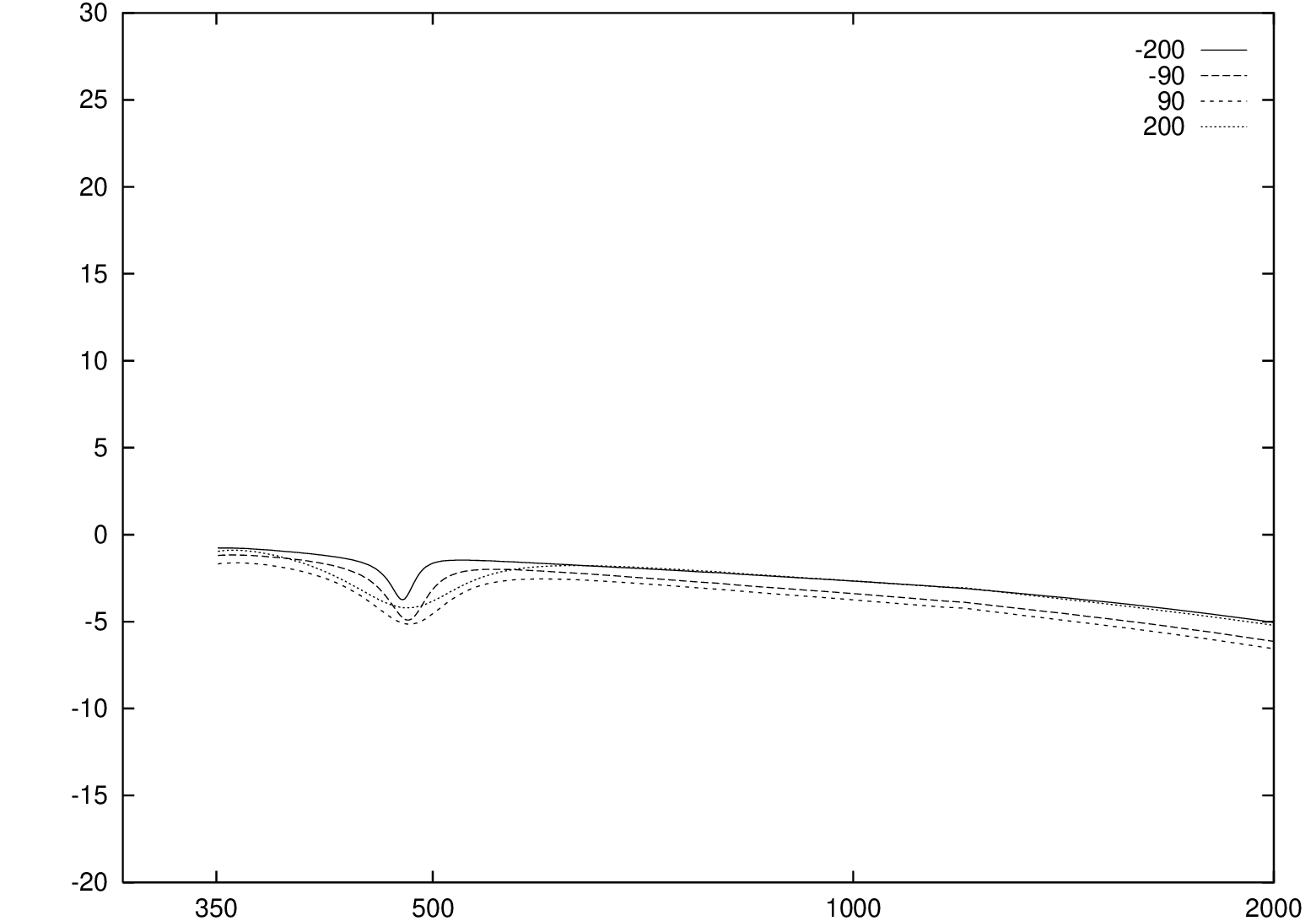}}
\put(3.25,7.25){\makebox(0,0){$\sqrt{\hat s}$ , GeV}}
\put(5.9,13.25){\makebox(0,0){$\mu:$}}
\put(3.25,13.25){\makebox(0,0){$\mstz=175$ GeV}}
\put(7.,7.25){\shit{7cm}{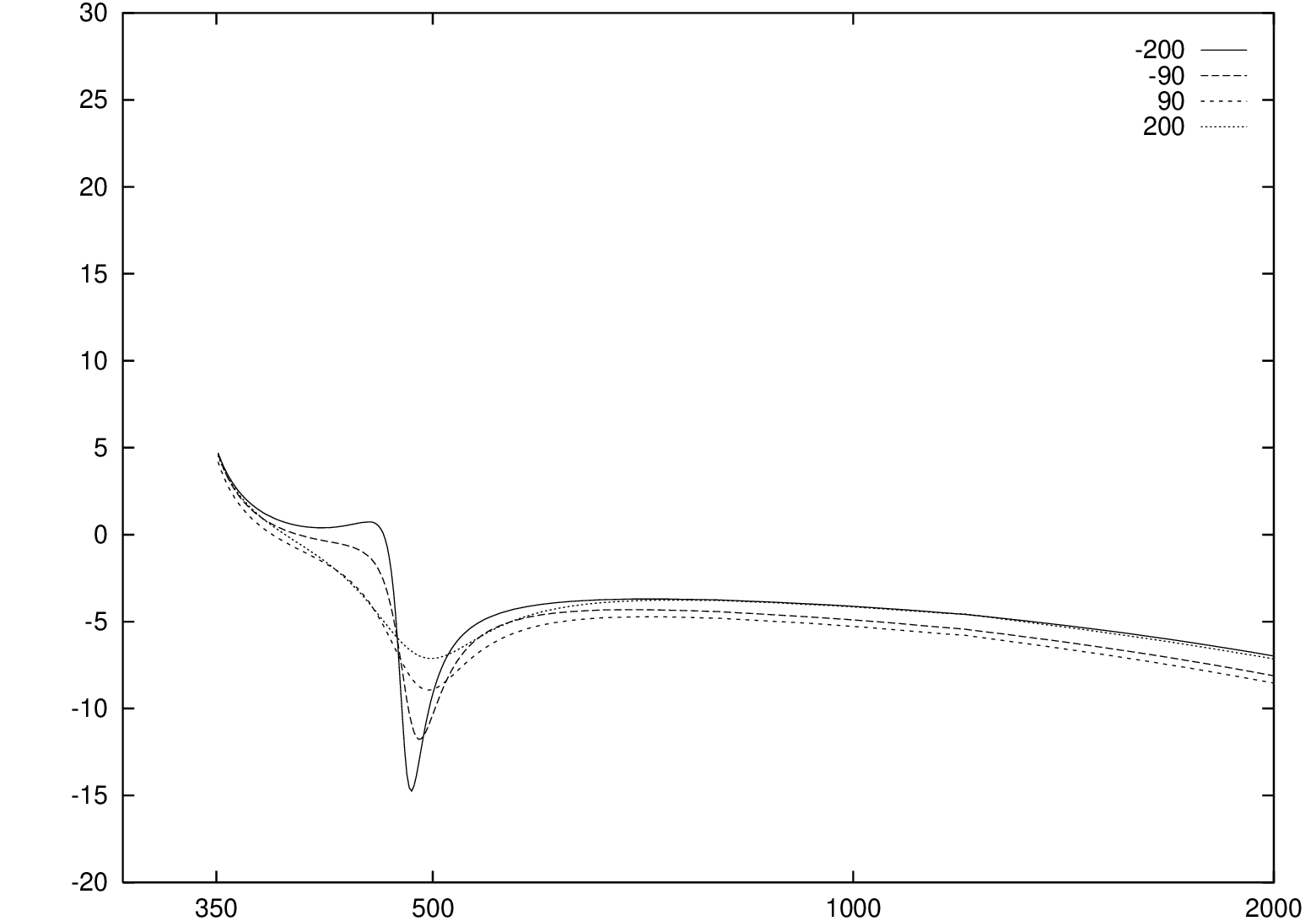}}
\put(12.5,7.25){\makebox(0,0){$\sqrt{\hat s}$ , GeV}}
\put(14.9,13.25){\makebox(0,0){$\mu:$}}
\put(12.5,13.25){\makebox(0,0){$\mstz=175$ GeV}}
\put(-1.75,0){\shit{7cm}{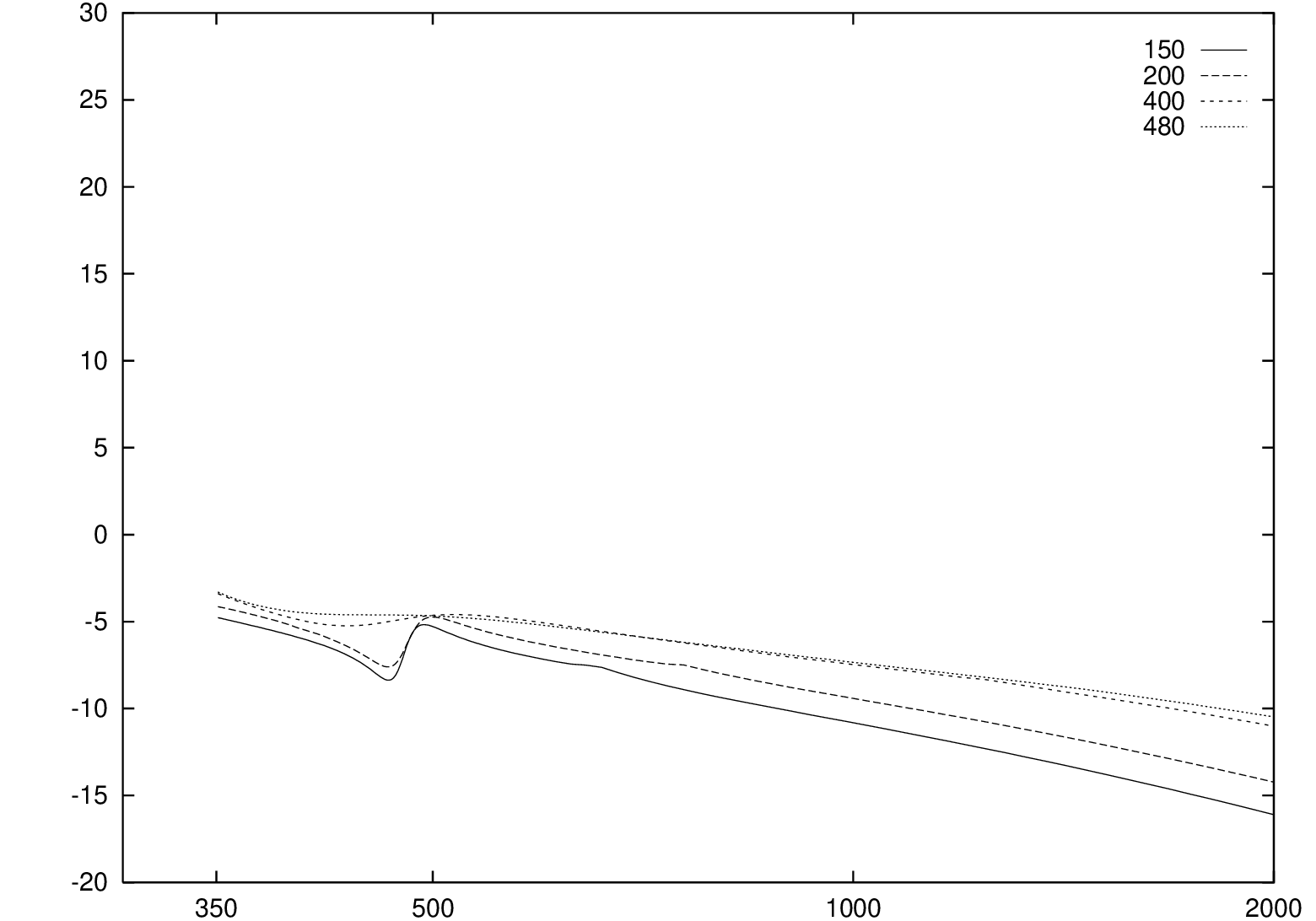}}
\put(3.25,0){\makebox(0,0){$\sqrt{\hat s}$ , GeV}}
\put(5.9,6){\makebox(0,0){$\msbe:$}}
\put(3.25,6){\makebox(0,0){$\mstz=75$ GeV}}
\put(7.,0){\shit{7cm}{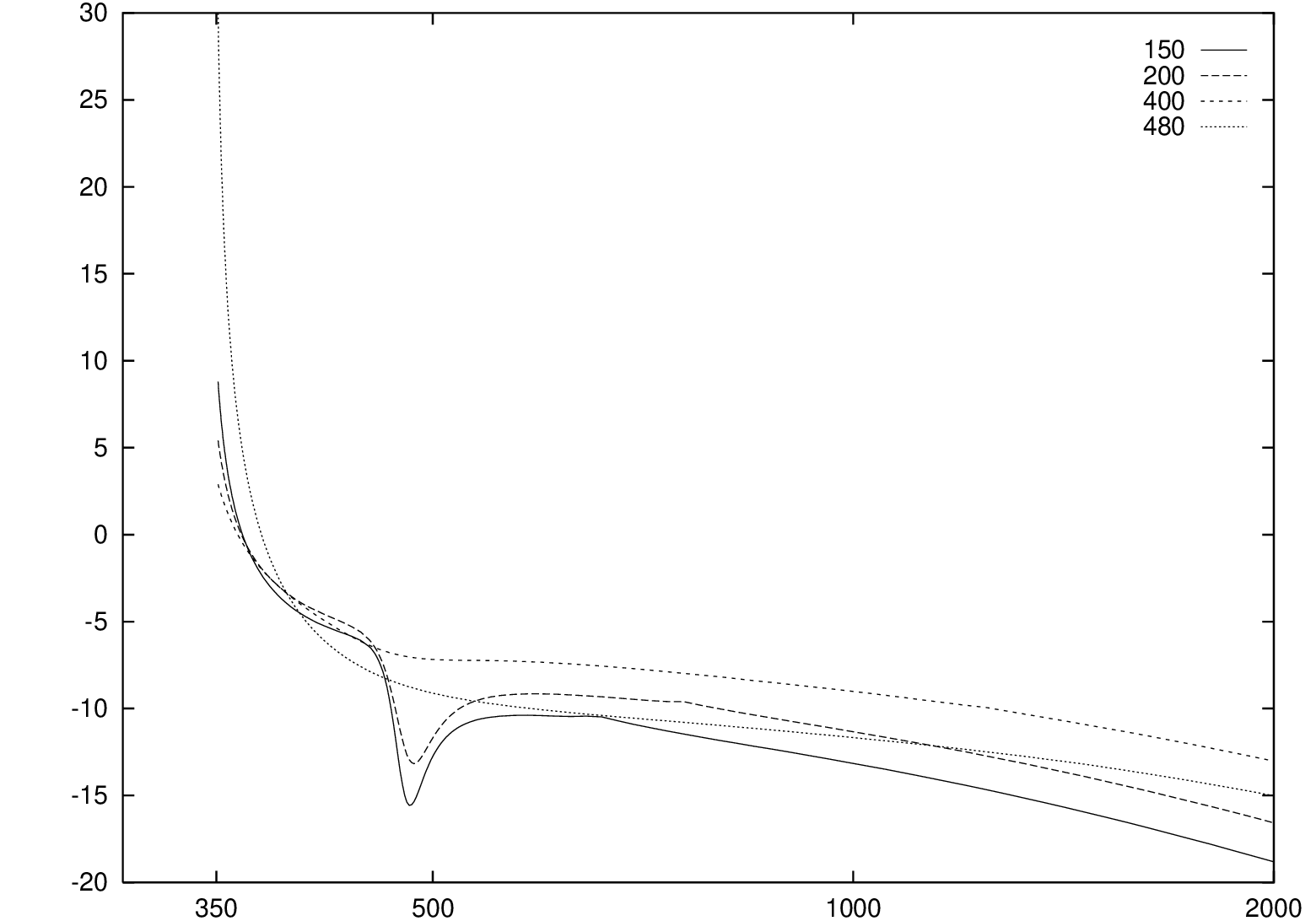}}
\put(12.5,0){\makebox(0,0){$\sqrt{\hat s}$ , GeV}}
\put(14.8,6){\makebox(0,0){$\msbe:$}}
\put(12.5,6){\makebox(0,0){$\mstz=75$ GeV}}
\end{picture}
\end{center}
\caption{The variation of the relative correction
$\Delta_{gg}$ with $\mu$ for two values of $\mstz$ ($\msbe=400$ GeV) and 
variation with $\msbe$ ($\mu=150$ GeV) 
(with $M_2=3|\mu|$, $\ma=450$ GeV, $\tanb=0.7$ and $\phist=\pi/4$).}
\end{figure}
Although the $s$-channel Higgs-exchange diagrams provide an interesting
Higgs-specific structure its significance will only preveal 
after having performed
the convolution of the partonic cross sections with the
parton distribution functions.
\subsection{The $p\overline{p},pp \rightarrow \ttbar X$ cross section
to ${\cal O}(\alpha\alpha_s^2)$}
The observable hadronic cross section is
obtained by convoluting the partonic cross sections of Eq.~(\ref{parton})
with parton distribution functions
\begin{equation}
\sigma(S) = \int_{\frac{4 \mt^2}{S}}^1 \frac{d\tau}{\tau}
\left(\frac{1}{S} \frac{dL_{\qqbar}}{d\tau} \; \hat s \hat \sigma_{\qqbar}(\hat s,\alpha_s(\mu))
+\frac{1}{S} \frac{dL_{gg}}{d\tau} \; \hat s \hat \sigma_{gg}(\hat s,\alpha_s(\mu))\right)
\end{equation}
with $\tau=x_1 x_2=\hat s/S$ and the parton luminosities
\begin{equation}
\frac{d L_{ij}}{d\tau}=\frac{1}{1+\delta_{ij}}
\int_{\tau}^1 \frac{dx_1}{x_1} \left[ f_i(x_1,Q)f_j(\frac{\tau}{x_1},Q)
+(1 \leftrightarrow 2)\right] \; .
\end{equation}
In the numerical evaluation
we use the MRSA set of parton distribution functions~\cite{mrsa}
with the factorization ($Q$) and renormalization scale ($\mu$) chosen to be
$Q=\mu=\mt$.

In order to avoid numerical instabilities and
to take into account that jets originating from the produced
top quarks at large scattering angles are better
distinguishable from the background we impose a cut on
the transverse momentum $p_t$ and the pseudo rapidity $\eta$ of the 
top quark in the cm frame:
$p_t>$20, 100 GeV (Tevatron, LHC) and $|\eta|<$ 2.5.
Imposing a transverse momentum cut also enhances
the relative correction at hadron level 
\[ \sigma(S) = \sigma_B(S)+ \delta \sigma(S)=\sigma_B (1+\Delta) \; ,\]
which again is introduced to reveal the numerical impact
of the ${\cal O}(\alpha)$ contribution
on the observable cross sections.
\begin{figure}[htb]
\begin{center}
\setlength{\unitlength}{1cm}
\setlength{\fboxsep}{0cm}
\begin{picture}(16,6)
\put(-1.75,0){\shit{7cm}{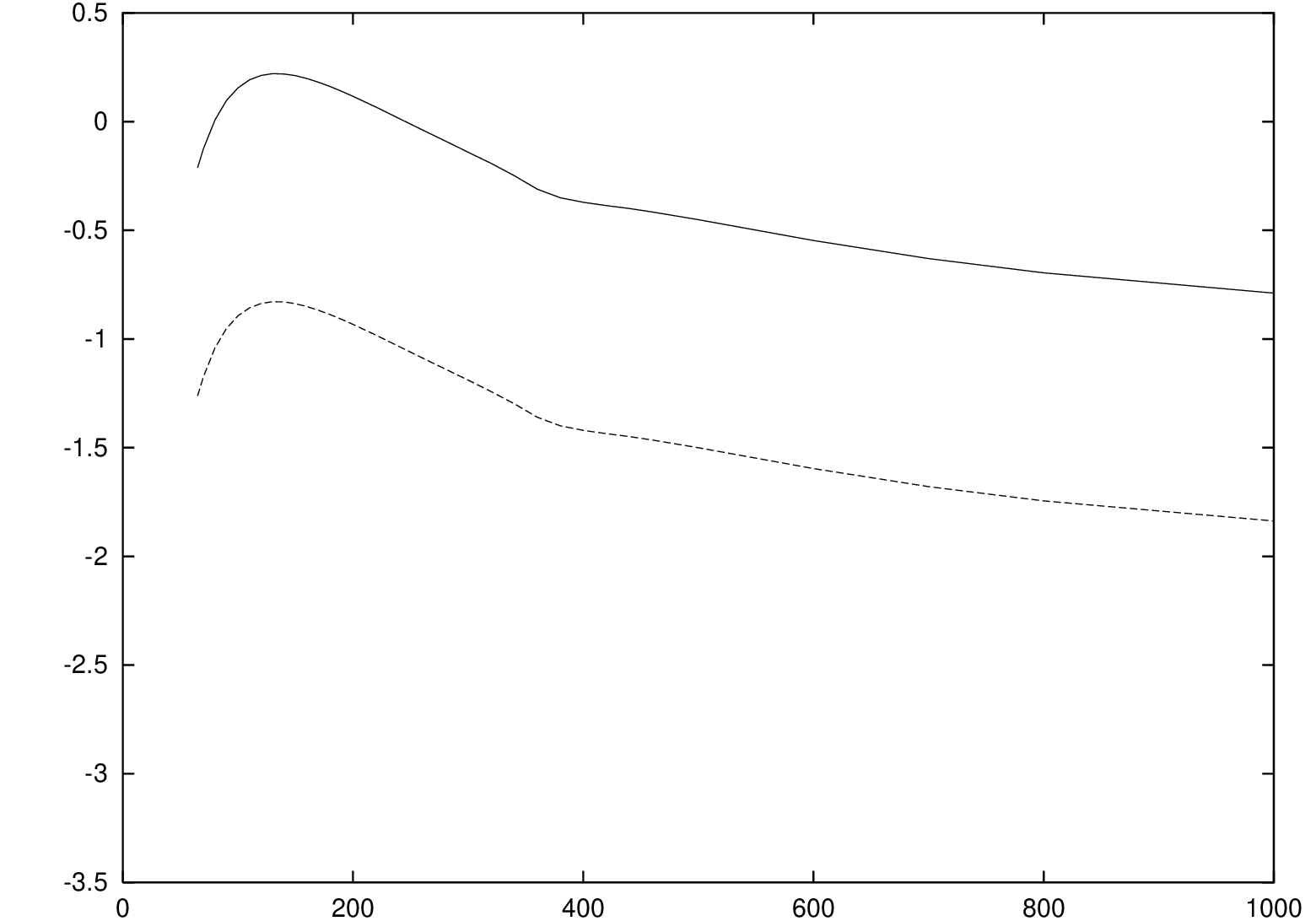}}
\put(0,7){\makebox(0,0){$\Delta , \%$}}
\put(3.25,0){\makebox(0,0){$M_H$ , GeV}}
\put(4.75,2.5){\makebox(0,0){gauge+Higgs}}
\put(4,7){\makebox(0,0){Tevatron}}
\put(7.,0){\shit{7cm}{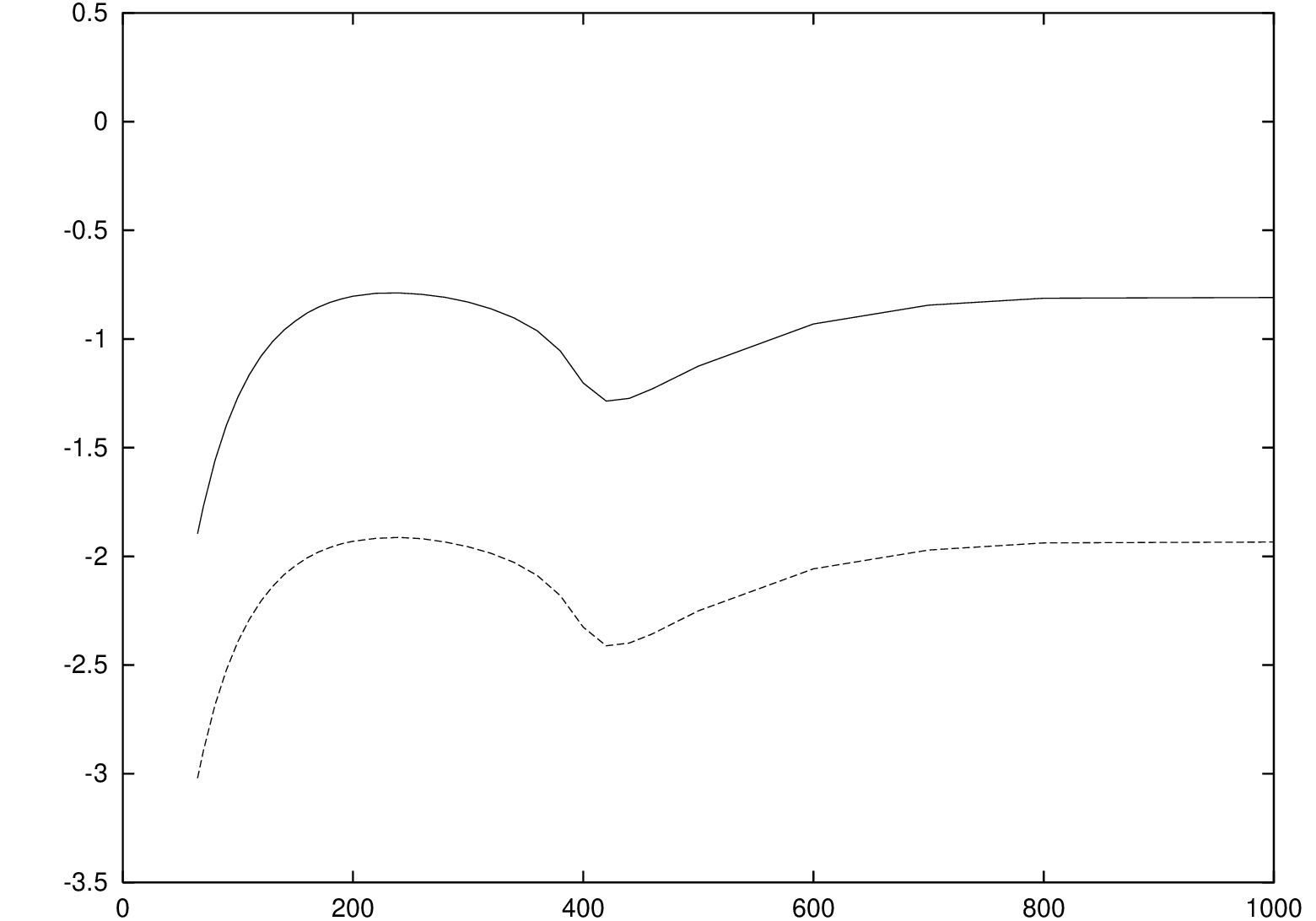}}
\put(8.5,7){\makebox(0,0){$\Delta , \%$}}
\put(12.5,0){\makebox(0,0){$M_H$ , GeV}}
\put(14,2){\makebox(0,0){gauge+Higgs}}
\put(12,7){\makebox(0,0){LHC}}
\end{picture}
\end{center}
\caption{The variation of the relative correction $\Delta$
with the MSM Higgs-boson mass $M_H$ at the Tevatron
($S= (2 \mbox{TeV})^2$) and at the LHC ($S= (14 \mbox{TeV})^2$).
The ${\cal O}(\alpha)$ contribution from the Higgs-sector is shown separately 
(solid line).}
\end{figure}
\begin{figure}[htb]
\begin{center}
\setlength{\unitlength}{1cm}
\setlength{\fboxsep}{0cm}
\begin{picture}(16,6)
\put(-1.75,0){\shit{7cm}{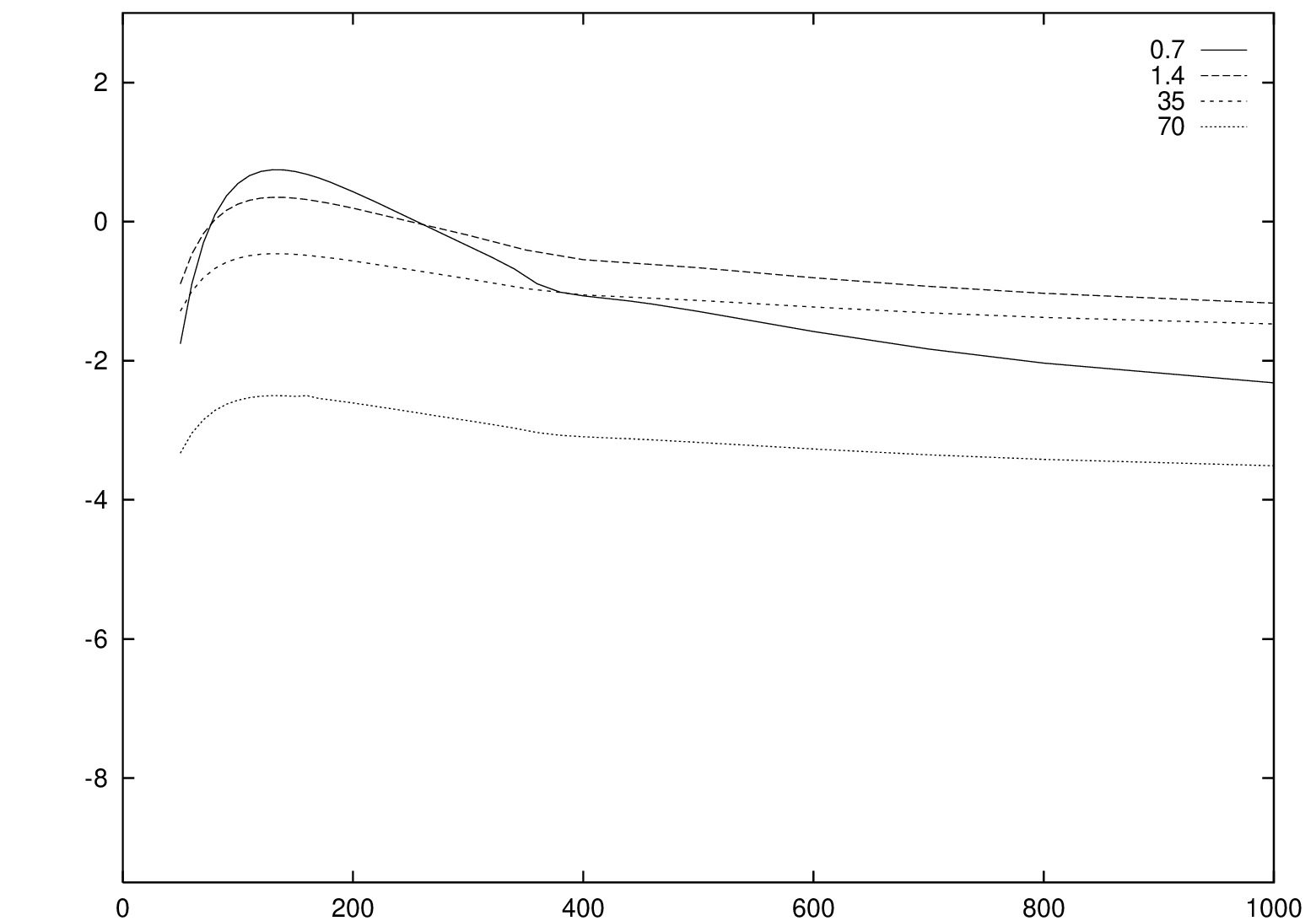}}
\put(0,7){\makebox(0,0){$\Delta , \%$}}
\put(3.25,0){\makebox(0,0){$M_{H^0}$ , GeV}}
\put(4,7){\makebox(0,0){Tevatron}}
\put(7.,0){\shit{7cm}{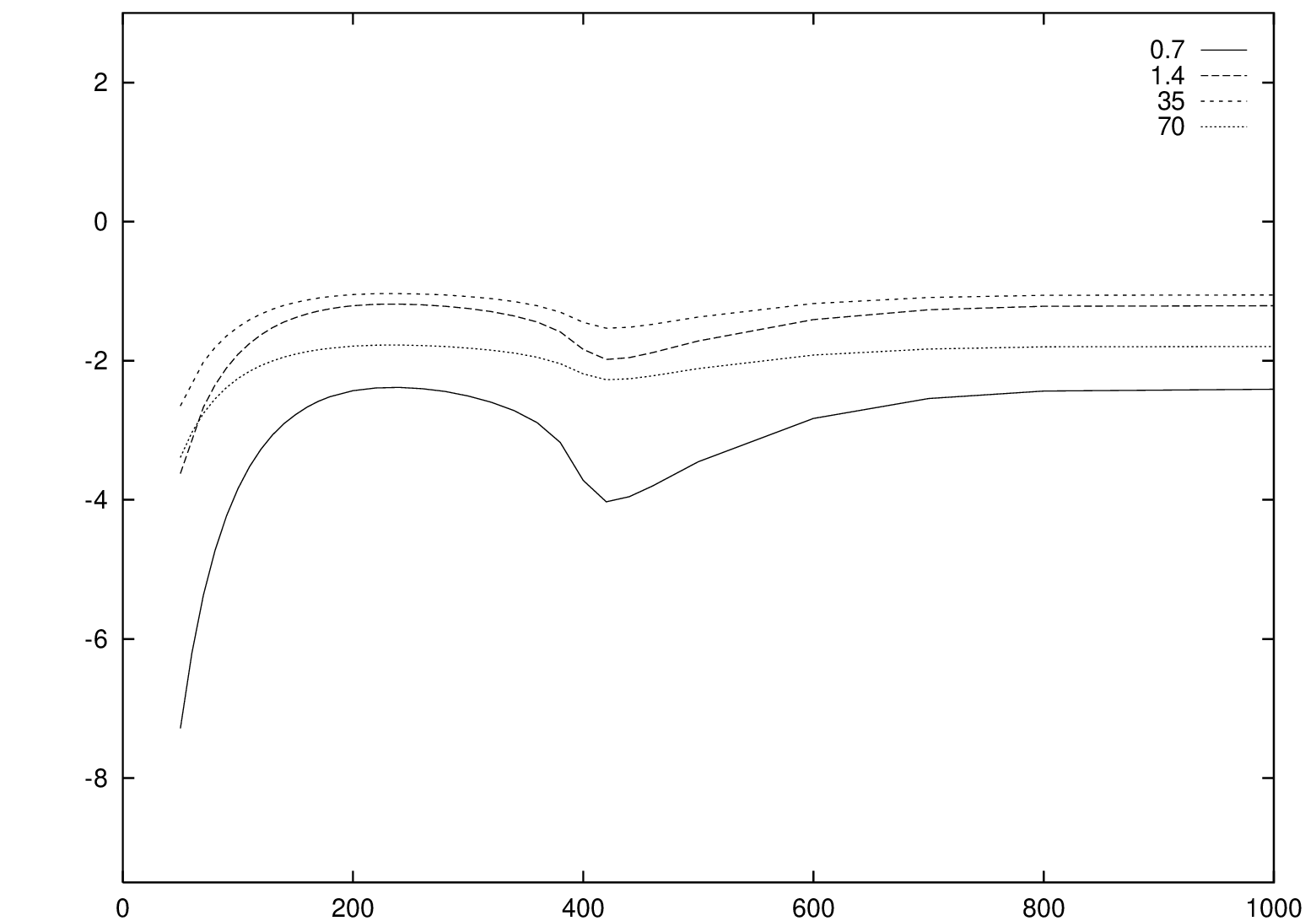}}
\put(8.5,7){\makebox(0,0){$\Delta , \%$}}
\put(12.5,0){\makebox(0,0){$M_{H^0}$ , GeV}}
\put(12,7){\makebox(0,0){LHC}}
\end{picture}
\end{center}
\caption{The variation of the relative correction $\Delta$
with $\mgh$ and $\tanb$ at the Tevatron ($S= (2 \mbox{TeV})^2$)
and at the LHC ($S= (14 \mbox{TeV})^2$) within the G2HDM
(with $\mkh=45$ GeV, $\ma=50$ GeV,
$\mhp=50$ GeV and $\alpha = \pi/2$).}
\end{figure}
\begin{figure}[htb]
\begin{center}
\setlength{\unitlength}{1cm}
\setlength{\fboxsep}{0cm}
\begin{picture}(16,6)
\put(-1.75,0){\shit{7cm}{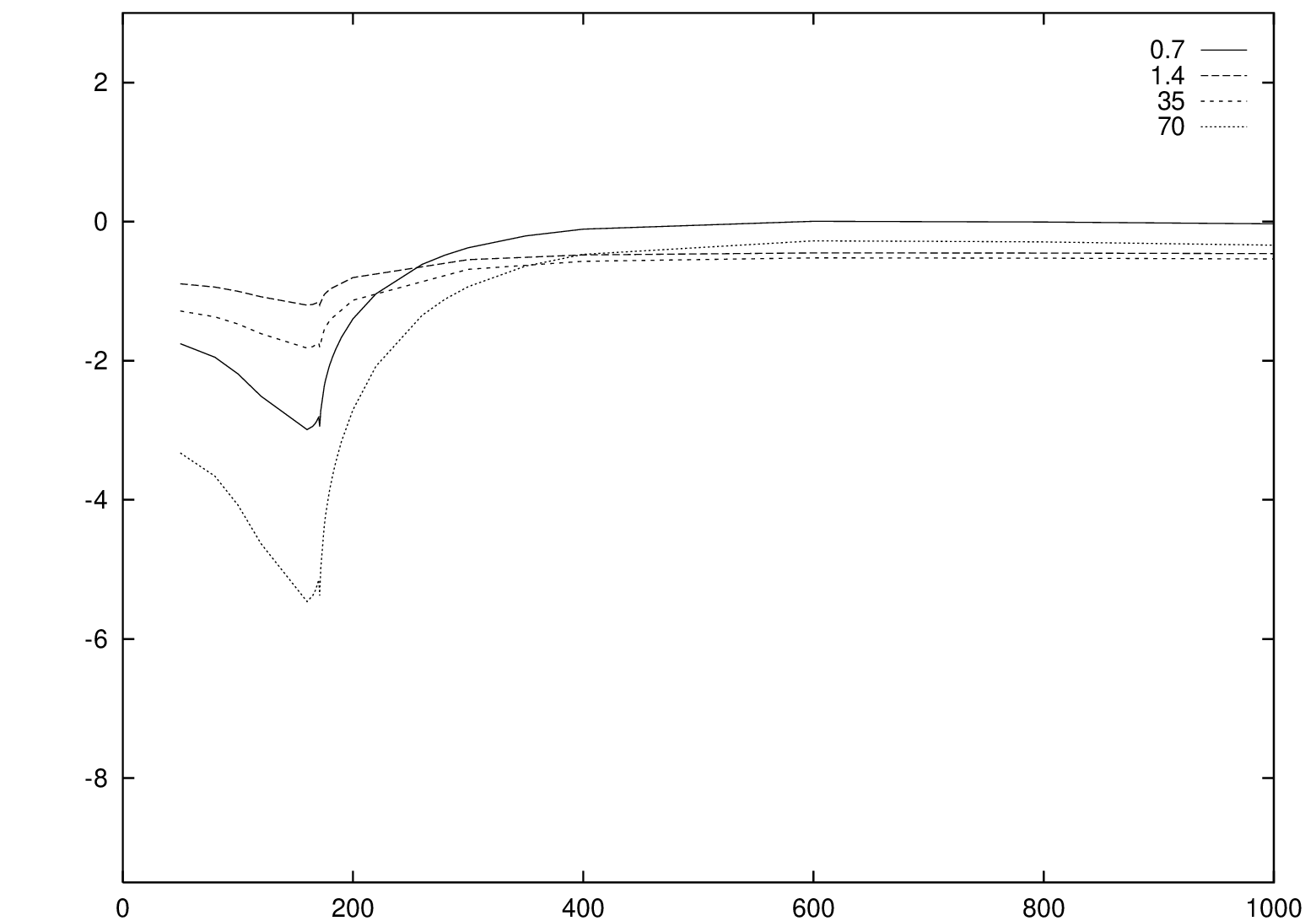}}
\put(0,7){\makebox(0,0){$\Delta , \%$}}
\put(3.25,0){\makebox(0,0){$M_{H^{\pm}}$ , GeV}}
\put(4,7){\makebox(0,0){Tevatron}}
\put(7.,0){\shit{7cm}{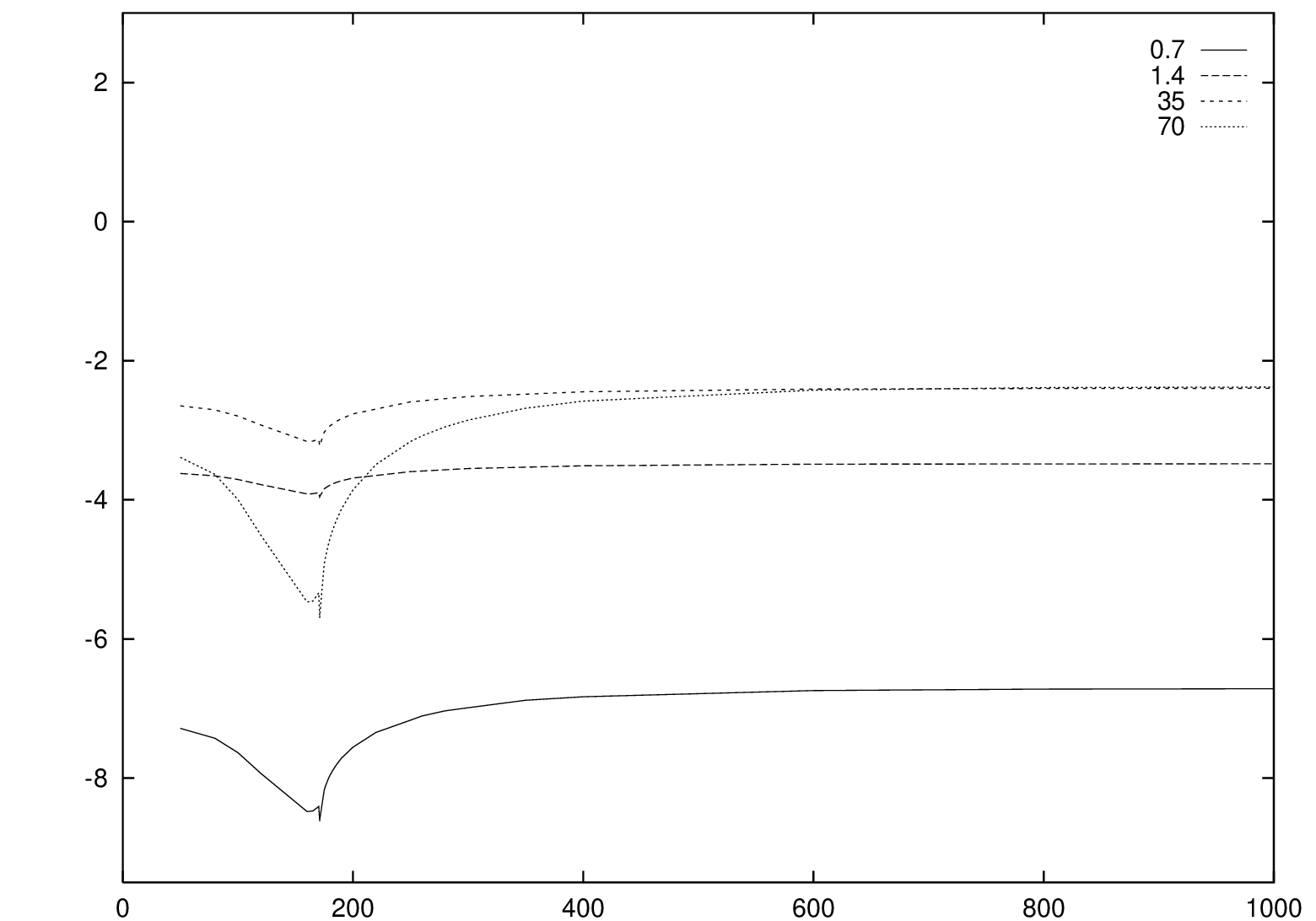}}
\put(8.5,7){\makebox(0,0){$\Delta , \%$}}
\put(12.5,0){\makebox(0,0){$M_{H^{\pm}}$ , GeV}}
\put(12,7){\makebox(0,0){LHC}}
\end{picture}
\end{center}
\caption{The variation of the relative correction $\Delta$
with $\mhp$ and $\tanb$ at the Tevatron ($S= (2 \mbox{TeV})^2$)
and at the LHC ($S= (14 \mbox{TeV})^2$) within the G2HDM
(with $\mkh=45$ GeV, $\ma=50$ GeV,
$\mgh=50$ GeV and $\alpha = \pi/2$).}
\end{figure}
Before we start the numerical discussion in detail we
describe the general characteristics of the
electroweak one-loop corrections:
at the parton level apart from a small region
close to the threshold $\sqrt{\hat s} \approx 2 \mt$,
where the ${\cal O}(\alpha)$ contribution
can be extremely large and positive,
the EW-like radiative corrections 
reduce the leading-order cross sections the more the larger
$\sqrt{\hat s}$.
At the hadron level, as a result of the interplay between
the partonic cross sections and the Bjorken-$x$ quark and gluon distributions
the relative corrections are predominantly negative,
only at the Tevatron where the
$\qqa$ subprocess dominates small positive
contributions arise as a remnant of the large corrections
arising in the threshold region.

We start the numerical discussion at hadron level with the MSM prediction.
In Fig.~10 the effects of the ${\cal O}(\alpha)$ Higgs- and
electroweak gauge boson contributions are shown separately.
Within the MSM the complete electroweak one-loop corrections
diminish the leading order observable cross sections at the Tevatron and
the LHC by up to $-2 \%$ and $-3 \%$, respectively.
As can be seen in Fig.~10 the consideration of the electroweak
gauge boson contribution yield a constant shift
of about $-1 \%$ at both colliders. In the following discussion
of non-minimal Standard Model implications
this effect is not explicitly included, since
it can apparently easily be taken into account.

In Fig.~11 we show the G2HDM prediction as a function of 
$\mgh$ for different values of $\tanb$.  
Within the G2HDM due to the enhancement of the Yukawa-couplings
by $\tanb$ the Born-cross sections can be
considerably diminished: at the Tevatron up to $-3.5 \%$ ($\tanb=70$
and $\mgh$ either very small or very high) and at the LHC up to
$-7.3 \%$ ($\tanb=0.7$ and $H^0$ very light).
Except for very small values of $\mgh$ the 
contribution only weakly depends on $\mgh$.
As already discussed at the parton level, the new increase
for $\tanb$=70 is due to the
enhancement of the top-Yukawa-coupling to the charged Higgs-boson. 
The possibility of a further
enhancement of the radiative corrections due to a discontinuity
in the derivative of the $B$-functions in the vicinity of the threshold
for top quark decay $t \rightarrow b+H^+$
is illustrated in Fig.~12. There we show
the relative corrections as a function of the charged
Higgs-boson mass $\mhp$ for different values of $\tanb$.

Within the MSSM the freedom of the G2HDM in choosing that set of parameters
which yield the maximum effect is limited by imposing 
supersymmetric constraints
on the 2-Higgs-Doublet Model. Thus,
the effects of the supersymmetric Higgs-sector are in general 
less pronounced than the one observed in the G2HDM. 
This is illustrated in Fig.~13, where the 
relative corrections obtained within the supersymmetric 2-Higgs-Doublet Model
are shown as a function of $\ma$ for different values of $\tanb$.
\begin{figure}
\begin{center}
\setlength{\unitlength}{1cm}
\setlength{\fboxsep}{0cm}
\begin{picture}(16,6)
\put(-1.75,0){\shit{7cm}{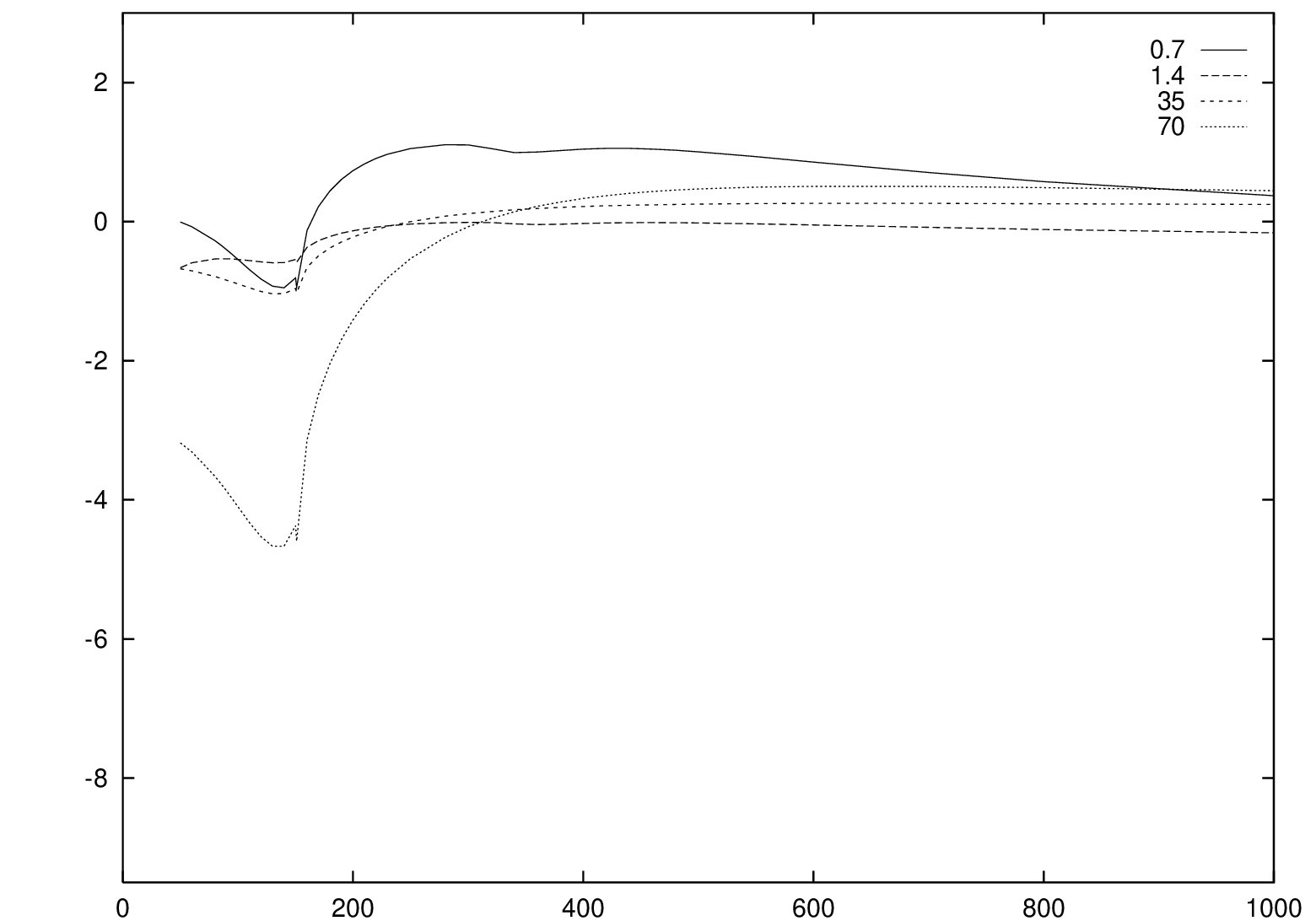}}
\put(0,7){\makebox(0,0){$\Delta , \%$}}
\put(3.25,0){\makebox(0,0){$M_{A^0}$ , GeV}}
\put(4,7){\makebox(0,0){Tevatron}}
\put(7.,0){\shit{7cm}{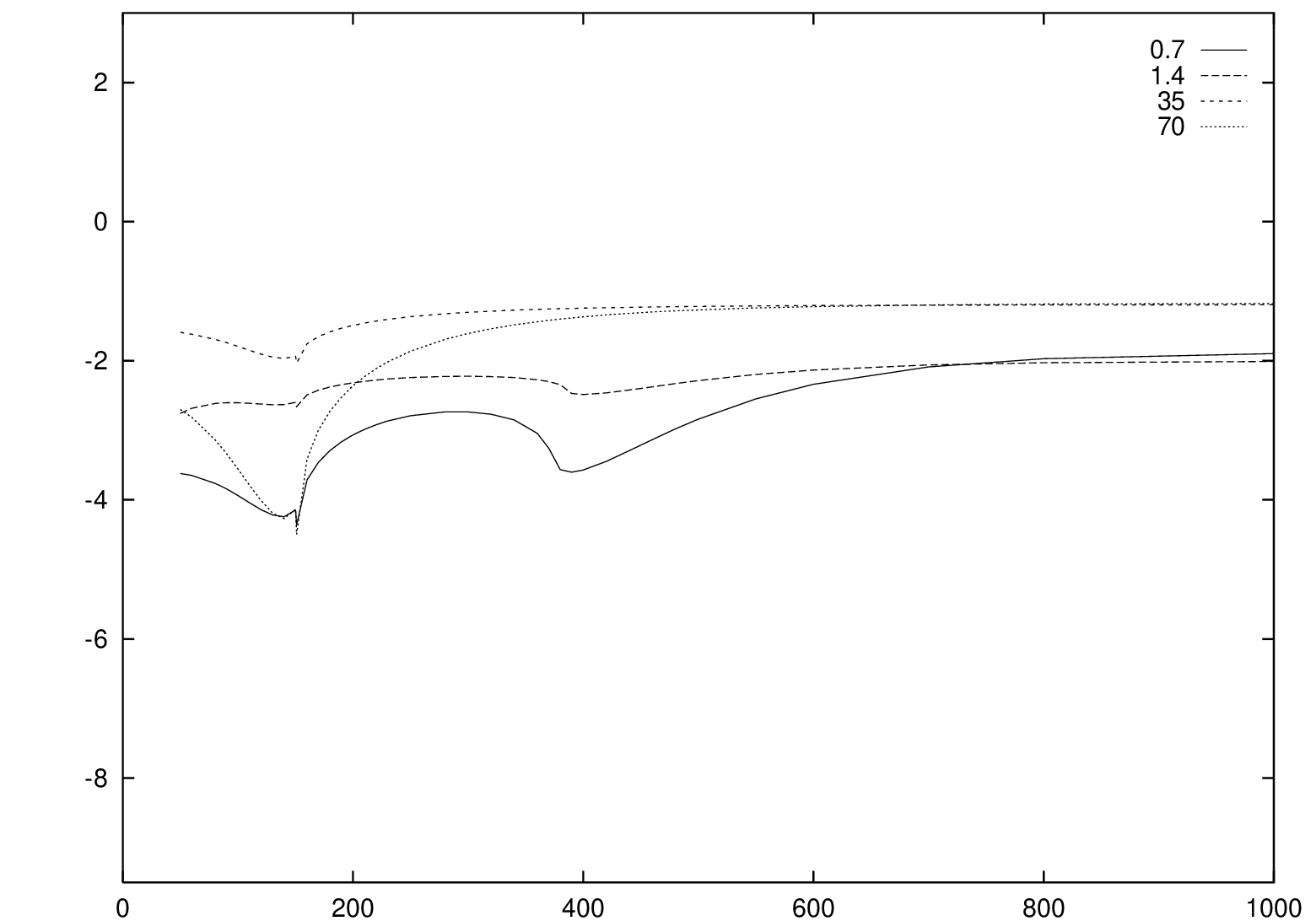}}
\put(8.5,7){\makebox(0,0){$\Delta , \%$}}
\put(12.5,0){\makebox(0,0){$M_{A^0}$ , GeV}}
\put(12,7){\makebox(0,0){LHC}}
\end{picture}
\end{center}
\caption{The variation of the relative correction $\Delta$
with $\ma$ and $\tanb$ at the Tevatron and at the LHC within the
supersymmetric 2-Higgs-Doublet Model
($\mstz=$75 GeV, $\msbe=$150 GeV, $\phist = \pi/4$ and $\mu=100$ GeV).}
\end{figure}
In Fig.~14 and Fig.~15 we illustrate the most pronounced effect
of the SUSY EW-like contributions: the enhancement of the 
radiative corrections when $m_t\approx \mstz+\mneut$. 
We show the relative corrections versus 
$\mstz$ for different values of $\tanb$ with and without mixing.
The absence of the second dip in Fig.~14 illustrates the possibility
of a cancellation
of the large contributions between the derivative of the $B_0$ and
$B_1$ functions depending on the choice of the $L,R$-mixing angle $\phist$
(see also~\cite{qcdew} (Tevatron)).
\begin{figure}[htb]
\begin{center}
\setlength{\unitlength}{1cm}
\setlength{\fboxsep}{0cm}
\begin{picture}(16,6)
\put(-1.75,0){\shit{7cm}{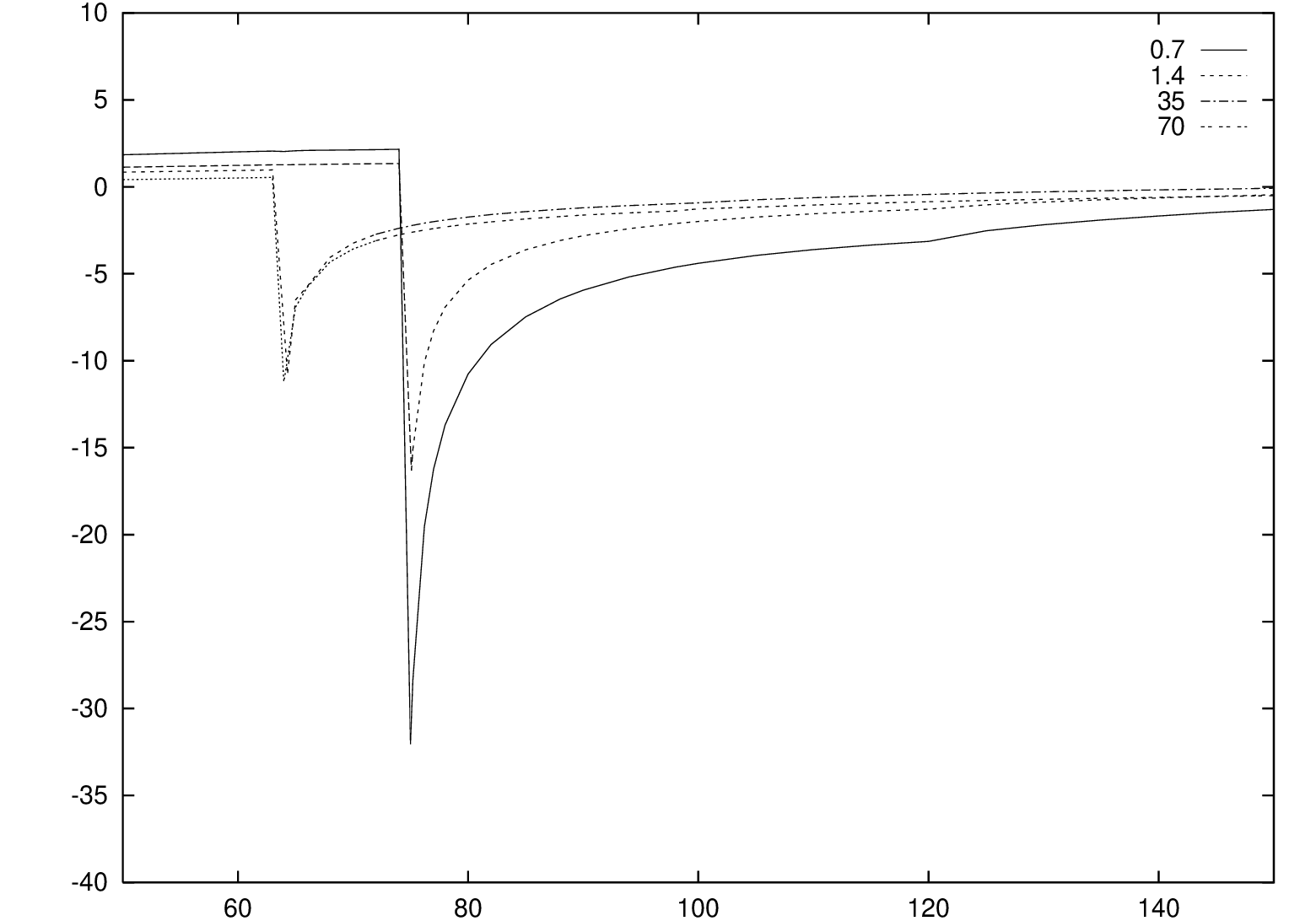}}
\put(0,7){\makebox(0,0){$\Delta , \%$}}
\put(3.25,0){\makebox(0,0){$\mstz$ , GeV}}
\put(4,7){\makebox(0,0){Tevatron}}
\put(7.,0){\shit{7cm}{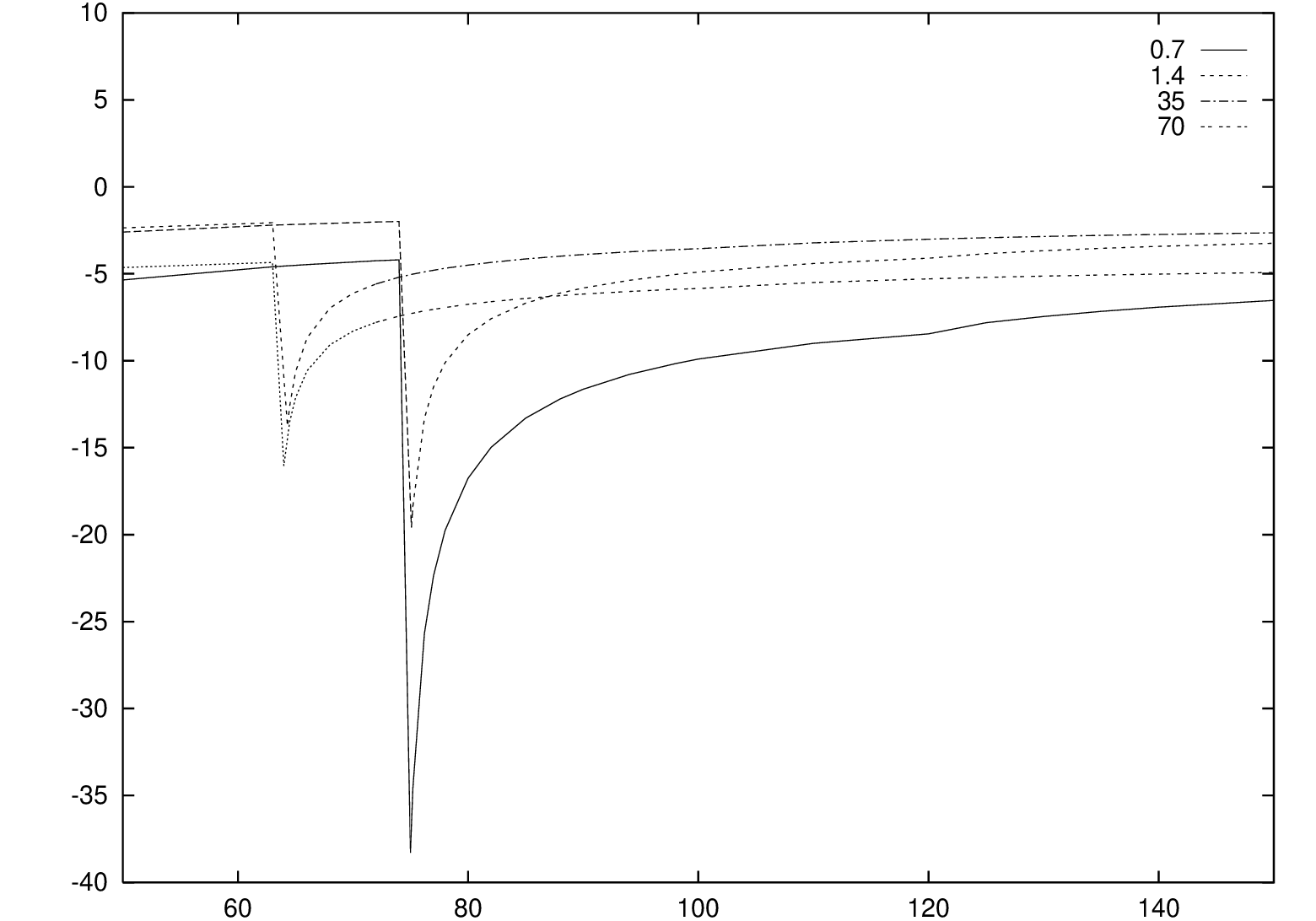}}
\put(8.5,7){\makebox(0,0){$\Delta , \%$}}
\put(12.5,0){\makebox(0,0){$\mstz$ , GeV}}
\put(12,7){\makebox(0,0){LHC}}
\end{picture}
\end{center}
\caption{The variation of the relative correction $\Delta$
with $\mstz$ and $\tanb$ at the Tevatron and at the LHC when
including only the SUSY EW-like one-loop corrections
(with $\msbe=150$ GeV, $\phist=\pi/4$, $\mu=100$ GeV and $M_2=3|\mu|$).}
\end{figure}
\begin{figure}[htb]
\begin{center}
\setlength{\unitlength}{1cm}
\setlength{\fboxsep}{0cm}
\begin{picture}(16,6)
\put(-1.75,0){\shit{7cm}{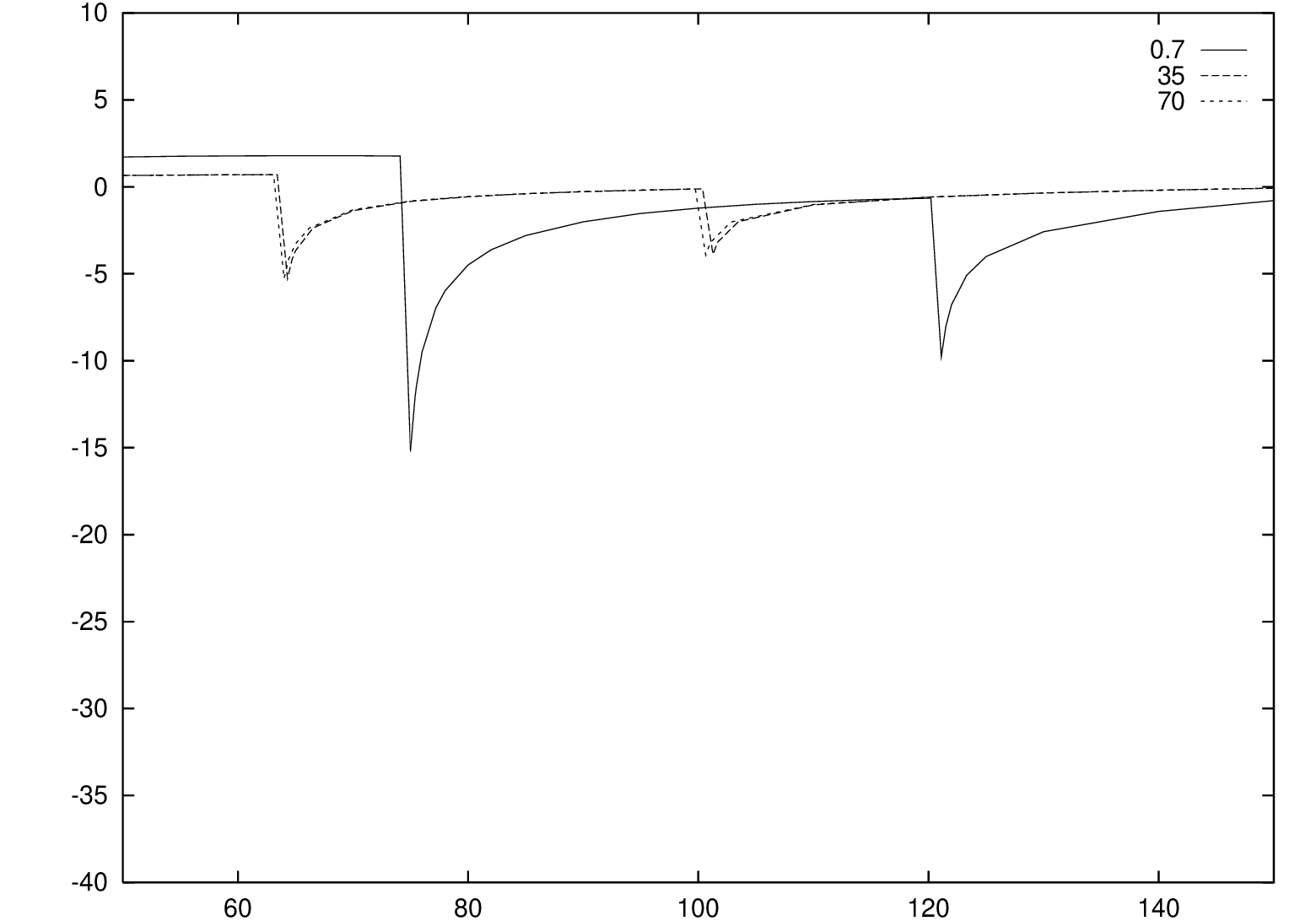}}
\put(0,7){\makebox(0,0){$\Delta , \%$}}
\put(3.25,0){\makebox(0,0){$\mstz$ , GeV}}
\put(4,7){\makebox(0,0){Tevatron}}
\put(7.,0){\shit{7cm}{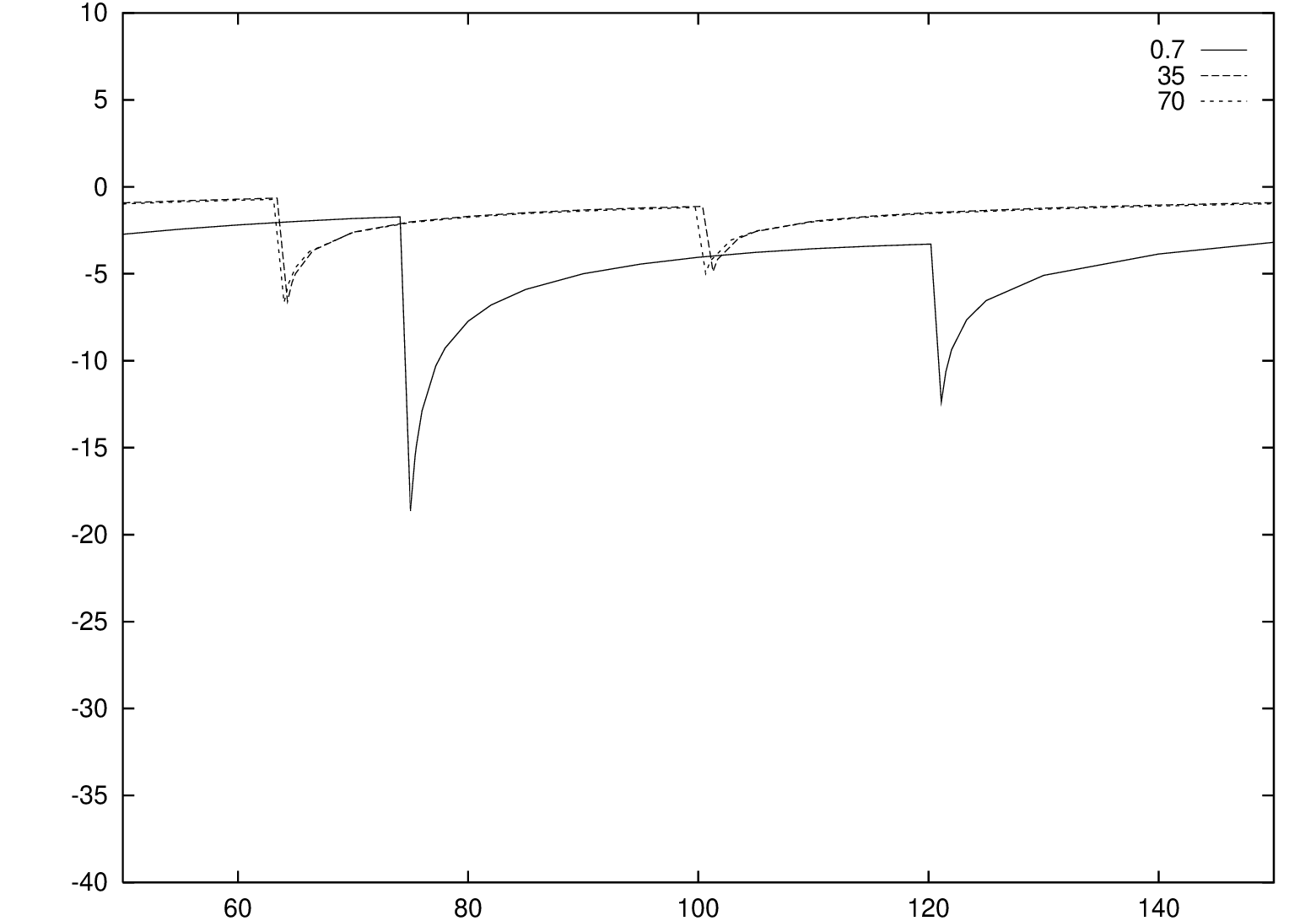}}
\put(8.5,7){\makebox(0,0){$\Delta , \%$}}
\put(12.5,0){\makebox(0,0){$\mstz$ , GeV}}
\put(12,7){\makebox(0,0){LHC}}
\end{picture}
\end{center}
\caption{The variation of the relative correction $\Delta$
with $\mstz$ and $\tanb$ at the Tevatron and at the LHC when including only the
SUSY EW-like one-loop corrections
(with $\mu=100$ GeV, $M_2=3 |\mu|$, $\msbe=1$ TeV and $\phist=0$).}
\end{figure}
Finally, we show in the Figs.~16,17,18
the impact of the complete ${\cal O}(\alpha)$ contribution
within the MSSM (including the $W,Z$ boson contribution)
on the hadronic cross sections for top pair production
at the Tevatron and the LHC.
There, we imposed additional constraints to account for
the experimental bounds on the supersymmetric particle mass spectrum
from the negative search at LEP
and only allow those parameter combinations which yield
\[ \mkh \;  > 45 \; \mbox{GeV}\]
\[ M_{\tilde \chi^{\pm}_{1,2}} > 65 \; \mbox{GeV}
 \; , \; \mneut^{lightest}> 24 \; \mbox{GeV}
\; ,\; \mneut^{next-to-lightest}>46 \; \mbox{GeV} \; .\]
We already emphasized the most interesting features of the
radiative corrections originating from the MSSM Higgs sector and
the SUSY EW-like contribution. To summarize we show the   
dependence of the relative corrections on the MSSM input parameters
\[\tanb, \ma, \mu, M_2, \msbe, \mstz, \phist\]
within the following MSSM scenarios:
\begin{itemize}
\item
Fig.~16:
variation with $\ma$ for different values of $\tanb$,
\item
Fig.~17:
variation with $\mu$ for different values of $M_2$,
\item
Fig.~18: variation with $\mstz$ for\\
a.) different values of $\tanb$ (with $\ma=450$ GeV, $\msbe=150$ GeV and
$\phist=\pi/4$), b.) different values of the $L,R$-mixing
angle $\phist$ (with $\ma=150$ GeV, $\tanb=0.7$ and $\msbe=500$ GeV)
and c.) different combinations of the values of
$(\msbe; \phist)$ (with $\ma=150$ GeV and $\tanb=0.7$).
\end{itemize}
\begin{figure}[htb]
\begin{center}
\setlength{\unitlength}{1cm}
\setlength{\fboxsep}{0cm}
\begin{picture}(16,6)
\put(-1.75,0){\shit{7cm}{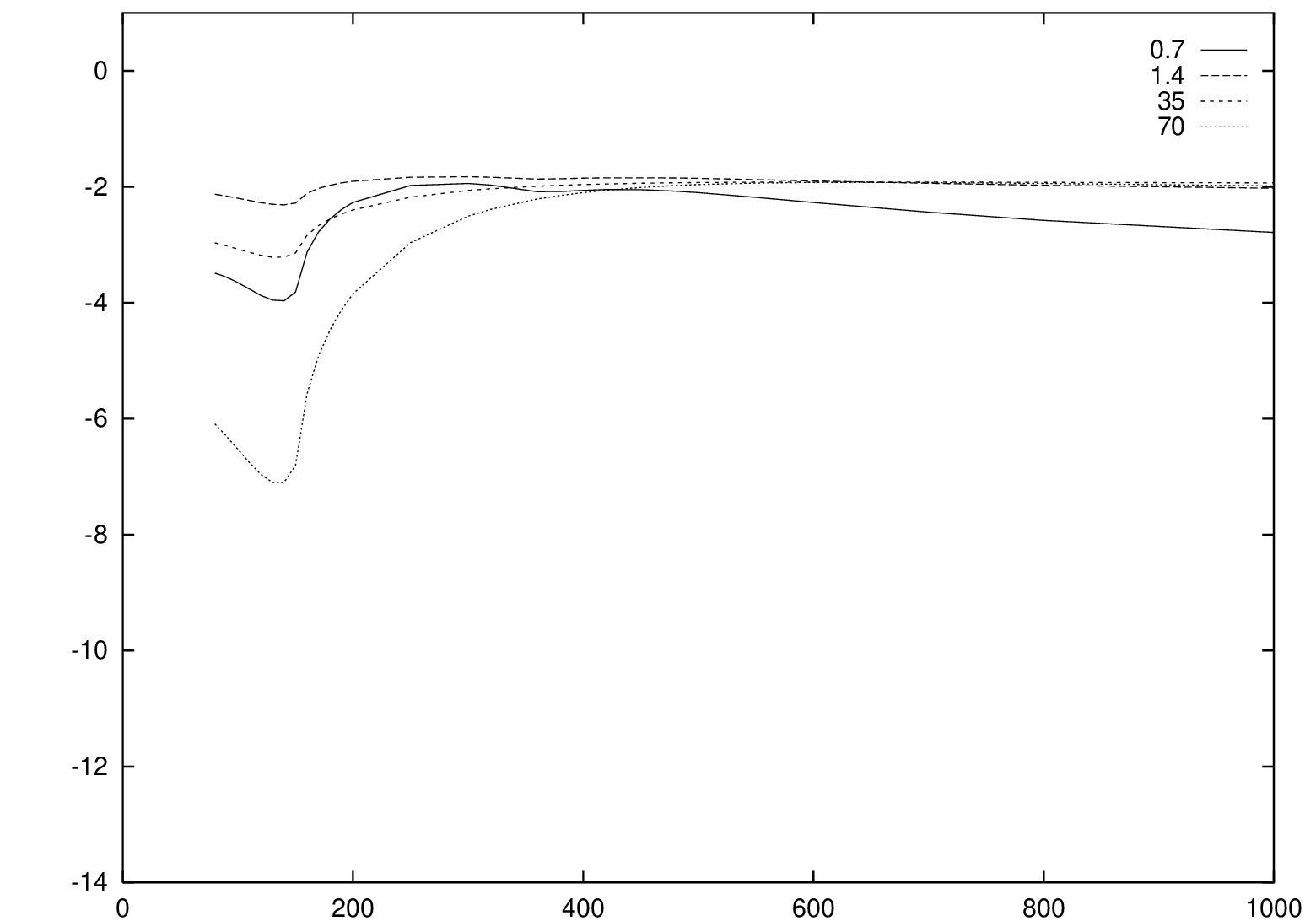}}
\put(0,7){\makebox(0,0){$\Delta , \%$}}
\put(3.25,0){\makebox(0,0){$\ma$ , GeV}}
\put(4,7){\makebox(0,0){Tevatron}}
\put(7.,0){\shit{7cm}{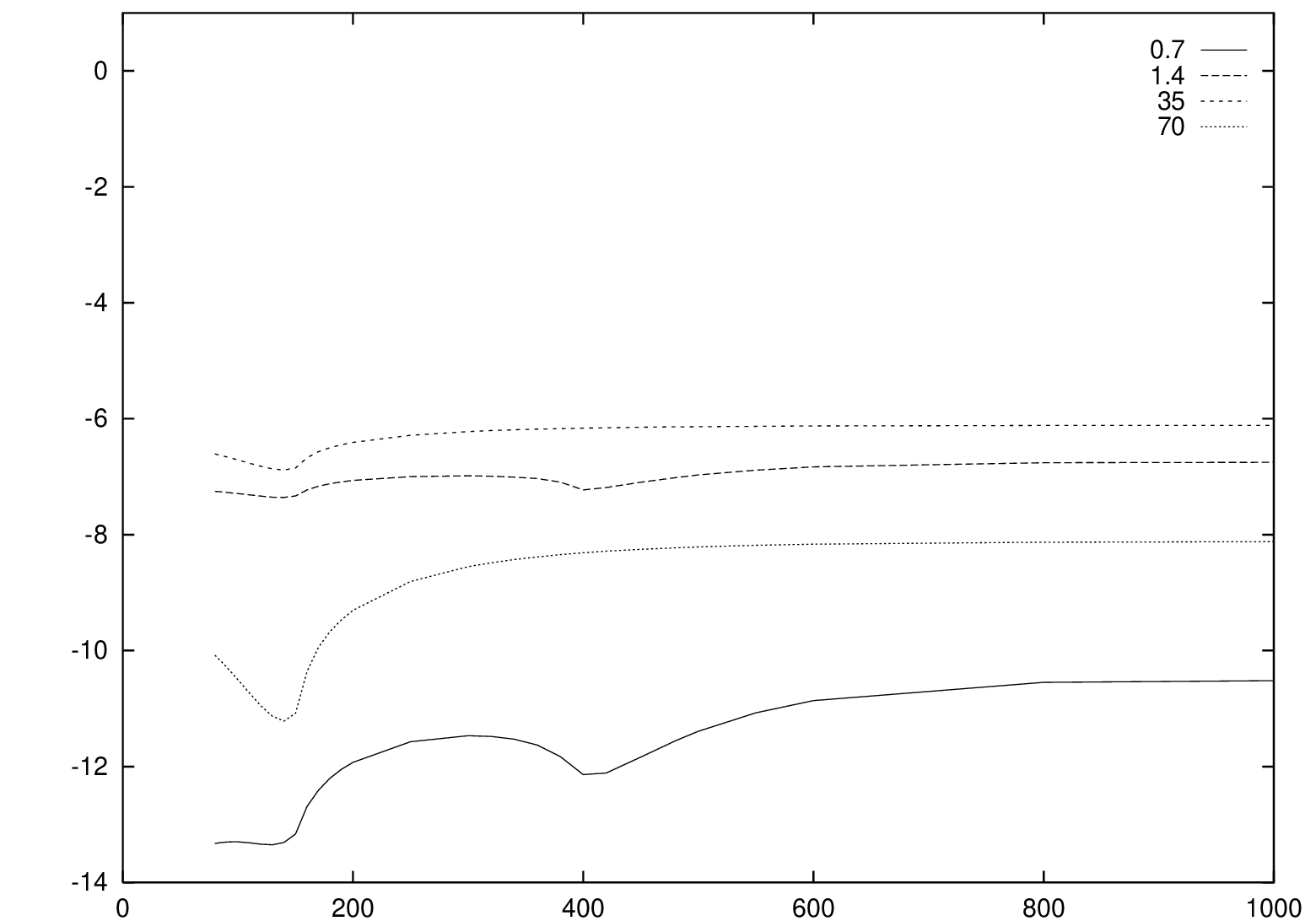}}
\put(8.5,7){\makebox(0,0){$\Delta , \%$}}
\put(12.5,0){\makebox(0,0){$\ma$ , GeV}}
\put(12,7){\makebox(0,0){LHC}}
\end{picture}
\end{center}
\caption{The variation of the relative correction $\Delta$
with $\ma$ and $\tanb$ at the Tevatron and at the LHC within the MSSM
(with $\mstz=75$ GeV, $\mu=-120$ GeV, $M_2=3 |\mu|$, $\msbe=150$ GeV
and  $\phist=\pi/4$).}
\end{figure}
\begin{figure}[htb]
\begin{center}
\setlength{\unitlength}{1cm}
\setlength{\fboxsep}{0cm}
\begin{picture}(16,6)
\put(-1.75,0){\shit{7cm}{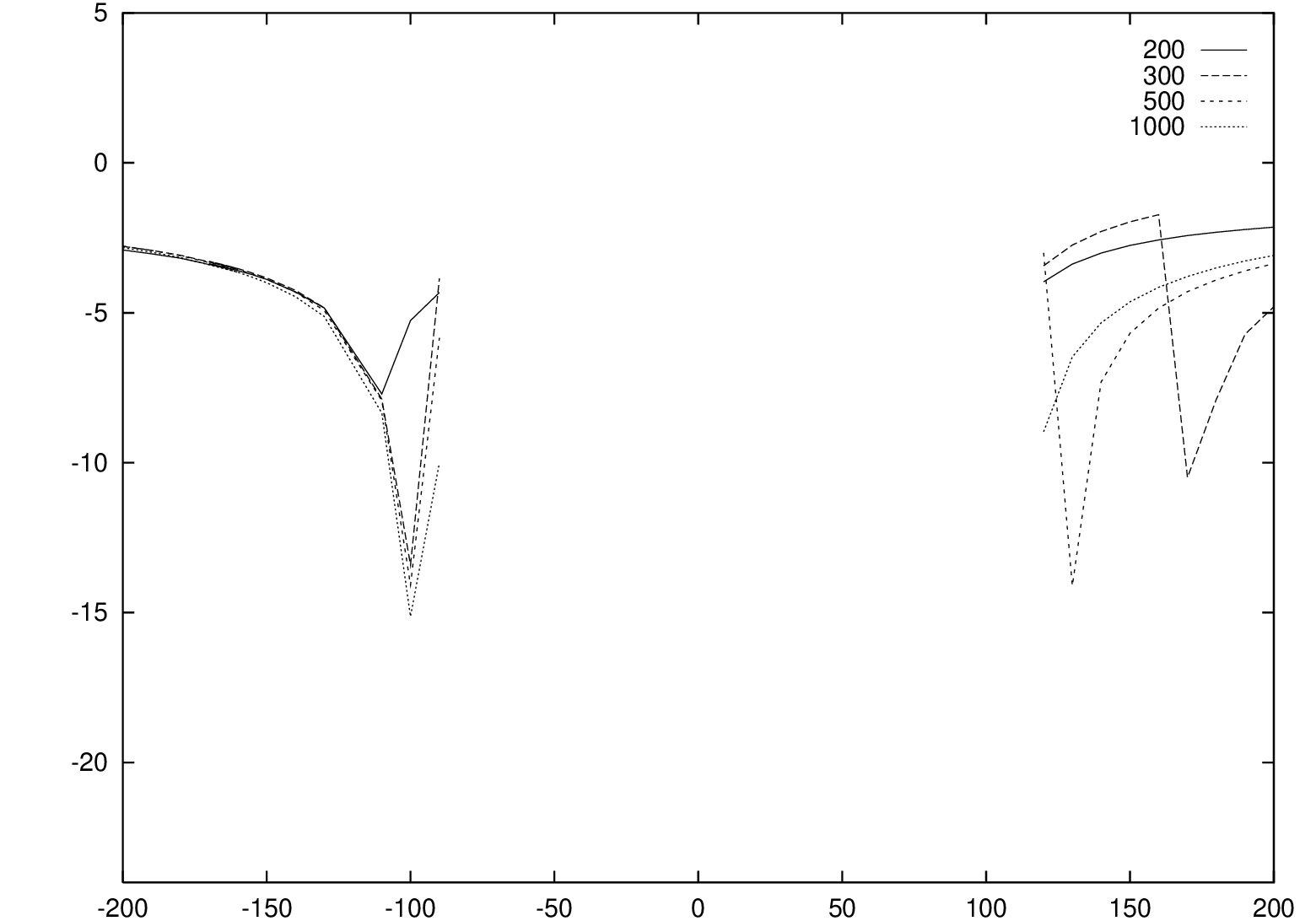}}
\put(0,7){\makebox(0,0){$\Delta , \%$}}
\put(3.25,0){\makebox(0,0){$\mu$ , GeV}}
\put(4,7){\makebox(0,0){Tevatron}}
\put(7.,0){\shit{7cm}{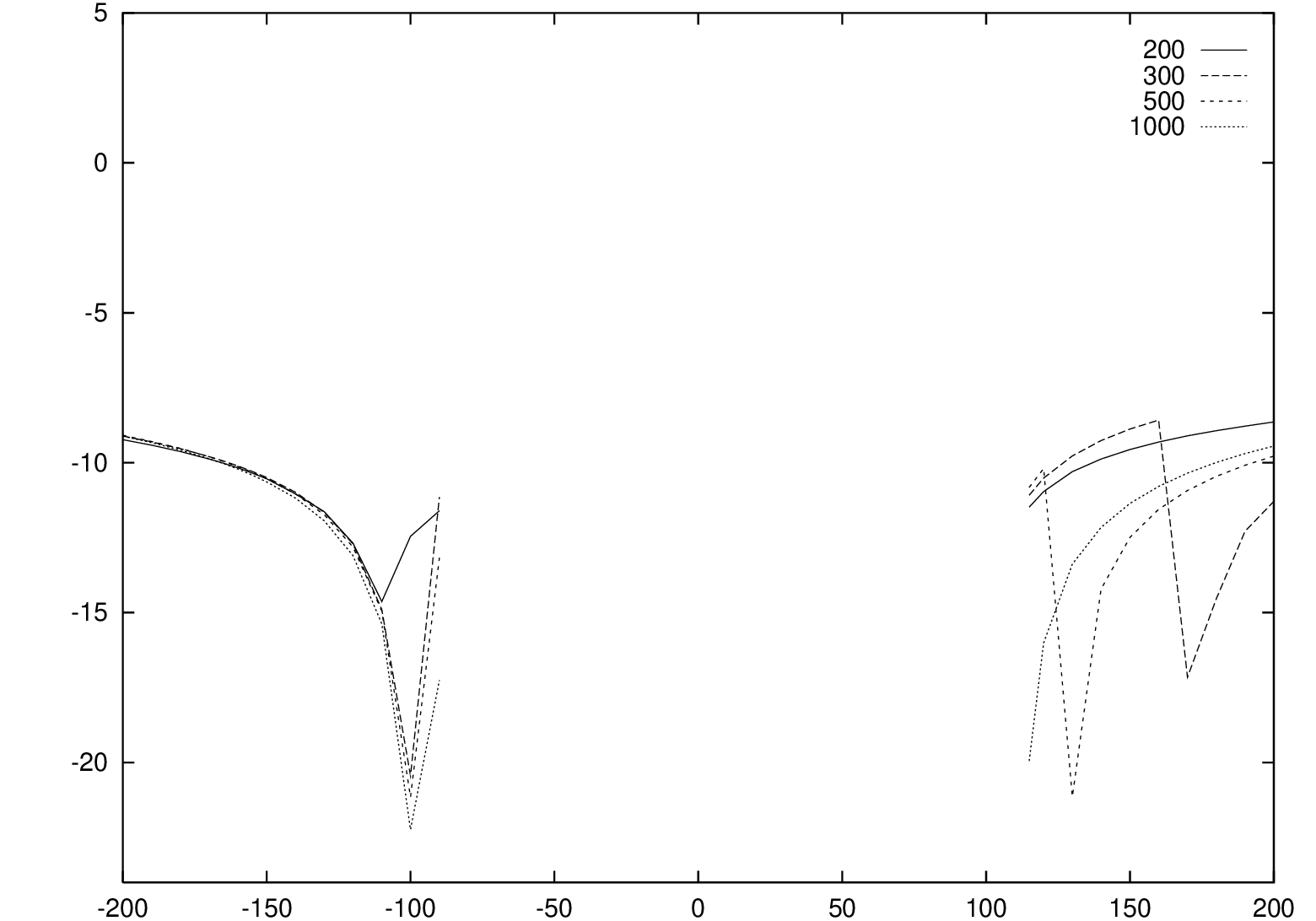}}
\put(8.5,7){\makebox(0,0){$\Delta , \%$}}
\put(12.5,0){\makebox(0,0){$\mu$ , GeV}}
\put(12,7){\makebox(0,0){LHC}}
\end{picture}
\end{center}
\caption{The variation of the relative correction $\Delta$
with $\mu$ and $M_2$ at the Tevatron and at the LHC within the MSSM
(with $\ma=150$ GeV, $\mstz=75$ GeV, $\msbe=1$ TeV, $\phist=0$
and $\tanb=0.7$).}
\end{figure}
\begin{figure}
\begin{center}
\setlength{\unitlength}{1cm}
\setlength{\fboxsep}{0cm}
\begin{picture}(16,20)
\put(-1.75,14.5){\shit{7cm}{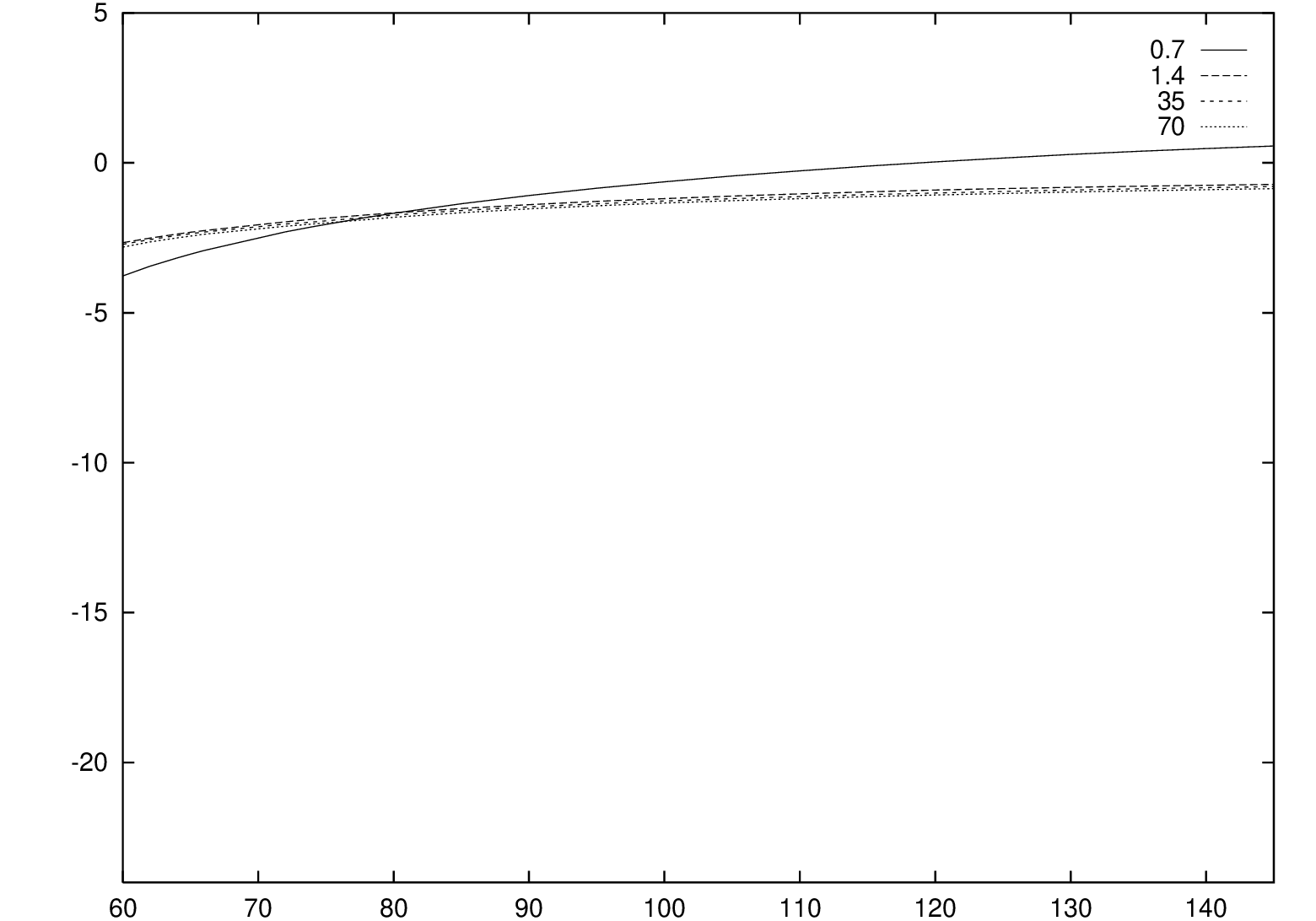}}
\put(0.5,21.5){\makebox(0,0){$\Delta , \%$ Tevatron}}
\put(3.25,14.5){\makebox(0,0){$\mstz$ , GeV}}
\put(3.25,15.5){\makebox(0,0){$a.)$}}
\put(7.,14.5){\shit{7cm}{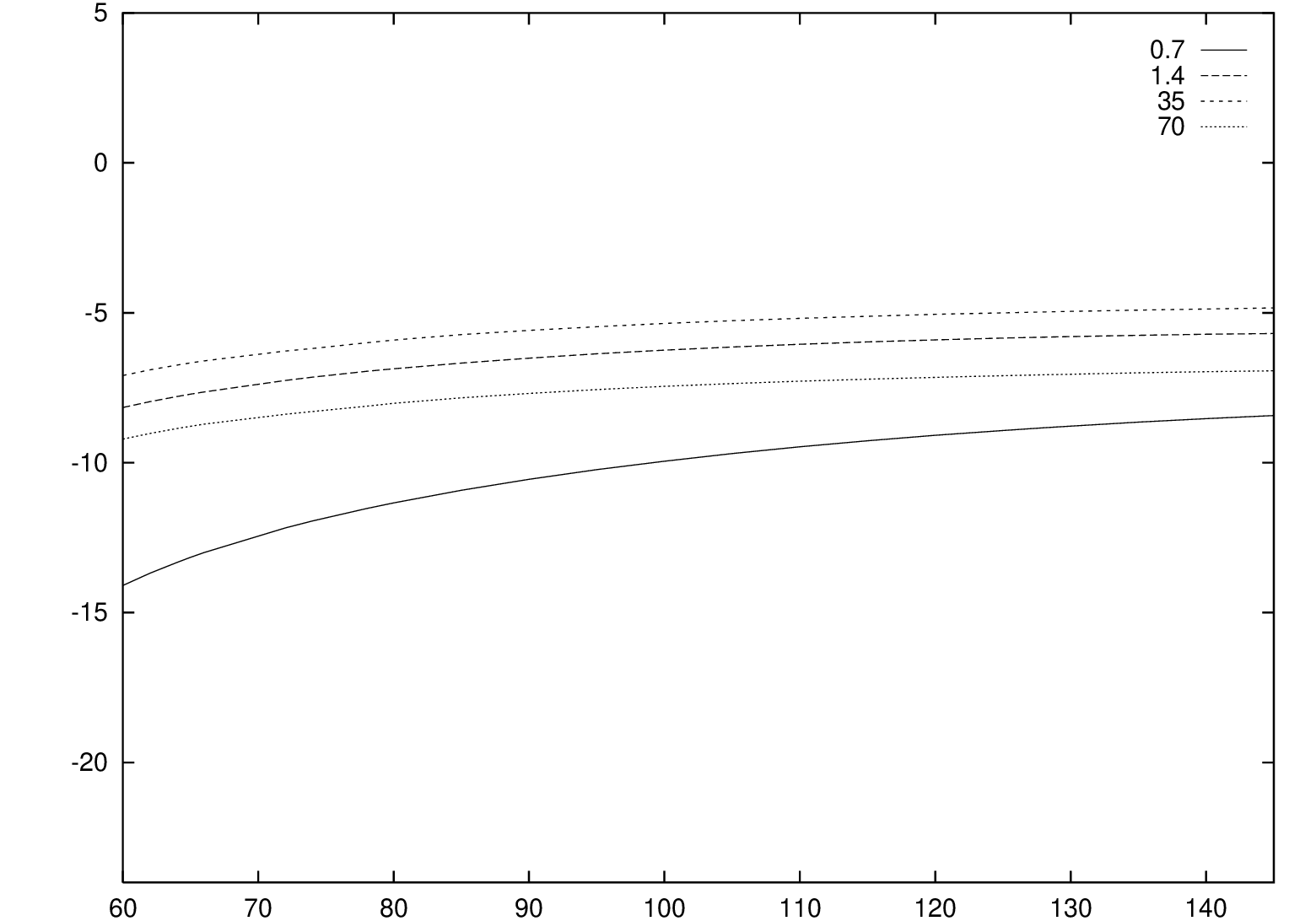}}
\put(9,21.5){\makebox(0,0){$\Delta , \%$ LHC}}
\put(12.5,14.5){\makebox(0,0){$\mstz$ , GeV}}
\put(-1.75,7.25){\shit{7cm}{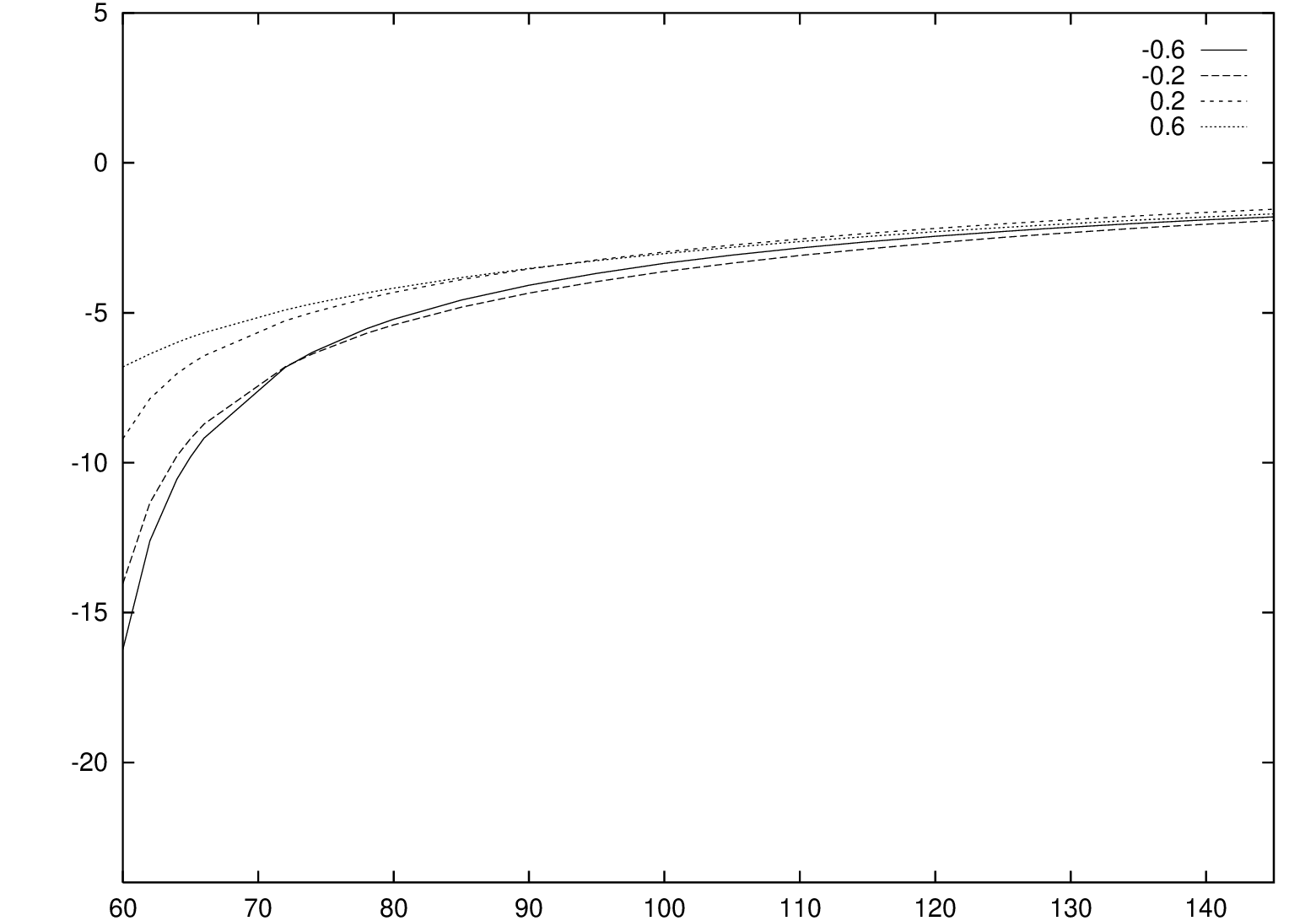}}
\put(3.25,7.25){\makebox(0,0){$\mstz$ , GeV}}
\put(3.25,8.25){\makebox(0,0){$b.)$}}
\put(7.,7.25){\shit{7cm}{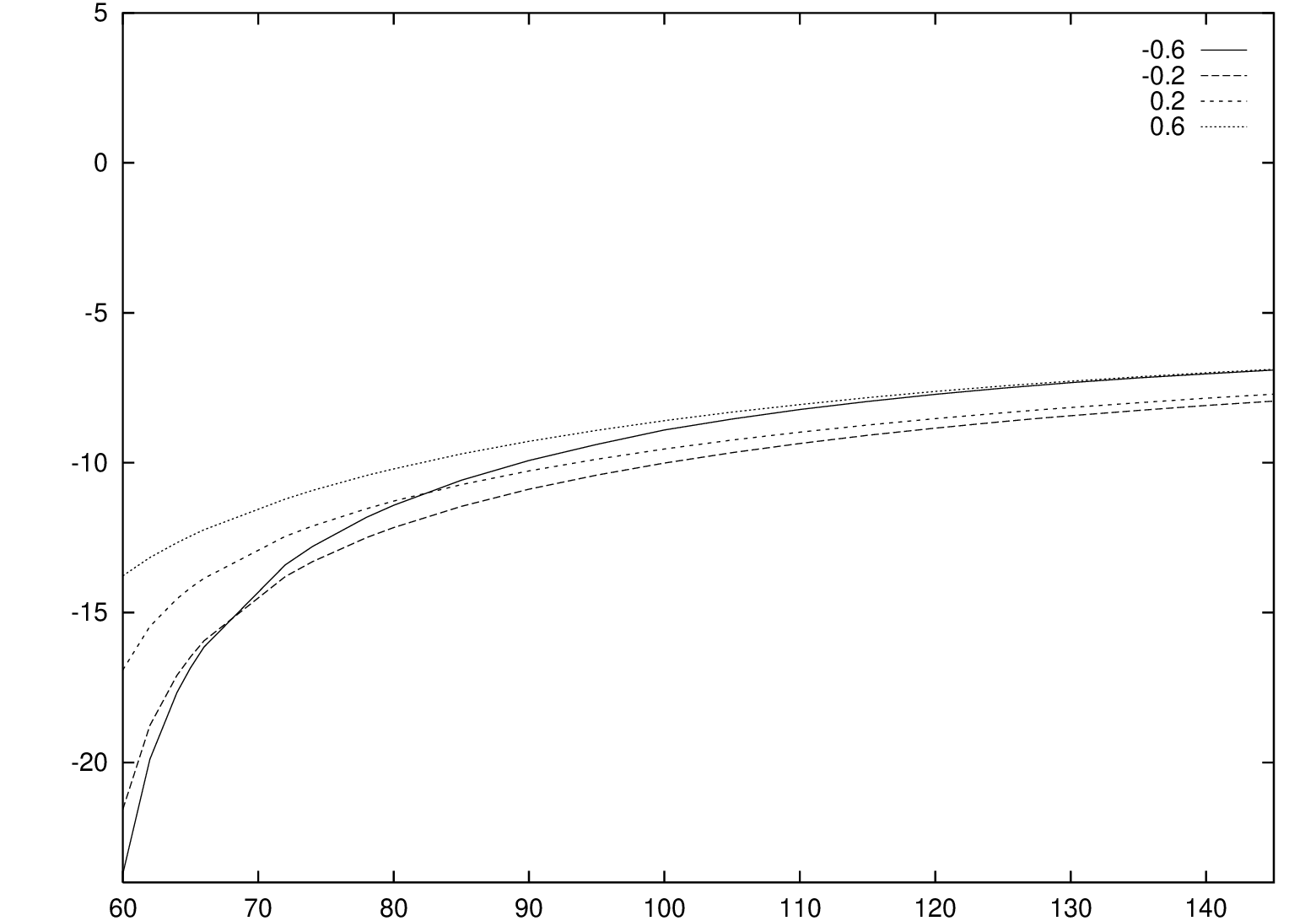}}
\put(12.5,7.25){\makebox(0,0){$\mstz$ , GeV}}
\put(-1.75,0){\shit{7cm}{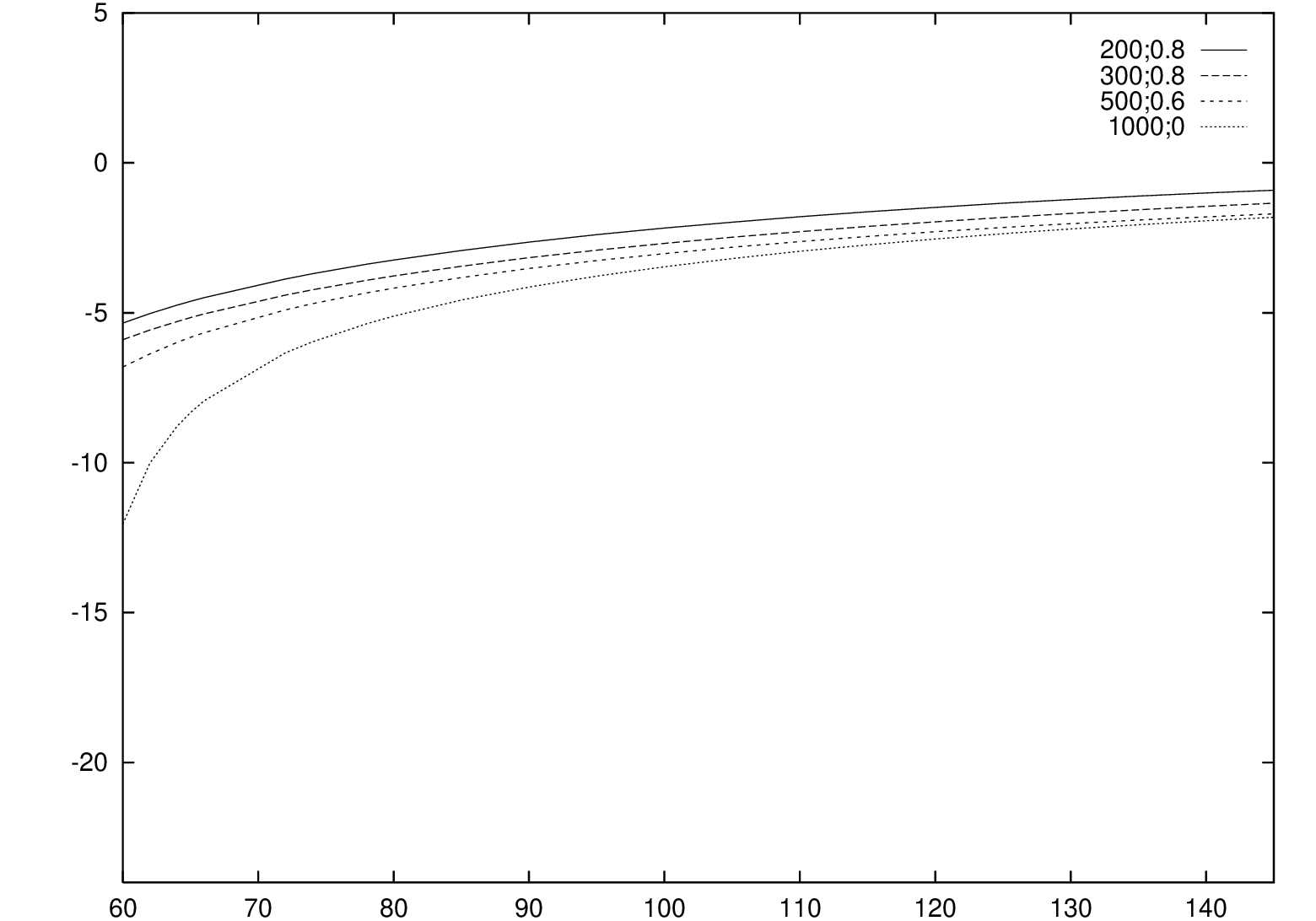}}
\put(3.25,0){\makebox(0,0){$\mstz$ , GeV}}
\put(3.25,1){\makebox(0,0){$c.)$}}
\put(7.,0){\shit{7cm}{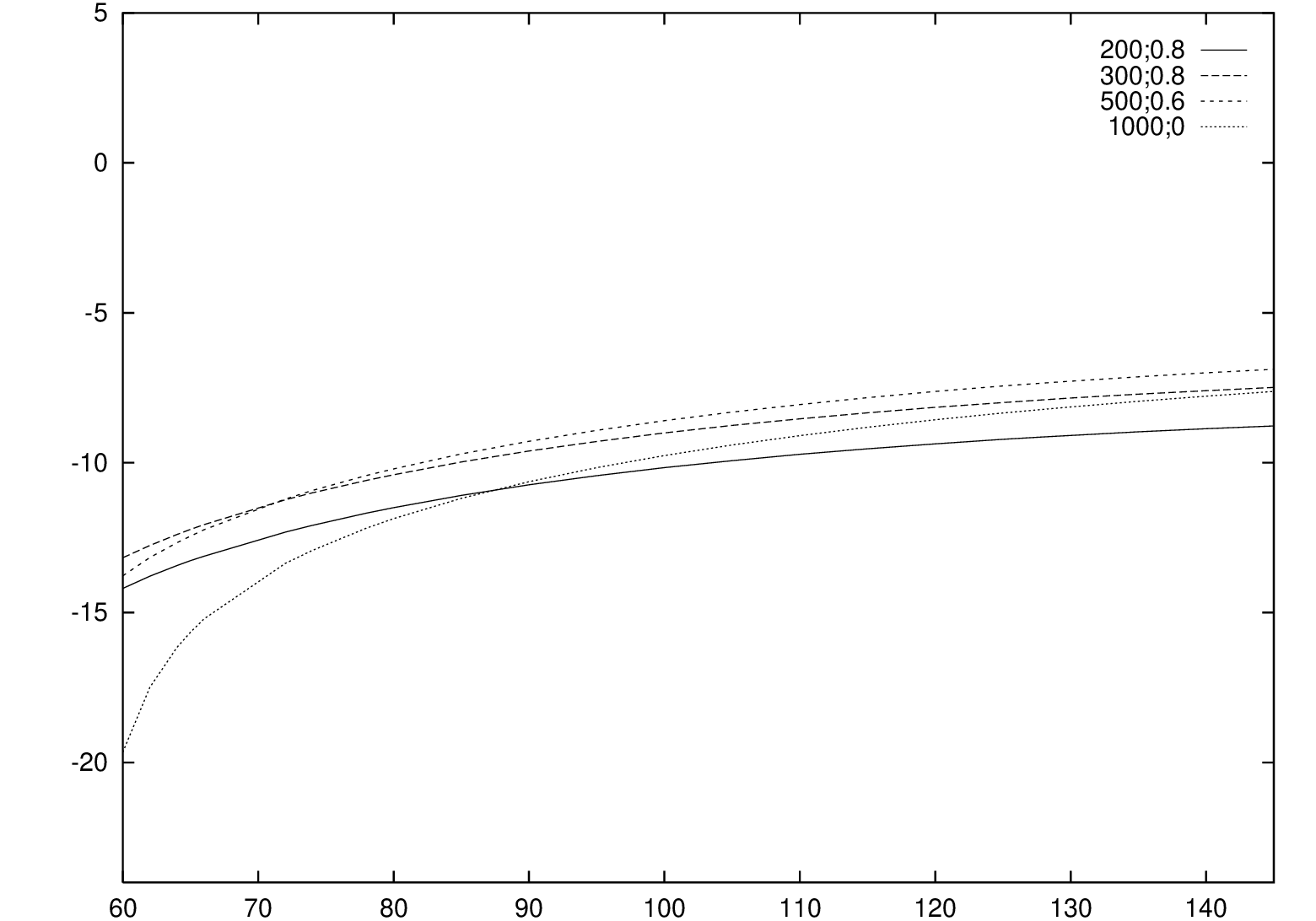}}
\put(12.5,0){\makebox(0,0){$\mstz$ , GeV}}
\end{picture}
\end{center}
\caption{The variation of the relative correction $\Delta$
with $\mstz$ at the Tevatron and at the LHC for different 
MSSM scenarios (with $\mu=-120$ GeV and $M_2=3 |\mu|$).}
\end{figure}
It is interesting to compare these effects with
the relative corrections obtained
when using the best $\chi^2$-fit results of a
global MSSM-fit to most recent electroweak precision
data~\cite{wolf}: 
$\Delta_{Tevatron}= -0.7 \%;-1.6 \%$ and
$\Delta_{LHC}= -4.3 \%;-3.2 \%$ with $\tanb = 1.6;35$, respectively.
As expected the effects are of the same order of magnitude
as the one obtained within the MSM. This is due to the decoupling behavior of
the MSSM in the limit of very heavy supersymmetric particles
which allow for the presence of supersymmetry in 
electroweak $Z$ pole observables measured at LEP 
and SLC even though at the present level of 
accuracy no convincing deviation from the MSM prediction
has been observed.

As an interesting alternative to the total top pair
production cross section we show in Fig.~19 and Fig.~20 the impact of
the G2HDM and the MSSM ${\cal O}(\alpha)$ contribution, respectively,
on the invariant mass distribution of
the produced top quark pair $d\sigma/d M_{\ttbar}$
\begin{equation}
\frac{d \sigma}{d M_{\ttbar}} = \sum_{ij=\qqbar,gg}\frac{2}{M_{\ttbar}}
\hat \sigma_{ij}(\hat s=\tau S)\;  \tau \frac{d L_{ij}}{d \tau}
\end{equation}
with $\tau=M_{\ttbar}^2/S$.
There we are especially interested in 
the distortion of the $\ttbar$ invariant mass due to the
$s$-channel Higgs-exchange diagrams in the gluon fusion subprocess
as a very characteristic
signature of the electroweak symmetry breaking sector
possibly observable at $pp$ colliders.
As can be seen in Figs.~19,20 at the LHC a significant distortion 
arises for Higgs-boson masses $\mgh>2 \mt$ for a sufficiently small
Higgs-decay width $\Gamma_{H^0}$.
\begin{figure}[htb]
\begin{center}
\setlength{\unitlength}{1cm}
\setlength{\fboxsep}{0cm}
\begin{picture}(16,14)
\put(-1.75,7.25){\shit{7cm}{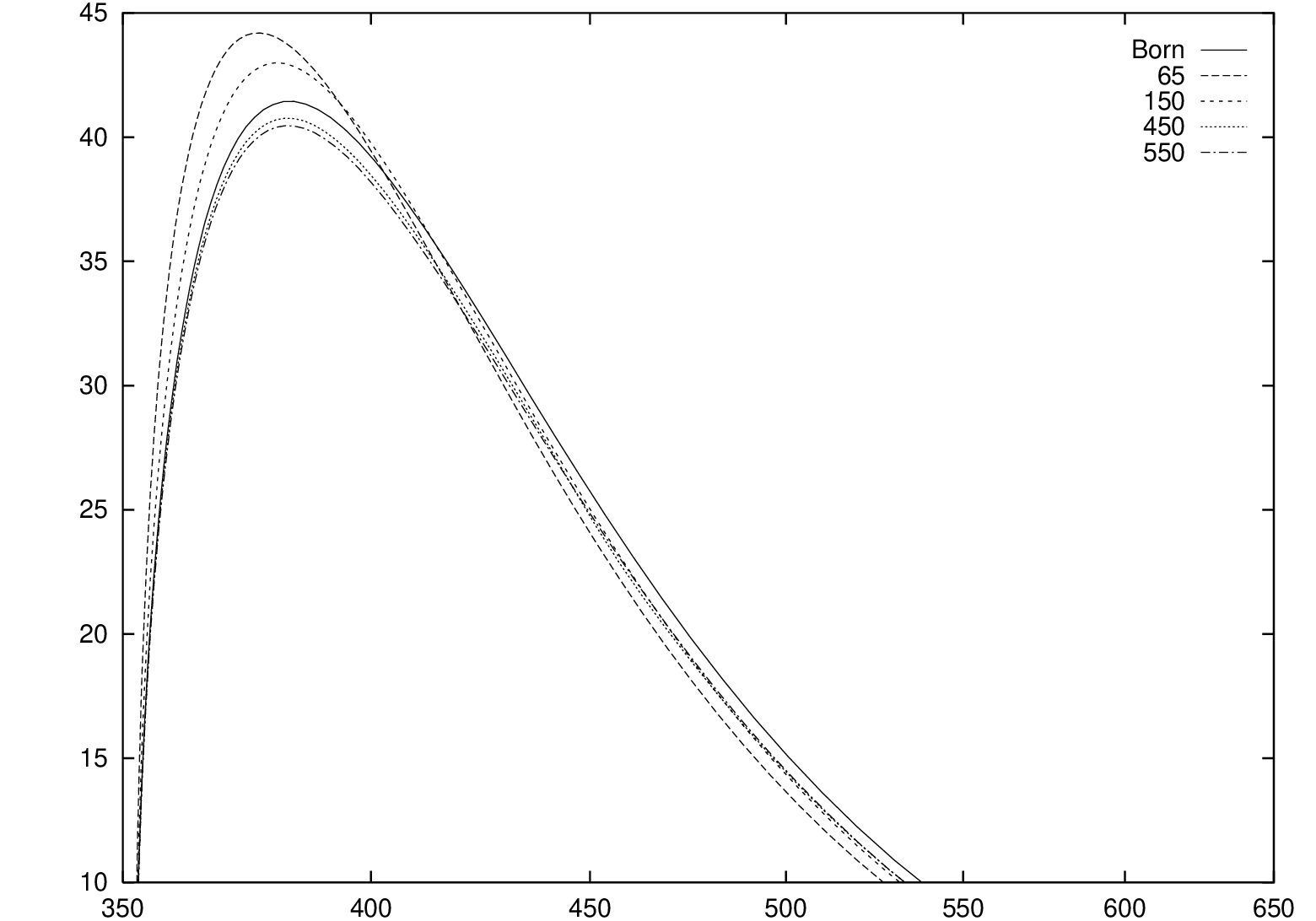}}
\put(3.25,7.25){\makebox(0,0){$M_{\ttbar}$ , GeV}}
\put(1,14.25){\makebox(0,0){$d\sigma / d M_{\ttbar} , 
\mbox{fb} \;\mbox{GeV}^{-1}$}}
\put(4,14.25){\makebox(0,0){Tevatron}}
\put(4,13.25){\makebox(0,0){$\tanb = 0.7$}}
\put(7.,7.25){\shit{7cm}{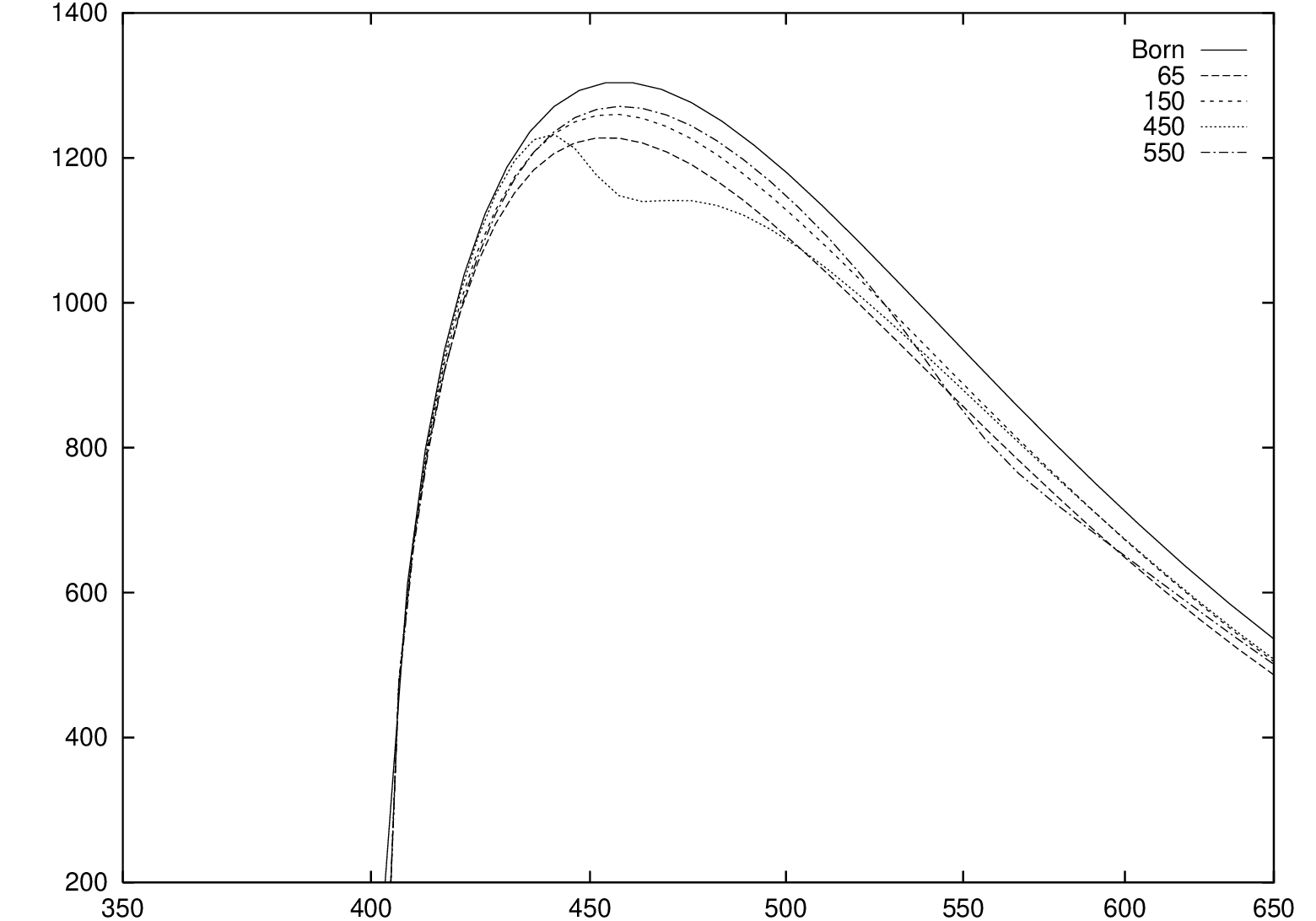}}
\put(12.5,7.25){\makebox(0,0){$M_{\ttbar}$ , GeV}}
\put(10,14.25){\makebox(0,0){$d\sigma / d M_{\ttbar} , 
\mbox{fb}\; \mbox{GeV}^{-1}$}}
\put(13,14.25){\makebox(0,0){LHC}}
\put(9.5,13.25){\makebox(0,0){$\tanb = 0.7$}}
\put(-1.75,0){\shit{7cm}{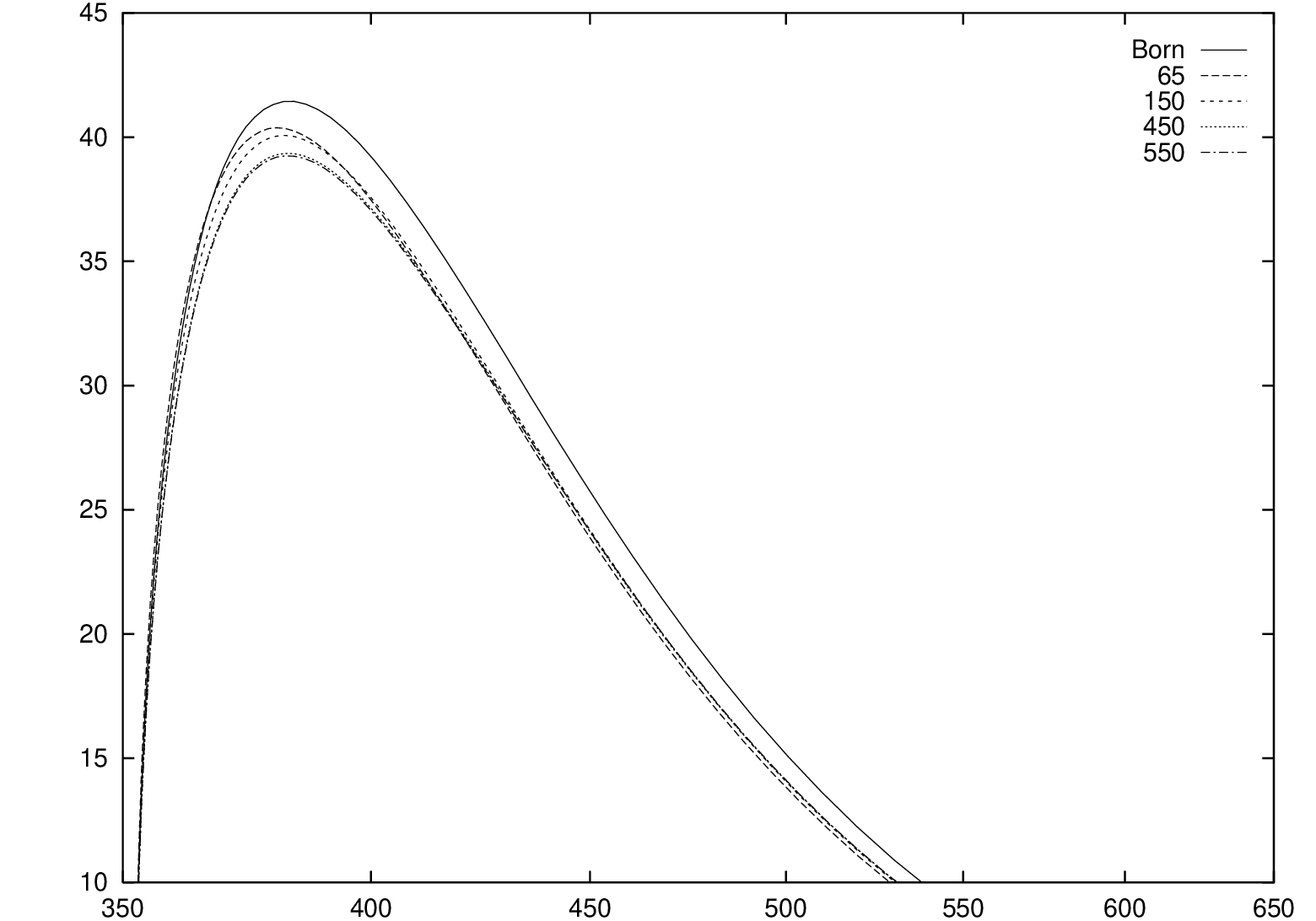}}
\put(3.25,0){\makebox(0,0){$M_{\ttbar}$ , GeV}}
\put(4,6){\makebox(0,0){$\tanb = 70$}}
\put(7.,0){\shit{7cm}{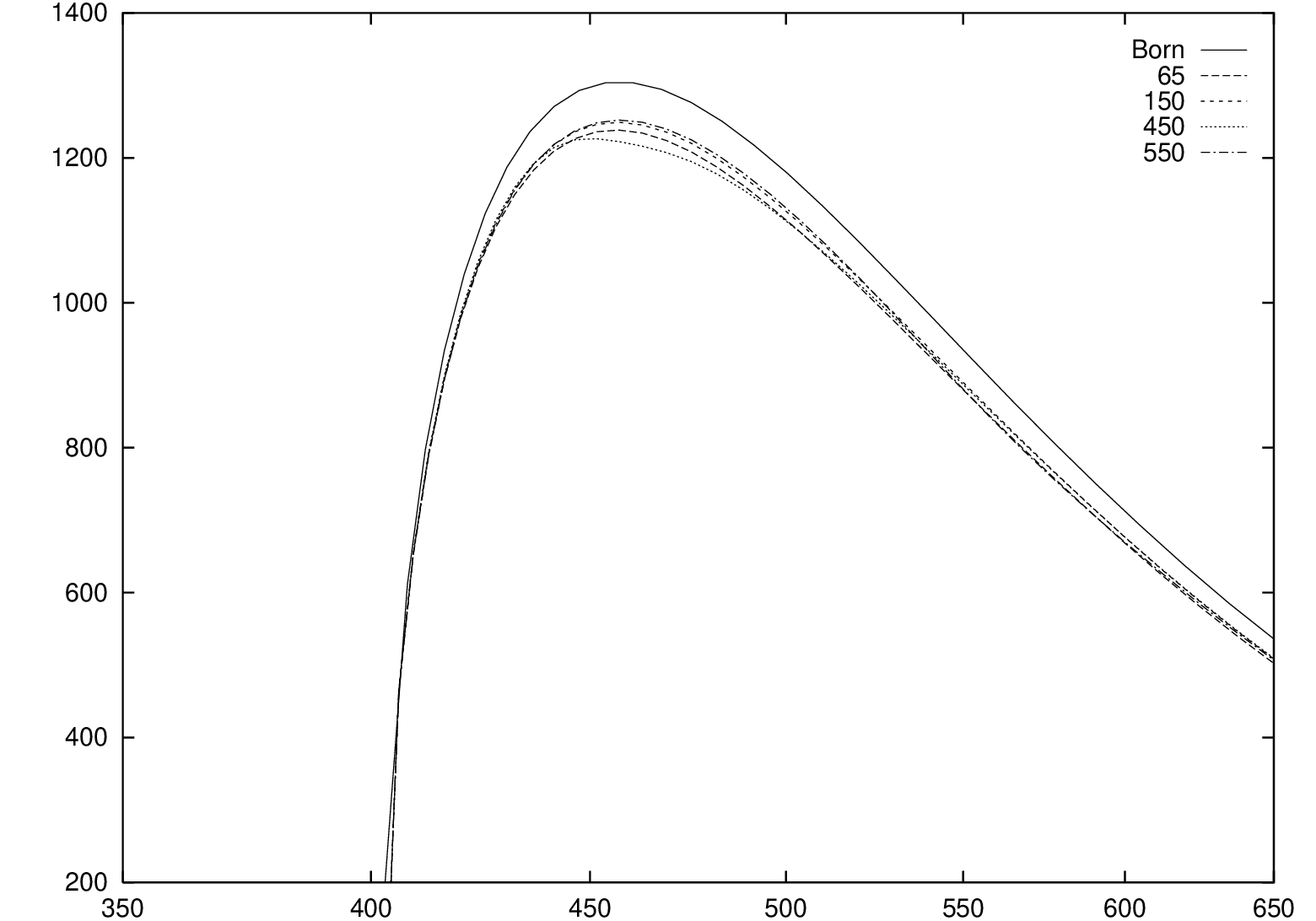}}
\put(12.5,0){\makebox(0,0){$M_{\ttbar}$ , GeV}}
\put(9.5,6){\makebox(0,0){$\tanb = 70$}}
\end{picture}
\end{center}
\caption{The invariant mass $M_{\ttbar}$ distribution within the G2HDM
for different values of $\mgh$ and $\tanb$ (with $\mkh=45$ GeV, $\ma=50$ GeV,
$\mhp=150$ GeV and $\alpha=\pi/2$).}
\end{figure}
\begin{figure}[htb]
\begin{center}
\setlength{\unitlength}{1cm}
\setlength{\fboxsep}{0cm}
\begin{picture}(16,14)
\put(-1.75,7.25){\shit{7cm}{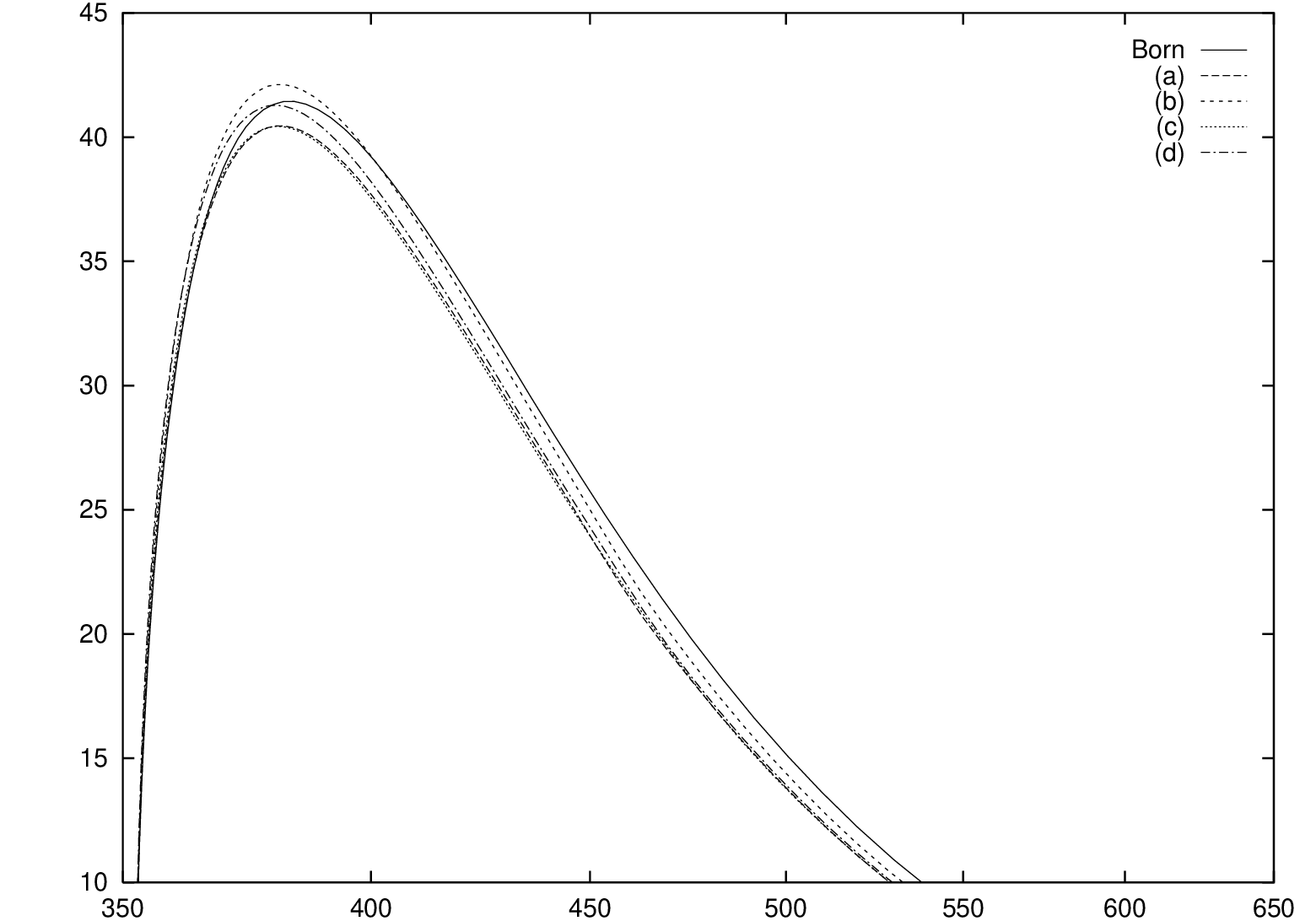}}
\put(3.25,7.25){\makebox(0,0){$M_{\ttbar}$ , GeV}}
\put(1,14.25){\makebox(0,0){$d\sigma / d M_{\ttbar} , 
\mbox{fb}\; \mbox{GeV}^{-1}$}}
\put(4,14.25){\makebox(0,0){Tevatron}}
\put(4,13.25){\makebox(0,0){$\tanb = 0.7$}}
\put(7.,7.25){\shit{7cm}{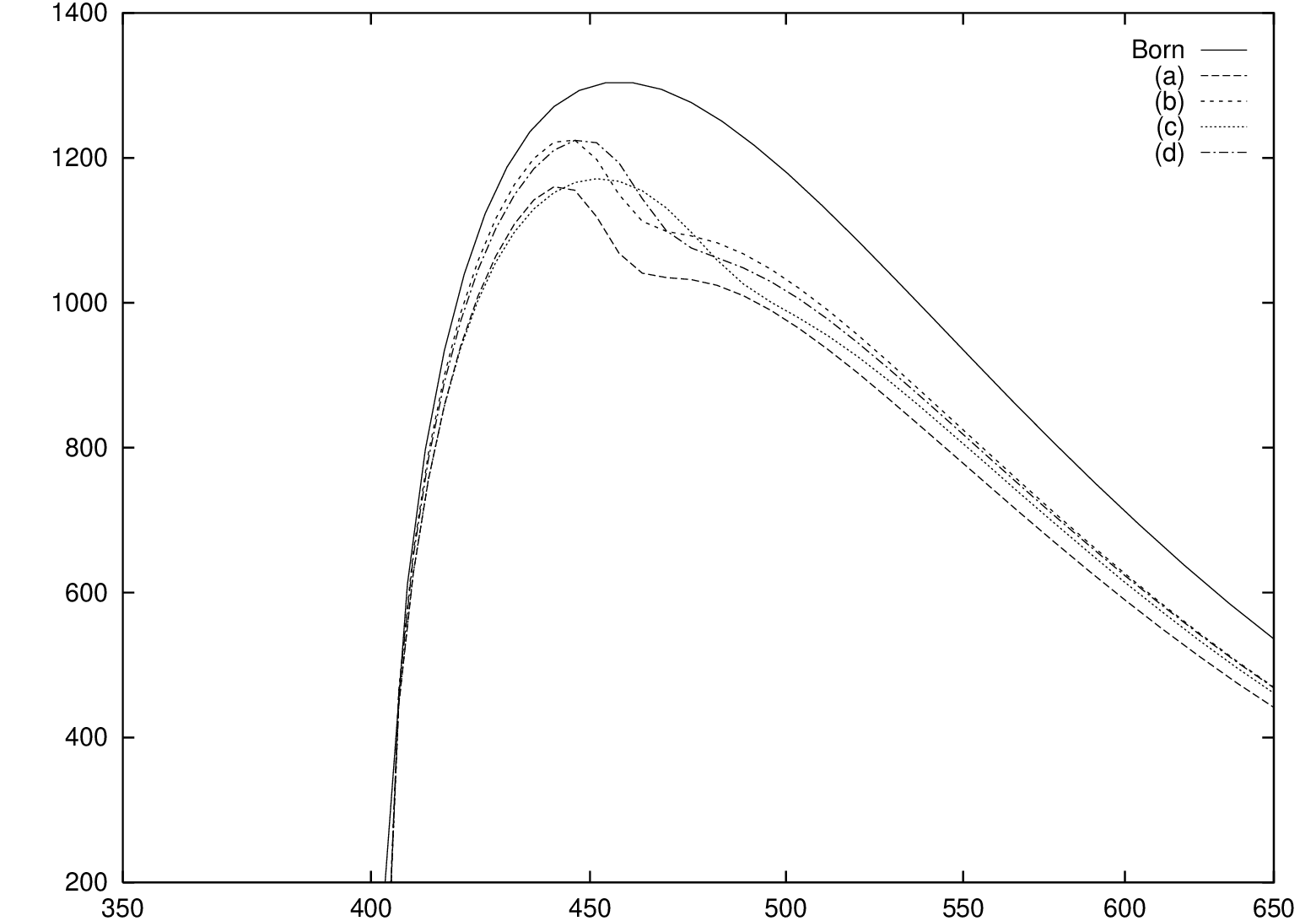}}
\put(12.5,7.25){\makebox(0,0){$M_{\ttbar}$ , GeV}}
\put(10,14.25){\makebox(0,0){$d\sigma / d M_{\ttbar} , 
\mbox{fb} \;\mbox{GeV}^{-1}$}}
\put(13,14.25){\makebox(0,0){LHC}}
\put(9.5,13.25){\makebox(0,0){$\tanb = 0.7$}}
\put(-1.75,0){\shit{7cm}{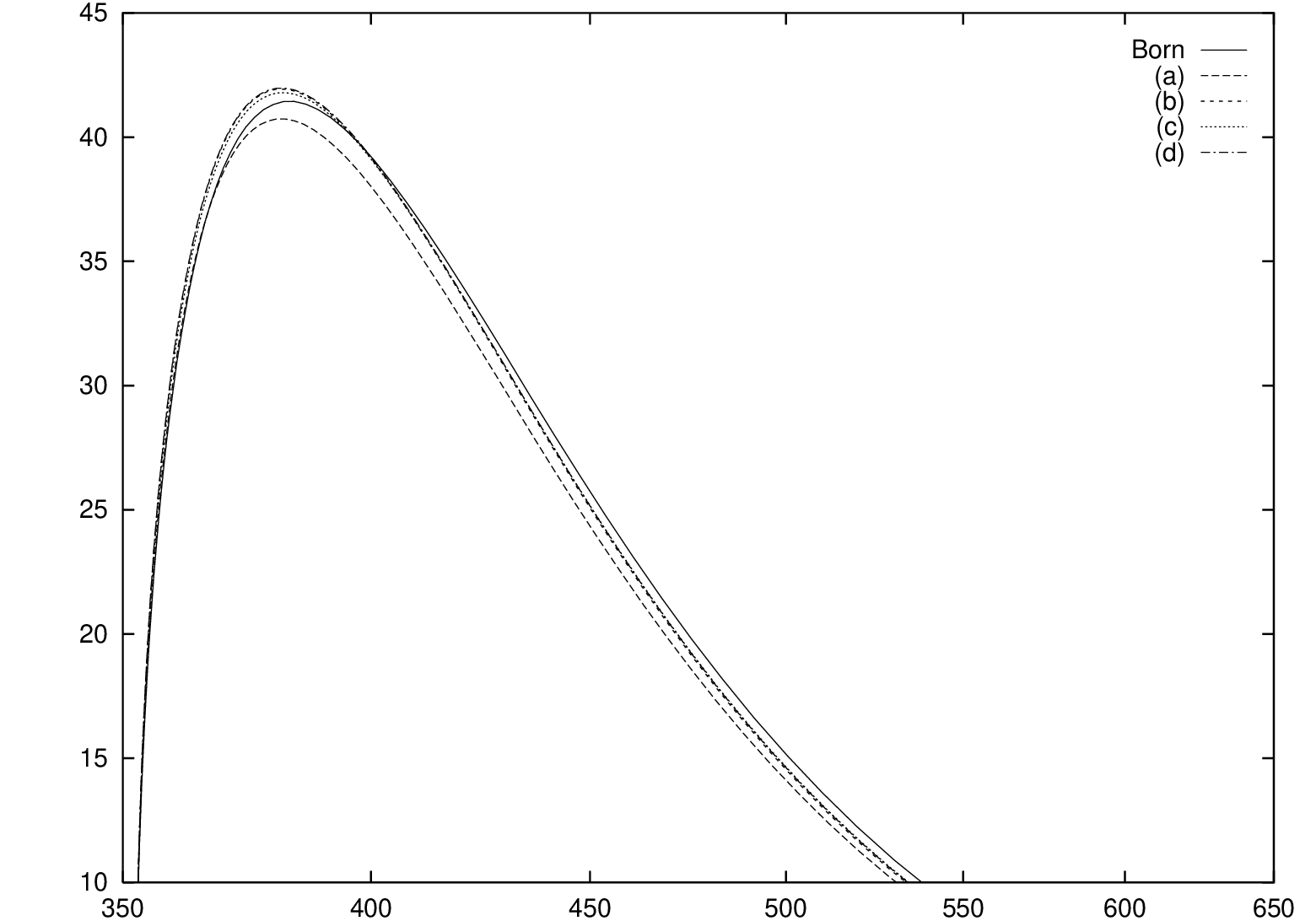}}
\put(3.25,0){\makebox(0,0){$M_{\ttbar}$ , GeV}}
\put(4,6){\makebox(0,0){$\tanb = 70$}}
\put(7.,0){\shit{7cm}{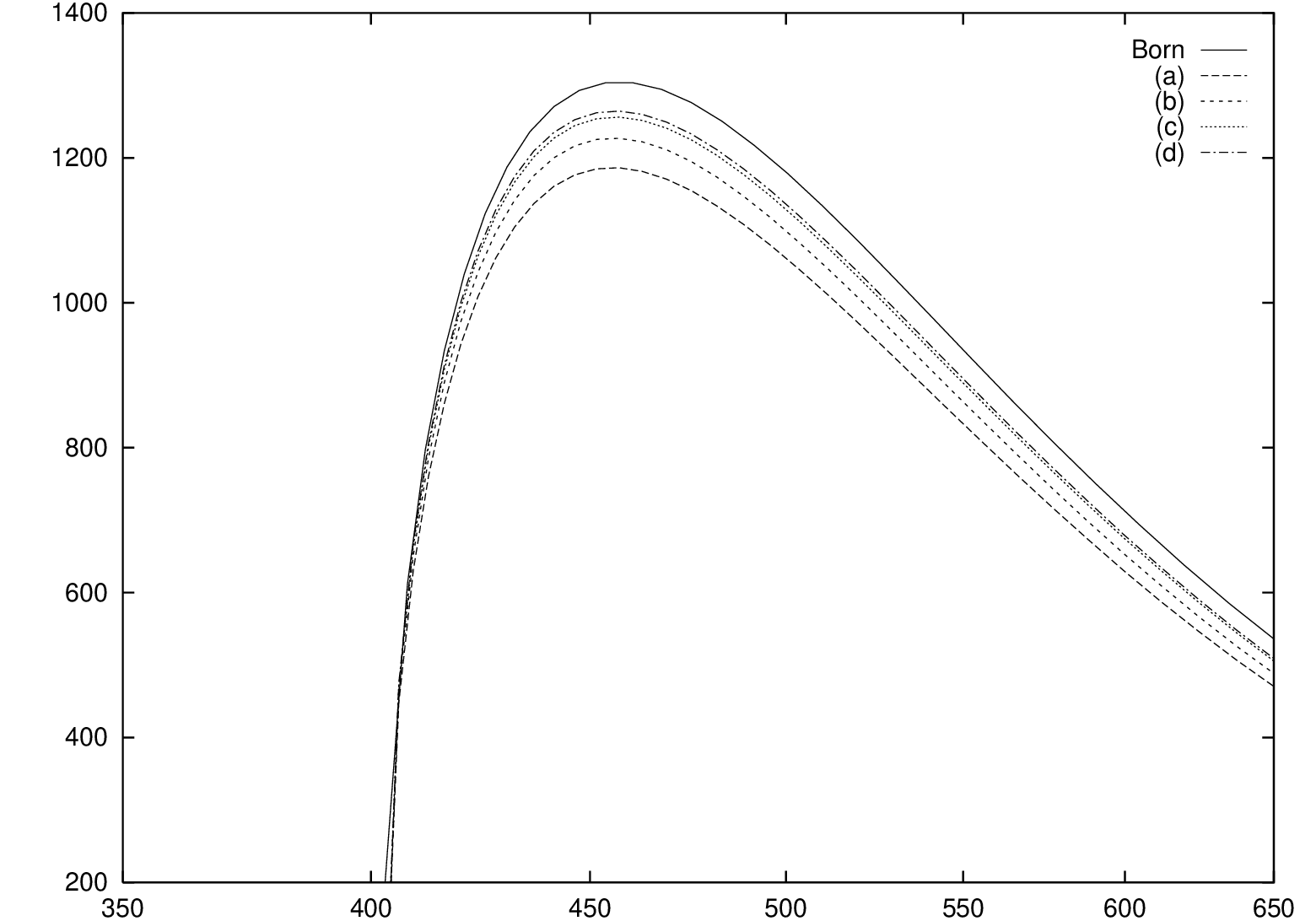}}
\put(12.5,0){\makebox(0,0){$M_{\ttbar}$ , GeV}}
\put(9.5,6){\makebox(0,0){$\tanb = 70$}}
\end{picture}
\end{center}
\caption{The invariant mass $M_{\ttbar}$ distribution within different MSSM
scenarios (with $\ma=450$ GeV, $M_2=3 |\mu|$ and 
with (a) $\mstz=50$, $\msbe=150$ GeV, $\phist=\pi/4$, $\mu=-120$ GeV,
(b) same as (a) but with $\mstz=75$ GeV, 
(c) $\mstz=75$, $\msbe=400$ GeV, $\phist=-\pi/4$, $\mu=150$ GeV
and (d) $\mstz=75$, $\msbe=800$ GeV, $\phist=0$, $\mu=150$ GeV).}
\end{figure}
\section{Conclusions}
We have studied the (virtual) effects of the complete ${\cal O}(\alpha)$
contribution within the G2HDM and the MSSM 
to the main top pair production mechanisms, $\qqa$ and gluon fusion, 
at future hadron colliders:
the upgraded Tevatron with $\sqrt{S} = 2$ TeV and the LHC with
$\sqrt{S} = 14$ TeV.
Typically the Born-cross sections are reduced
by the order of several percent but in exceptional regions of the
parameter space, that is when $\mt\approx \mstz+\mneut ; m_b+\mhp$,
the radiative corrections can be considerably enhanced: up
to about $40 \%$ and $30 \%$ at the upgraded Tevatron and the
LHC, respectively.
In the gluon fusion subprocess an interesting Higgs-specific
signature occurs when due to the $s$-channel Higgs-exchange diagrams
the $\ttbar$ invariant mass distribution at the LHC 
can be significantly distorted 
for Higgs-boson masses $\mgh > 2 \mt$ and a sufficiently small 
Higgs-decay width $\Gamma_{H^0}$.  
We conclude that 
provided the intrinsic QCD uncertainties can be considerably
reduced there is potential for electroweak precision studies
in strong processes at future hadron colliders.
\acknowledgements
W.M.~M.~gratefully acknowledges the opportunity to participate
in the summer visitors program of the Theoretical Physics 
Department of FNAL and the kind hospitality extended to him 
during the visit when parts of this work were done. 
 
The Fermi National Accelerator Laboratory is operated by 
the Universities Research Association, Inc., under contract
DE-AC02-76CHO3000 with the United States Department of Energy.


\begin{references}
%

\bibitem{cdf}
F.~Abe {\it et al.} (CDF collaboration), 
Phys.~Rev.~Lett. {\bf 74}, 2626 (1995).

\bibitem{dzero}
S.~Abachi {\it et al.} (D\O\ collaboration), 
Phys.~Rev.~Lett. {\bf 74}, 2632 (1995).

\bibitem{msm}
A.~Salam, in {\em Elementary Particle Theory}, 8th Nobelsymposium,
Wiley N.Y. (1969);
S.~Weinberg, Phys.~Rev.~Lett. {\bf 19}, 1264 (1967);
S.L.~Glashow, Nucl.~Phys. {\bf 22}, 579 (1961);
S.L.~Glashow, J.~Iliopoulos and L.~Maiani, Phys.~Rev. {\bf D2}, 1285 (1970).

\bibitem{hig}
P.W.~Higgs, Phys.~Lett. {\bf 12}, 131 (1964),
Phys.~Rev.~Lett. {\bf 13}, 508 (1964) and
Phys.~Rev.~Lett. {\bf 145}, 1156 (1966);
T.W.B.~Kibble, Phys.~Rev. {\bf 155}, 1554 (1967);
R.~Brout and F.~Englert, Phys.~Rev.~Lett. {\bf 13}, 321 (1964).

\bibitem{wine}
P.C.~Bhat, for the D\O\ collaboration, talk presented at the
{\em Wine and Cheese} Seminar at Fermilab, February 1997.

\bibitem{ewwg}
The LEP Electroweak Working Group and the SLD Heavy Flavour
Group, LEPEWWG/97-01, April 1997.

\bibitem{cdfnew}
D.S.~Kestenbaum, for the CDF collaboration, talk presented
at the {\em 16th International Conference on Physics in Collision} (PIC96),
Mexico City, June 1996, published in the proceedings,
Fermilab-Conf-97/016-E.

\bibitem{berg}
E.L.~Berger and H.~Contopaganos, Phys.~Rev. {\bf D54}, 3085 (1996).

\bibitem{cat}
S.~Catani, M.L.~Mangano, P.~Nason and L.~Trentadue,
Phys.~Lett. {\bf B378}, 329 (1996).

\bibitem{laen}
E.~Laenen, J.~Smith and W.L.~van Neerven, Nucl.~Phys. {\bf B369},
543 (1992) and
Phys.~Lett. {\bf B321}, 254 (1994).

\bibitem{diplpubsm}
W.~Beenakker, A.~Denner, W.~Hollik, R.~Mertig, 
T.~Sack and D.~Wackeroth, Nucl.~Phys. {\bf B411}, 343 (1994).

\bibitem{topacc}
Report of the {\em tev2000} Study Group, D.~Amedei and R.~Brock (eds.),
FERMILAB-Pub-96/082, April 1996.

\bibitem{hunters}
J.F.~Gunion, H.E.~Haber, G.L.~Kane and S.~Dawson,
{\em The Higgs Hunter's Guide},
Addison-Wesley Reading, MA (1990) (Erratum: hep-ph/9302272, February 1993).

\bibitem{susy}
For a review see, e.g.~,~\cite{haka} and
J.~Wess and J.~Bagger, {\em Supersymmetry and Supergravity}, Princeton University
Press 1983;
P.~Fayet and S.~Ferrara, Phys.~Rep. {\bf 32}, 249 (1977) and
H.-P.~Nilles, Phys.~Rep. {\bf 110}, 1 (1984).

\bibitem{ttsm}
C.~Kao, G.A.~Ladinsky and C.P.~Yuan, FSU-HEP-940508, June 1994, published in
the {\em DPF94 Conference} proceedings.

\bibitem{ttmssm}
A.~Stange and S.~Willenbrock, Phys.~Rev. {\bf D49}, 1354 (1994).

\bibitem{ttthdm}
H.Y.~Zhou, C.S.~Li and Y.P.~Kuang, Phys.~Rev. {\bf D55}, 4412 (1997).

\bibitem{dipl}
D.~Wackeroth, Diploma Thesis, Technische Universit\"at M\"unchen, April 1992.

\bibitem{susyqcd}
S.~Alam, K.~Hagiwara and S.~Matsumoto, Phys.~Rev. {\bf D55}, 1307 (1997);
Z.~Sullivan, hep-ph/9611302, November 1996, to be published in Phys.Rev.D;
C.S.~Li, H.Y.~Zhou, Y.L.~Zhu and J.M.~Yang, Phys.~Lett. {\bf B379}, 135 (1996);
C.S.~Li, B.Q.~Hu, J.M.~Yang and C.G.~Hu, Phys.~Rev. {\bf D52}, 5014 (1995)
(Erratum: Phys.~Rev. {\bf D53}, 4112(E) (1996)).

\bibitem{qcdew}
J.~Kim, J.L.~Lopez, D.V.~Nanopoulos and R.~Rangarajan, Phys.~Rev. {\bf D54},
4364 (1996).

\bibitem{sqcdlhc}
H.Y.~Zhou and C.S.~Li, Phys.~Rev. {\bf D55}, 4421 (1997).

\bibitem{ewlike}
J.M.~Yang and C.S.~Li, Phys.~Rev. {\bf D54}, 4380 (1996)
(Erratum: Phys.~Rev. {\bf D54}, 3671(E) (1996)).

\bibitem{higexp}
J.P.~Martin, talk presented at the ICHEP96, Warsaw, 25-31 July 1996,
to appear in the proceedings. 

\bibitem{higtheo}
B.W.~Lee, C.~Quigg and H.B.~Thacker, Phys.~Rev. {\bf D16}, 1519 (1977).

\bibitem{pdg}
The Particle Data Group, Phys.~Rev. {\bf D54}, no.1 (1996).

\bibitem{wmass}
M.~Demarteau, talk presented at the {\em DPF96 Conference}, Minneapolis,
August 1996, to appear in the proceedings.

\bibitem{mrel}
J.~Ellis, G.~Ridolfi, F.~Zwirner, Phys.~Lett. {\bf B262}, 477 (1991) and
Phys.~Lett. {\bf B257}, 83 (1991).

\bibitem{mcharged}
M.A.~Diaz and H.E.~Haber, Phys.~Rev. {\bf D45}, 4246 (1992).

\bibitem{guka}  
J.F.~Gunion and H.E.~Haber, Nucl.~Phys. {\bf B272}, 1 (1986),
Nucl.~Phys. {\bf B278}, 449 (1986) and Nucl.~Phys. {\bf B307}, 445 (1988)
(Errata: hep-ph/9301205, January 1993).

\bibitem{haka}  
H.E.~Haber and G.L.~Kane, Phys.~Rep. {\bf 117},75 (1985).

\bibitem{mrsa}
A.D.~Martin, R.G.~Roberts and W.J.~Stirling, Phys.~Rev. {\bf D50}, 6734 (1994).

\bibitem{wolf}
W.de~Boer, A.~Dabelstein, W.~Hollik, W.M.~M\"osle and U.~Schwickerath,
IEKP-KA/96-08, hep-ph/9609209, November 1996, to be published in Phys.Rev.D.

\end{references}
\end{document}